\documentclass[11pt]{article}
\usepackage{jheppub}
\usepackage{epstopdf}
\usepackage{enumitem}
\usepackage{tcolorbox}

\usepackage{jheppub}
\usepackage{mathrsfs}
\usepackage{psfrag}
\usepackage{color}
\usepackage{hyperref}

\usepackage{slashed}
\usepackage{feynmp-auto}
\usepackage{simplewick}
\usepackage{cancel}

\usepackage{todonotes}

\usepackage{amsmath}
\usepackage{amsfonts}
\usepackage{graphicx}
\usepackage{amssymb}
\usepackage{xcolor}

\usepackage{mathtools}

\usepackage{amsmath,bbm,array,amsfonts,graphicx,wrapfig,arydshln,lscape,float,multirow,longtable,rotating,makecell}
\usepackage{url}

\DeclareMathAlphabet\mathbfcal{OMS}{cmsy}{b}{n}

\newcommand{\ra}{\rangle}

\newcommand{\p}{\partial}
\newcommand{\ve}{\varepsilon}
\newcommand{\la}{\langle}

\newcommand{\eq}{\begin{equation}}
\newcommand{\eqe}{\end{equation}}
\newcommand{\eqa}{\begin{eqnarray}}
\newcommand{\eqae}{\end{eqnarray}}
\newcommand{\nn}{\nonumber}
\newcommand{\bn}{\begin{enumerate}}
\newcommand{\en}{\end{enumerate}}

\newcommand{\eqc}[1]{(\ref{#1})}

\def\identity{{\rlap{1} \hskip 1.6pt \hbox{1}}}
\def\iden{\identity}

\def\CO{{\mathcal O}}

\def\IC{\mathbb{C}}

\def\CA{{\mathcal A}}
\def\CB{{\mathcal B}}

\def\CG{{\mathcal G}}

\def\CL{{\mathcal L}}

\def\CO{{\mathcal O}}
\def\CP{{\mathcal P}}

\def\a{\alpha}
\def\b{\beta}
\def\g{\gamma}
\def\e{\epsilon}
\def\ve{\varepsilon}
\def\z{\zeta}
\def\th{\theta}

\def\k{\kappa}
\def\l{\lambda}
\def\m{\mu}
\def\n{\nu}

\def\s{\sigma}

\def\G{\Gamma}

\def\O{\Omega}

\def\half{\frac{1}{2}}

\def\p{\partial}

\def\identity{{\rlap{1} \hskip 1.6pt \hbox{1}}}

\def\det{{\rm det}}

\newcommand{\bfig}{\begin{figure}}
\newcommand{\efig}{\end{figure}}

\def\lan{{\langle}}
\def\abs#1{{\left| #1 \right|}}

\def\bl#1\el{\begin{align} #1 \end{align}}
\def\bg#1\eg{\begin{gather} #1 \end{gather}}
\newcommand{\fig}[1]{figure \ref{#1}}
\def\bld#1\eld{\begin{aligned} #1 \end{aligned}}
\def\bgd#1\egd{\begin{gathered} #1 \end{gathered}}

\newcommand{\bra}[1]{\langle{#1}|}
\newcommand{\ket}[1]{|{#1}\rangle}

\newcommand{\fbra}[1]{ ({#1} |}
\newcommand{\fket}[1]{ | {#1} )}

\newcommand{\sbra}[1]{ [{#1} |}
\newcommand{\sket}[1]{ | {#1} ]}

\renewcommand{\tt}{\texttt}
\renewcommand{\bf}{\textbf}

\newcommand{\AB}[1]{\langle #1 \rangle}
\newcommand{\SB}[1]{[ #1 ]}
\newcommand{\MixLeft}[3]{\langle #1 | #2 | #3 ]}
\newcommand{\BS}[1]{\boldsymbol{#1}}
\newcommand{\RN}[1]{%
  \textup{\uppercase\expandafter{\romannumeral#1}}%
}

\newcommand{\ThreePTwo}{\MixLeft{3}{p_1}{2}}
\newcommand{\SpDept}{ \AB{\BS{4}3}\SB{\BS{1}2} + \AB{\BS{1}3}\SB{\BS{4}2} }
\newcommand{\sm}{(s- m^2)}
\newcommand{\um}{(u- m^2)}

\title{The simplest massive S-matrix: from minimal coupling to Black Holes }
\author{Ming-Zhi Chung$^{1}$}
\author{Yu-tin Huang$^{1,2}$}
\author{Jung-Wook Kim$^{3}$}
\author{Sangmin Lee$^{3,4,5}$}

\affiliation{$^1$ Department of Physics and Astronomy, National Taiwan University, Taipei 10617, Taiwan}
\affiliation{$^2$ Physics Division, National Center for Theoretical Sciences, National Tsing-Hua University, No.101, Section 2, Kuang-Fu Road, Hsinchu, Taiwan}
\affiliation{$^3$ Department of Physics and Astronomy, Seoul National University, Seoul 08826, Korea
}
\affiliation{$^4$ Center for Theoretical Physics, Seoul National University, Seoul 08826, Korea
}
\affiliation{$^5$ College of Liberal Studies, Seoul National University, Seoul 08826, Korea
}

\emailAdd{dchung0741@gmail.com}
\emailAdd{yutinyt@gmail.com}
\emailAdd{jwkonline@snu.ac.kr}
\emailAdd{sangmin@snu.ac.kr}

\abstract{In this paper, we explore the physics of electromagnetically and gravitationally coupled massive higher spin states from the on-shell point of view. Starting with the three-point amplitude, we focus on the simplest amplitude characterized by matching to minimal coupling in the UV. In the IR, for charged states this leads to $g=2$ for arbitrary spin, and the leading deformation corresponds to the anomalous magnetic dipole moment. We proceed to construct the (gravitational) Compton amplitude for generic spins via consistent factorization. We find that in  gravitation couplings, the leading deformation leads to inconsistent factorization. This implies that for systems with Gauge$^2=$ Gravity relations, such as perturbative string theory, all charged states must have $g=2$. It is then natural to ask for generic spin, what is the theory that yields such minimal coupling. By matching to the one body effective action,  we verify that for large spins the answer is Kerr black holes. This identification is then an on-shell avatar of the no- hair theorem. Finally using this identification as well as the newly constructed Compton amplitudes, we proceed to compute the spin-dependent pieces for the classical potential at 2PM order up to degree four in spin operator of either black holes. }

\begin{document}
\begin{flushright}
\vspace{10pt} \hfill{NCTS-TH/1817} \vspace{20mm}
\end{flushright}
\maketitle

\section{Introduction}
Recently a new formalism for four-dimensional massive scattering amplitude was introduced by one of the authors~\cite{Arkani-Hamed:2017jhn} that manifests the covariance of the SU(2) massive Little group. Through such formalism, many fundamental properties of interacting systems become manifest, including Weinberg-Witten theorem, limits on the spin of fundamental point like particles, as well as Higgs mechanisms as the natural infrared unification. Furthermore, the new formalism also allows one to streamline computations such as anomalous magnetic dipole moment as well as classical electric and gravitational potentials~\cite{Cachazo:2017jef, Guevara:2017csg}.

Given its utility in making physical properties manifest, it is natural to pose the following question to such formalism: what is the simplest massive scattering amplitude? A similar question was posed for the massless case long ago~\cite{ArkaniHamed:2008gz}, for which remarkable properties of $\mathcal{N}=8$ supergravity amplitudes were unmasked. Here we expect the lessons to be equally, if not more, interesting. For one, the space of massive theories is much richer than that of massless ones. It includes not only fundamental particles, but monopoles, BPS states, and infinite tower of string resonances. Indeed, recently such on-shell approach was utilized for extremal (half-BPS) black holes in $\mathcal{N}=8$ supergravity~\cite{Caron-Huot:2018ape}, which demonstrated the absence of perihelion precession. 

We answer this question by starting with the three point amplitude describing a spin-$s$ state coupled to either a photon or a graviton. As discussed in~\cite{Arkani-Hamed:2017jhn}, this is given by an $\{2s\}\otimes\{2s\}$ symmetric SL(2,C) tensor, with $2s{+}1$ distinct structures. Assigning the massive legs to be $1$ and $2$ with equal mass, the general three point amplitude is parametrized by $\lambda_3$ (along with $\epsilon_{\alpha\beta}$ shares the responsibility of carrying the SL(2,C) indices) and $x$, defined as:
\eq
x\lambda_{3\alpha}=\frac{p_{1\alpha\dot{\alpha}}\tilde{\lambda}_3^{\dot\alpha}}{m}\,,
\eqe 
and $m$ is the mass of the massive legs. The simplest amplitude then corresponds to that comprises of $x$ and $\epsilon_{\alpha\beta}$ solely. This amplitude is identified as minimal coupling in the sense that in the high energy limit, the amplitude matches the minimal massless amplitude that has the least number of derivatives. For the case of charged particles this also matches with that of classical magnetic dipole moment 2 for any spin, and deformation of the form $\lambda_3^2$ represents $g{-}2$. Interestingly when extended to gravitational coupling $\lambda_3^2$ deformations are forbidden on the grounds of general covariance. Note that for systems in which the gravitation coupling is given by the square of vector couplings, such as perturbative string theories, this immediately leads to the conclusion that the charged particles must have $g=2$. 

Given that the minimal coupling has special properties both in the UV and IR, it is natural to ask for generic spin-$s$, which theory leads to such minimal coupling. Na\"ive expectation would be the leading trajectory states of open and closed string theories, since from the world-sheet CFT point of view, their vertex operators are the simplest. It turns out, the answer is quite the contrary, as we demonstrate that the leading trajectory massive spin states are the \textit{maximal} non-minimal coupling, reflected in the fact that all $2s{+}1$ tensor structure are present. Allowing ourselves to take the classical values of spin, i.e. $s\gg1$, we show by matching to the one-body effective action of a point particle coupled to gravitational background, minimal coupling matches on to that of a Kerr black hole. Thus the matching between minimal coupling and Kerr black hole, is the on-shell way of stating a consequence of the no-hair theorem.

Given the importance of minimal coupling, we explore the four-point (gravitational) Compton amplitude for general spin, by constructing an ansatz whose residues match that of products of minimal coupling. This leads to potential polynomial ambiguities. For $s\leq2$, such ambiguities are identified as finite size effects, as they are accompanied with additional $\frac{1}{m}$ factors. For $s>2$ the polynomial terms in general can be of the same order in $\frac{1}{m}$ as the pole terms, reflecting the inherent non-fundamental nature of such higher spin particles. We also consider four-point amplitudes with deformations from minimal coupling, demonstrating that consistent factorisation bans $\lambda_3^2$ terms in the three-point coupling with graviton. This provides an on-shell origin of inconsistencies of $\lambda_3^2$ terms in gravitational coupling alluded to earlier. 

Equipped with the identification of minimal coupling with Kerr black holes, as well as its Compton scattering amplitude for $s\leq2$, an immediate application is to compute the classical contributions to long-range gravitational interactions at 2 post-Minkowskian(PM) order, or $G^2$ order where $G$ is the Newton constant. It has been known for some time that quantum field theory (QFT) loop effects are not entirely quantum, but includes classical effects as well~\cite{Holstein:2004dn}. Such effects have been computed by various authors~\cite{Iwasaki:1971vb,Feinberg:1988yw,Holstein:2008sw,Holstein:2008sx}, and the results have become important in the era of gravitational wave astronomy where gravitational wave sources undergo hundreds to thousands of revolutions before their merger, which is long enough to push the small corrections of inverse-square-law to the detectable range~\cite{Goldberger:2007hy}.

Recently there has been tremendous activity in applying advanced developments in perturbative QFT computations to the computation of such classical effects, commonly referred to as classical potentials. These include generalized unitarity methods~\cite{Bern:1994zx, Bern:1994cg}, double copy relations~\cite{Goldberger:2016iau, Luna:2016due, Shen:2018ebu, Kosower:2018adc}, and    spinor-helicity variables~\cite{Neill:2013wsa,Bjerrum-Bohr:2013bxa,Vaidya:2014kza,Cachazo:2017jef,Guevara:2017csg,Bjerrum-Bohr:2018xdl}. Following Cachazo and Guevara~\cite{Cachazo:2017jef, Guevara:2017csg}, we compute the spin-dependent pieces of the 2PM classical potential to cubic and quartic in either Black Hole's spin. Such corrections, to the best of authors' knowledge, have not been presented in the literature before.

This paper is organized as follows. First, we start with a brief review of the massive spinor helicity formalism in section 2 and set up the the 3pt amplitudes. In section 3, we will analyze the physical implications of the 3pt amplitudes from section 2 for photons and gravitons.  Then in section 4, we take the graviton minimal coupling amplitude to the infinite spin limit and match with the effective action of a Kerr black hole. In section 5, we start to construct the Compton amplitudes with these 3pt amplitudes via constraints from consistent factorization. We discuss the high energy behaviour of these 4pt amplitudes and the polynomial ambiguities in our amplitude. In section 6, we start to calculate the classical potential at 1 PM with the leading singularity technique up to quartic order in spin. Finally, in section 7, we start with a review of the 1-loop leading singularity. Then we predict new results up to quartic order in spin. Then, we will use the consistent condition of the classical potential to fix some of the polynomial ambiguities in the higher spin Compton amplitude.

At the final stage of this work we were informed of the draft~\cite{Guevara:2018wpp} that has some overlap with the content in this work. 

\section{Review: on-shell formalism}
Scattering amplitudes are Lorentz invariant but Little group covariant quantities. This means that the amplitude must reflect the Little group representation of each external leg.  As we will be interested in four dimensions, the Little group in interest will be $U(1)$ and $SU(2)$ for massless and massive states respectively. Representations of $U(1)$ are simply labeled by the helicity weight $h$, while for $SU(2)$ instead of introducing a reference $z$-direction and label the states by its eigenvalue for $J_z$, we will represent a spin-$s$ state as a rank $2s$ symmetric tensor. As an example a four point amplitude with two massless and two massive states should be represented as:
\eq
M^{\{I_{1}, I_{2},\cdots I_{2s_1}\}, h_2, h_3, \{J_{1}, J_{2},\cdots J_{2s_4}\} }
\eqe
where the massive legs (1 and 4) are of spin-$s_1$ and $s_4$ respectively and the massless legs 2 and 3 have helicity $h_2,h_3$. The curly bracket indicates that one is symmetrizing over the $2s$ $SU(2)$ indices $I, J$, taking value in $1,2$.

Since amplitudes are covariant quantities, it should be a function of objects that are not singlets under the Little group, i.e. objects that carry little group indices. In the usual textbook approach, one introduces external line factors or polarization tensors which serve the purpose of converting Lorentz representations into Little group representations. Since except for scalars the size of the two representations are distinct, doing so introduces large amount of redundancy, which is the underlying reason for the complexity in the usual Feynman diagram approach. In contrast, the spinor helicity formalism introduces bosonic spinor variables that transform under the fundamental representation of the Little group, while directly comprising the kinematic data, the momenta. This allows us to remove the redundancy and dramatically reduce the complexity of the final answer. Furthermore, as we will see, such \emph{on-shell} approach will render many physical properties, such as high the energy behaviour, transparent and straightforward. 
 
\subsection{The massless/massive spinor helicity formalism}
We begin by introducing $SL(2,\IC)$ representations. A Lorentz vector, such as the momenta, is written as a bi-fundamental tensor under $SL(2,\IC)$:
\eq
p^\mu \rightarrow p_{\alpha\dot{\alpha}}
\eqe  
where $\alpha, \dot{\alpha}=1,2$. The usual Lorentz invariant inner products are then mapped to the contraction of these tensors with the $2\times 2$ Levi-Cevita tensor:
\eq
p_i^\mu p_{j\nu}=\frac{1}{2}\epsilon^{\alpha\beta}\epsilon^{\dot{\alpha}\dot{\beta}}\,p_{i\alpha\dot{\alpha}}\,p_{j\beta\dot{\beta}}\,.
\eqe
From the above one sees that $p^2={\rm det}p^{\alpha\dot{\alpha}}$. Thus for massless momenta, the $2\times 2$ tensor $p^{\alpha\dot{\alpha}}$ is of rank one and one has:\footnote{For real future-directed momenta with Minkowski signature, we have $\tilde\lambda=(\lambda)^*$. For complex momenta or $(2,2)$ signature the two spinors are independent. It will sometimes be convenient to consider complexified momenta when discussing the analytic properties of the scattering amplitude.}
\eq
p_{\alpha\dot{\alpha}}=\lambda_\alpha \tilde{\lambda}_{\dot{\alpha}}\,.
\eqe
The relation between the bosonic spinor variables and the momenta is invariant under the following transformation:
\eq
\lambda\rightarrow e^{-i\frac{\theta}{2}} \lambda,\quad \tilde\lambda\rightarrow e^{i\frac{\theta}{2}} \tilde\lambda
\eqe
Note that this is precisely the definition of the Little group! Thus we identify the spinors $\lambda, \tilde{\lambda}$ as having $(-\frac{1}{2}, +\frac{1}{2})$ Little group weight respectively. Using these bosonic spinors it is then convenient to define the following Lorentz invariant, Little group covariant building blocks:
\eq
\langle i j \rangle \equiv\lambda_i^\alpha\lambda_j^\beta\epsilon_{\alpha\beta},\quad [ij]\equiv\tilde\lambda_{i\dot\alpha}\tilde\lambda_{j\dot\beta}\epsilon^{\dot\alpha\dot\beta}\,.
\eqe
In terms of these blocks, the usual Mandelstam variables are given as $2p_i\cdot p_j=\langle ij\rangle[ji]$.

For massive momenta, $p_{\alpha\dot{\alpha}}$ has full rank and we have
\eq
p_{\alpha\dot{\alpha}}=\lambda^{I}_{\alpha}\tilde{\lambda}_{I\dot{\alpha}}\,,
\eqe
where $I=1,2$. The index $I$ indicate that they form a doublet under the $SU(2)$ massive Little group. Indeed the momentum is invariant under the following transformations:
\eq
\lambda^{I\alpha}\rightarrow U^I\,_J\,\lambda^{J\alpha},\quad \tilde\lambda^{I\dot\alpha}\rightarrow U^I\,_J\,\lambda^{J\alpha}\,,
\eqe
where $U$ is an element of $SU(2)$. One can convert between the two spinors via
\eq
p_{\alpha\dot\alpha}\tilde{\lambda}^{I\dot{\alpha}}=m\lambda^I_{\alpha},\quad p_{\alpha\dot\alpha}\lambda^{I\alpha}=-m\tilde\lambda^I_{\dot\alpha}\,. \label{eq:SHvarDiracEq}
\eqe
A detailed description of spinor-helicity formalism is given in appendix \ref{sec:spin-helconventions}.

An important property of the Little group is that it is defined for each individual momenta separately. In other words, only the spinor variables of a given leg can carry its Little group index. This implies that without lost of generality we can pull out overall factors of $\lambda_i^I$ from the amplitude, 
\eq\label{ChiralDef}
M_n^{\cdots\{I_1,I_2,\cdots,I_{2s_i} \}\cdots}=\lambda^{I_1}_{i\alpha_1}\lambda^{I_2}_{i\alpha_2}\cdots \lambda^{I_{2s_i}}_{i\alpha_{2s_i}}M_n^{\cdots\{\alpha_1,\alpha_2,\cdots,\alpha_{2s_i} \}\cdots}\;.
\eqe  
leaving behind a function that is symmetric in $SL(2,\IC)$ indices instead. We will refer to this representation as the chiral basis, reflecting the fact that we are using the un-dotted $SL(2,\IC)$ indices. One can equally use the anti-chiral basis, and the two can be converted to each other by contracting with $\frac{p^{\alpha\dot{\alpha}}}{m}$. This separation will be useful when considering suitable basis for all possible three-point interactions as we will now see.  
\subsection{General structure of the three-point amplitude}
We now consider the most general form of the three-point amplitude for one massless and two equal mass legs with spin $s$. Without loss of generality, the momenta $p_1$ and $p_2$ can be taken to be massive, and the amplitude takes the form:
\eq
M_3^{h,\{\alpha_1,\cdots,\alpha_{2s}\},\{\beta_1,\cdots,\beta_{2s}\}}\,.
\eqe
where $h$ is the helicity of the massless leg. Now we have a $\{ 2s \} \otimes \{ 2s \}$ $SL(2,\IC)$ tensor, and we are interested in the general structure of all possible couplings. This entails the need of a basis to span the two-dimensional space. It is preferable to use the kinematic variables of the problem to serve as a basis, thus it is natural to introduce
\eq
\lambda_{3\alpha}, \; \epsilon_{\alpha\beta}
\eqe   
as the expansion basis. 

Since $\lambda_3$ carries helicity weight $-\frac{1}{2}$ of the massless leg $3$, in order to represent general amplitudes, one should also have a variable that carries positive weights. This variable is introduced by noting that for equal mass kinematics,\footnote{Here $\langle i|p_j |k]=\lambda_i^\alpha p_{j\alpha\dot{\alpha}} \tilde{\lambda}^{\dot{\alpha}}_k$.} 
\eq
2 p_3\cdot p_1=\langle 3|p_1|3]=0\,,
\eqe 
and hence the spinor $\lambda_3^\alpha$ must be proportional to $\tilde{\lambda}_{3\dot{\alpha}}p_{1}^{\dot{\alpha}\a}$. Through this proportionality, we introduce a new variable $x$ defined as:
\eq
x\lambda_3^\alpha = \tilde{\lambda}_{3\dot{\alpha}} \frac{p_{1}^{\dot{\alpha}\alpha}}{m}\,.
\eqe
where $p^2_1=m^2$. Note that the above equality tells us that $x$ is dimensionless and carries ${+}1$ helicity of leg $3$. Using an auxiliary spinor $\xi$, we can represent $x$ as 
\eq
x=\frac{[3|p_1|\xi\rangle}{m\langle 3\xi\rangle}\,.
\eqe
The above shows that $x$ can be nicely written in terms of polarization vectors:
\eq
mx=\frac{1}{\sqrt{2}}\varepsilon^{(+)}\cdot(p_1-p_2) 
\eqe
with the polarization vector $\varepsilon^{(+)}_{\alpha\dot{\alpha}}=\sqrt{2}\frac{\tilde{\lambda}_{3\dot{\alpha}}\xi_\alpha}{\langle 3\xi\rangle}$, and the auxiliary spinor is identified with the reference spinor of the polarization vector.

Equipped with the new variable, we can write down the general structure of a three point amplitude for two spin $s$ and a helicity $h$ state:
\eqa
M_3^{h,\{\alpha_1,\cdots,\alpha_{2s}\},\{\beta_1,\cdots,\beta_{2s}\}}&=&(m x)^{h}\left[g_0\epsilon^{2s}+g_1\epsilon^{2s{-}1}x\frac{\lambda_3\lambda_3}{m}+\cdots +\left(x\frac{\lambda_3\lambda_3}{m}\right)^{2s}\right]^{\{\alpha_1,\cdots,\alpha_{2s}\},\{\beta_1,\cdots,\beta_{2s}\}}\nonumber\\
&=& (m x)^{h}\left[\sum_{a=0}^{2s} g_a\epsilon^{2s-a}\left(x\frac{\lambda_3\lambda_3}{m}\right)^a\right]^{\{\alpha_1,\cdots,\alpha_{2s}\},\{\beta_1,\cdots,\beta_{2s}\}}\,,
\eqae
 where the $2s\otimes 2s$ separately symmetrized $SL(2,\IC)$ indices are distributed across the Levi-Cevita tensors $\epsilon$ and $\lambda_3$s. Thus we see that there are in total $2s{+}1$ structures for spin $s$ states, and we've normalized the couplings such that the $g_i$s are dimensionless.

Note that the above classification is purely kinematic in nature, and does not correspond to the classification of local operators in the usual derivative expansion. Indeed in the usual Lagrangian language, there may be a large number of operators at a given derivative order simply due to the different ways the derivative can contract. Furthermore, operators at the same derivative order may behave very differently in the high-energy limit. For example, consider the following Lagrangian for a charged spin-$s$ field: 
\eq\label{TempL}
\mathcal{L} = (-1)^s D^\nu \phi^{(s)} D_{\nu} \bar{\phi}^{(s)}+\cdots 
\eqe
where $\phi^{(s)}$ is the short hand notation for a rank $s$ field, the Lorentz indices of $\phi$ is contracted with $\bar{\phi}$, and $\cdots$ represents additional terms needed to ensure that through equations of motion, $\phi^{(s)}$ and $\bar{\phi}^{(s)}$ are symmetric, traceless and transverse\footnote{For massive higher spin fields, transversality will be defined as having no time-like polarisations. This means transverse polarisations in this manuscript will include degrees of freedom referred to as longitudinal polarisations in the literature.}. Consider the three-point amplitude from the leading term, given by:
\eq
\varepsilon_3\cdot(p_1{-}p_2)\;\varepsilon^{(s)}_1\cdot\varepsilon^{(s)}_2\,.
\eqe
To convert to our \emph{chiral} basis, we strip-off the polarization tensors and convert the dotted indices into un-dotted indices, by contracting with $\frac{p}{m}$:
\eq
\varepsilon_3\cdot(p_1{-}p_2)\,\mathbf{O}_{\alpha_1\beta_1} \mathbf{O}_{\alpha_2\beta_2}\cdots \mathbf{O}_{\alpha_s\beta_s} \, \epsilon_{\alpha_{s{+}1}\beta_{s{+}1}}\cdots \epsilon_{\alpha_{2s}\beta_{2s}},\quad \mathbf{O}_{\alpha\beta}\equiv \frac{p_{1\alpha}\,^{\dot\alpha}p_{2\beta\dot\alpha}}{m^2}
\eqe
 Using the identity $\frac{p_{1\alpha}\,^{\dot\alpha}p_{2\beta\dot\alpha}}{m^2}=\epsilon_{\alpha\beta}-x\frac{\lambda_{3\alpha}\lambda_{3\beta}}{m}$, we find that in the chiral basis the leading coupling in eq.(\ref{TempL}) written as:
 \eq
 mx\left[\prod_{i=1}^s\left(\epsilon-x\frac{\lambda_{3}\lambda_{3}}{m}\right)_{\alpha_i\beta_i}\right]\left[\prod_{k=s{+}1}^{2s}\epsilon_{\alpha_{k}\beta_{k}}\right]+sym\{\alpha_1\cdots\alpha_{2s}\}sym\{\beta_1\cdots\beta_{2s}\}\,.
 \eqe
Here we have all $g_i\neq0$ for all $i\leq n$. In other words, a single local operator in the Lagrangian is expressed as a sum of many terms in such on-shell basis. The reason that there is such dramatic difference is because the on-shell basis is completely determined from kinematics, and thus each term in the expansion is distinct in a purely kinematic way. On the other hand, operators in a Lagrangian can often be related through integration by parts or field redefinitions, and each operator can contain several kinematically distinct pieces. In fact, as we will see in the next section, by expressing the three-point amplitude on such on-shell basis, we will be able to cleanly separate terms that behave poorly in the UV, allowing us to define in a physically meaningful way what \emph{minimal coupling} is.

It will be convenient to make connection with the amplitudes computed from the usual Feynman diagram approach. For this, we simply put back the $\lambda_i^I$ factors that was pulled out that defined the chiral basis in eq.\eqc{ChiralDef}. For example,
\bl
M_3^{h,s,s} &= (mx)^h \left[ g_0 \frac{\la \bold{2} \bold{1} \ra^{2s}}{m^{2s-1}} + g_1 x \frac{\la \bold{2} \bold{1} \ra^{2s-1} \la \bold{2} 3 \ra \la 3 \bold{1} \ra}{m^{2s}} + \cdots + g_{2s} x^{2s} \frac{\la \bold{2} 3 \ra^{2s} \la 3 \bold{1} \ra^{2s}}{m^{4s-1}} \right]\,.\label{eq:poshel3pt}
\el
Note that we have suppressed the massive Little group indices, and simply ``bolding" the massive spinors with the understanding that its Little group indices are symmetrised.  Taking the conjugate, one obtains the anti-chiral representation: 
\bl
M_3^{h,s,s} &= \frac{x^h}{m^{h}} \left[ \bar{g}_0 \frac{[ \bold{2} \bold{1} ]^{2s}}{m^{2s-1}} + \frac{\bar{g}_1}{x} \frac{[ \bold{2} \bold{1} ]^{2s-1} [ \bold{2} 3 ] [ 3 \bold{1} ]}{m^{2s}} + \cdots + \frac{\bar{g}_{2s}}{x^{2s}} \frac{[ \bold{2} 3 ]^{2s} [ 3 \bold{1} ]^{2s}}{m^{4s-1}} \right] \,.\label{eq:neghel3pt}
\el
The coefficients in the anti-chiral basis are of course linearly related to that in the chiral basis. Indeed using the following identities;
\bg
\la \bold{2} \bold{1} \ra = [\bold{2} \bold{1}] + \frac{[\bold{2} 3][3 \bold{1}]}{mx} = \sbra{\bold{2}} \left( \iden + \frac{\sket{3} \sbra{3}}{mx} \right) \sket{\bold{1}} \label{eq:3ptkin1}
\\ \la \bold{2} 3 \ra \la 3 \bold{1} \ra = -\frac{[\bold{2} 3][3 \bold{1}]}{x^2} = \sbra{\bold{2}} \left( - \frac{\sket{3} \sbra{3}}{x^2} \right) \sket{\bold{1}}\,, \label{eq:3ptkin2}
\eg
one can show that eq.\eqc{eq:poshel3pt} can be recast into the anti-chiral basis, where the coupling constants $\bar{g}_n$s are given as
\bl\label{barg}
\bar{g}_m &= \sum_{n=0}^{m} \left( -1 \right)^{n} \left( \begin{gathered} 2s-n \\ m-n \end{gathered} \right) g_n\,.
\el

As an illustration of how these interaction arrises from terms of a local Lagrangian, consider the coupling of Maxwell field to Dirac spinors; $\CL_{\text{int}} = e A_\m \bar{\Psi} \g^\m \Psi$. The 3pt amplitude for this interaction term with the convention $\Psi (p_1)$ incoming, $A_\m (k_3)$ incoming positive helicity, and $\bar{\Psi} (p_2)$ outgoing is
\bl
M_3^{+1,\mathbf{\frac{1}{2}},\mathbf{\frac{1}{2}}} &=  e \bar{u}(p_2)\displaystyle{\not} \varepsilon^{+}_3 u(p_1) =  \sqrt{2} e \frac{-[\bold{2} 3] \la \z \bold{1} \ra + \la \bold{2} \z \ra [ 3 \bold{1} ]}{\la 3 \z \ra}\,,
\el
which, with help of three particle kinematics we can write $[\bold{2} 3]=- \langle \bold{2}|p_1|3]/m$ and $[ 3 \bold{1} ]=[3|p_1|\bold{1}\rangle/m$. Substituting into the last equality we find that: 
\bl
M_3^{+1,\mathbf{\frac{1}{2}},\mathbf{\frac{1}{2}}} &=  \sqrt{2} e \frac{\sbra{3} p_1 \ket{\z}}{m \la 3 \z \ra} \la \bold{2} \bold{1} \ra = \sqrt{2} ex \la \bold{2} \bold{1}\ra \,,\label{eq:DiracMaxwell3pt}
\el
which, corresponds to the first term in eq.\eqc{eq:poshel3pt} after normalization. Similarly, for the Pauli term; $\CL_{\text{int}} = - \frac{e}{M} F_{\m\n} \bar{\Psi} \g^{\m\n} \Psi$ with $\g^{\m\n} := \frac{i}{4} \g^{[\m} \g^{\n]}$,
\bl
M_3^{+1,\mathbf{\frac{1}{2}},\mathbf{\frac{1}{2}}} &= - i\frac{e}{M} \bar{u}(p_2) \g^{\m\n} \left( \left. - i k_{3} \right._{[\m} \varepsilon^{+}_{3,\n]}\right) u(p_1) = - \frac{\sqrt{2} e}{M} [ \bold{2} 3 ][ 3 \bold{1} ]=x^2\frac{\sqrt{2} e}{M} \langle \bold{2} 3 \rangle\langle 3 \bold{1} \rangle\,,
\el
where we have used the eq.\eqc{eq:3ptkin2} for the last equality. This gives the second term of the expansion eq.\eqc{eq:poshel3pt}, with $s=\frac{1}{2}$.

\subsection{Classical spin-operators from amplitudes}
It will be useful to view the three-point amplitude as an operator acting on the Hilbert space of $SL(2,\IC)$ irreps. More precisely, since $M^{+h,s,s}$ is basically a $\{2s\}\otimes \{2s\}$ Lorentz tensor contracted between $2s$ $\lambda_1$ and $2s$ $\lambda_2$s, it can be viewed as an operator that maps the spin-$s$ representation in the Hilbert space of particle 1 to that of particle 2. In other words, schematically we have:
\eq
M^{+h,s,s}= \frac{1}{m^{2s-1}}\lambda^{I_1\alpha_1}_1\cdots \lambda^{I_{2s}\alpha_{2s}}_1  \mathcal{O}_{\{\alpha_1,\cdots,\alpha_{2s}\},\{\beta_1,\cdots,\beta_{2s}\}} \lambda^{J_1\beta_1}_2\cdots \lambda^{J_{2s}\beta_{2s}}_2
\eqe 
We know that the operator $\mathcal{O}_{\{\alpha_1,\cdots,\alpha_{2s}\},\{\beta_1,\cdots,\beta_{2s}\}} $ is a linear combination of polynomials in $\epsilon_{\alpha\beta}$ and $\lambda_{3\alpha}\lambda_{3\beta}$. The former naturally can be identified as the identity operator, while the latter is the \emph{spin-operator} which we now show. 

Let's start with the spin vector $S^\mu$ defined as the Pauli-Lubanski pseudo-vector $S^\mu=-\frac{1}{2m}\epsilon^{\mu\nu\rho\sigma}p_{1\nu}J_{\rho\sigma}$. For the Lorentz generator $J_{\mu\nu}$, we will be interested in its action on $SL(2,\IC)$ irreps. For spin-$s$, we write:
\eq
(J_{\mu\nu})_{\alpha_1\alpha_2\cdots\alpha_{2s}}\,^{\beta_1\beta_2\cdots\beta_{2s}}=\sum_{i}(J_{\mu\nu})_{\alpha_i}\,^{\beta_i}\,\bar{\mathbb{I}}_i=2s(J_{\mu\nu})_{\alpha_1}\,^{\beta_1}\,\bar{\mathbb{I}}_1,\quad (J_{\mu\nu})_{\alpha}\,^{\beta}=\frac{i}{2}\sigma_{[\mu}\bar{\sigma}_{\nu]}\,,
\eqe
where $\bar{\mathbb{I}}_i=\delta_{\alpha_1}^{\beta_1}\cdots \delta_{\alpha_{i{-}1}}^{\beta_{i{-}1}}\delta_{\alpha_{i{+}1}}^{\beta_{i{+}1}}\cdots \delta_{\alpha_{2s}}^{\beta_{2s}}$, and the last equality reflects the fact that the irreps are symmetric tensors of $2s$ indices. Using this, we find that
\bl
\bld
m \left( S_\m \right)_{\a}^{~\b} &= \frac{1}{4} \left[ \s_\m (p_1 \cdot \bar{\s}) - (p_1 \cdot \s) \bar{\s}_\m \right]_{\a}^{~\b}
\\ m \left( S_\m \right)^{\dot{\a}}_{~\dot{\b}} &= - \frac{1}{4} \left[ \bar{\s}_\m (p_1 \cdot \s) - (p_1 \cdot \bar{\s}) \s_\m \right]^{\dot{\a}}_{~\dot{\b}}\,.
\eld \label{eq:spinopangsqdefs}
\el
Finally, dotting into the massless momenta $p_3$ we arrive at:
\bl
\bld
(p_3\cdot S)_{\a}^{~\b} &= \frac{x}{2} \lambda_{3\a} \lambda_3^\b\equiv \frac{x}{2} |3\rangle\langle3| 
\\ (p_3\cdot S)^{\dot\a}_{~\dot\b} &=-\frac{\lambda_3^{\dot\a} \lambda_{3\dot\b}}{2x}\equiv-\frac{|3][3|}{2x}\,.
\eld \label{PSDef}
\el
From this result we see that the operator $\mathcal{O}_{\{\alpha_1,\cdots,\alpha_s\},\{\beta_1,\cdots,\beta_s\}}$ is comprised of identity operators and the spin vector operator, projected along the direction of the massless momenta. For example, for $s=1$, we have
\eq
\mathcal{O}_{\{\alpha_1\alpha_2\}}\,^{\{\beta_1\beta_2\}}=\left(g_0\mathbb{I}_{\{\alpha_1}^{~\beta_1}\mathbb{I}_{\alpha_2\}}^{~\beta_2}+2g_1\frac{\mathbb{I}_{\{\alpha_1}^{~\beta_1}(p_3\cdot S)_{\a_2\}}^{~\b_2}}{m}+4g_2\frac{(p_3\cdot S)_{\{\a_1}^{~\b_1}}{m}\frac{(p_3\cdot S)_{\a_2\}}^{~\b_2}}{m}\right)\,.
\eqe  
\section{The simplest three-point amplitude}

In the previous section, we've seen that for a massive spin-$s$ particle, whether it is fundamental or composite, the emission of a photon or graviton can in general be parameterized by eq.(\ref{eq:poshel3pt}). This parameterization is unique in the sense that the expansion basis is defined on kinematic grounds unambiguously. The expansion is organized in terms of powers of $\frac{1}{m}$, with higher order terms hinting at potential problems in the UV, i.e. the massive amplitude does not have a smooth $m\rightarrow 0$ limit. In other words, this parameterization manifests the high energy behaviour for a given interaction. To illustrate this feature in more detail, take for example the Lagrangian in eq.(\ref{TempL}) with spin-1, which is known to lead to four-point amplitudes that violate tree-unitarity at high energies and is not removable via the presence of an extra Higgs. Indeed this can be seen already at the three-point level, where in our parameterization is given as:
\eq
mx\frac{\langle \bf{1}\bf{2}\rangle^2}{m^2}-mx^2\frac{\langle \bf{1}\bf{2}\rangle\langle \bf{1}3\rangle\langle3\bf{2}\rangle}{m^3}\,.
\eqe  
We see that while in the Lagrangian the interaction is given by a single local operator, in our on-shell parameterization, it is comprised of two pieces, with the latter behaving worst in the high-energy limit compared to the first. Indeed, if we consider the three-point amplitude of a photon with W-bosons,  we will only find the leading piece at tree level.

Consider an amplitude with only the leading term in eq.(\ref{eq:poshel3pt}). The above discussion would indicate that not only is the amplitude simple in the number of terms involved, but is also simple in the sense of having the best UV behavior. At high energies we only have massless states, and we can ask what amplitude in the UV does this pure $x$-piece matches to. Note that simply ``unbolding" the spinors might lead to ill defined limit, as while the denominator of $\frac{1}{m}$ tends to zero in the limit, the angle brackets in the numerator can tend to zero as well even in complex kinematics. Thus we would like to have a controlled way of approaching the high energy limit for eq.(\ref{eq:poshel3pt}).  To this end, let us decompose the massive spinors onto the helicity basis of the massless limit:
\eq
\lambda^I_\alpha=\lambda_\alpha \xi^{-I}+\eta_\alpha \xi^{+I}\,,\quad \tilde\lambda^I_{\dot{\alpha}}=\tilde{\eta}_{\dot{\alpha}} \xi^{-I}+\tilde{\lambda}_{\dot{\alpha}} \xi^{+I}\,,
\eqe
where $\epsilon_{IJ}\xi^{+I}\xi^{-J}=1$ and $\langle \lambda \eta\rangle=[\tilde{\lambda}\tilde{\eta}]=m$. The $SU(2)$ spinors $\xi^{\pm I}$ are the eigenstates of spin-$\frac{1}{2}$ for $J_z$ in a given frame. In the $m\rightarrow0$ limit we see that the finite contribution correspond to taking the $\xi^{\pm I}$ of the two massive legs to have opposite helicity. In other words we the two massive spin-$s$ states will translate into a $+s$ and $-s$ helicity state separately at high energies. To avoid a singular piece we must have $\lambda_1\sim \lambda_2\sim \lambda_3$. Choosing leg $1$ to be the positive helicity, we then have 
\eq
\left.(m x)^h\left(\frac{\langle \bf{1}\bf{2}\rangle}{m}\right)^{2s}\right|_{m\rightarrow 0} =\left.\left(\frac{[3|p_1|\xi\rangle}{\langle 3\xi\rangle}\right)^h\left(\frac{\langle \eta_12\rangle}{m}\right)^{2s}\right|_{m\rightarrow 0} \,.
\eqe  
Since the $\lambda_i$s are proportional to each other, we introduce proportionality factors $y_1, y_2$ defined via $\lambda_1 = y_1 \lambda_3$ and $\lambda_2 = y_2 \lambda_3$. Momentum conservation then fixes:
\bl
y_1= \frac{[23]}{[12]}\,,\quad y_2 = - \frac{[13]}{[12]}
\el
Using that $\frac{\langle \eta_12\rangle}{m}= - \frac{[13]}{[23]}\frac{\langle \eta_11\rangle}{m}=\frac{[13]}{[23]}$, this leads to 
\eq
\left.\left(\frac{[3|p_1|\xi\rangle}{\langle 3\xi\rangle}\right)^h\left(\frac{\langle \eta_12\rangle}{m}\right)^{2s}\right|_{m\rightarrow 0} =\left(\frac{[23][31]}{[12]}\right)^h\left(\frac{[13]}{[23]}\right)^{2s}\,.
\eqe
We see that in the high-energy limit, the pure $x$-piece will become that of the minimal coupling: the minimal mass dependence for a three point amplitude with $h_3>0$ and $|h_1|=|h_2|=s$ states.\footnote{For more detail see appendix \ref{sec:HE}}

One can straightforwardly check that subleading terms in eq.(\ref{eq:poshel3pt}) match to higher derivative couplings in the UV. Thus minimal coupling in the UV uniquely picks out 
\eq\label{MinimalCoupling}
\framebox[4cm][c]{$\displaystyle (m x)^h\left(\frac{\langle \bf{1}\bf{2}\rangle}{m}\right)^{2s}$}
\eqe
from all possible low energy couplings. For this reason it is natural to refer to the choice of setting all coupling constants except $g_0$($\bar{g}_0$) to zero as \emph{minimal coupling}. Once again, we stress that our minimal coupling is defined through kinematics solely. As we will see in the next section, this will be minimal in a very precise sense in the IR as well! In the following, we will study this simplest amplitude in more detail for photons and gravitons separately.

\subsection{Photon minimal coupling and $g=2$}
Let us first consider the case where the minimal coupling involves the massive states coupled to a photon, $|h|=1$. The coupling we are interested in will then be:
\eq
e m x\, \left(\frac{\langle \bf{2}\bf{1}\rangle}{m}\right)^{2s}\,,
\eqe
where we've included the charge $e$ and made an overall sign choice for a better interpretation as operators. Since we are considering coupling to photon that is sensitive to its spin, a natural quantity of interest would be its magnetic dipole moment. Recall that in the non-relativistic limit, the magnetic dipole moment is defined through the Zeeman coupling: 
\bl
V_Z &:= - \vec{\m} \cdot \vec{B} = - \frac{g e}{2 m} \vec{S} \cdot \vec{B}\,.
\el
In the rest frame of the charged particle with momentum $p_1$, the magnetic field $\vec{B}$ can be written in the following Lorentz covariant form:
\bl
B^\m &:= \frac{1}{2m} \e^{\m\n\rho\s} p_{1\n} F_{\rho\s}\,.
\el
The expression for the Zeeman coupling then has the following Lorentz invariant form: 
\bl
\bld
V_Z &= -\frac{g e}{2 m} \vec{S} \cdot \vec{B} = \frac{g e}{4 m} J^{\mu\nu} F_{\mu\nu} + \frac{g e}{2 m^3} p_1^{\tau} F_{\tau\eta} J^{\eta\chi} p_{1\chi}\,.
\eld \label{eq:ZeemanDef}
\el
Substitute $F_{\m\n} = - i \sqrt{2} ( p_{3\m} \ve^{\pm}_\n - p_{3\n} \ve^{\pm}_\m)$\footnote{The normalisation factor of $\sqrt{2}$ may seem unconventional, but introduction of this factor simplifies the analysis: The scalar potential coupling $V_{\text{S}} = e\phi$ can be covariantly written as $\frac{P \cdot A}{m}$, and setting $A_\m = \sqrt{2} \ve^{\pm}_\m$ results in $V^{+}_{\text{S}} = ex$ and $V^{-}_{\text{S}} = e \bar{x}$.\label{fn:scalarpotential}} into the Zeeman coupling equation eq.\eqc{eq:ZeemanDef} for $s=\frac{1}{2}$ in the dotted frame. For plus helicity photon this results in:
\bl
\bld
\left( V_Z^{+} \right)^{\dot\beta}_{~\dot\alpha} &= \frac{g e}{4 m} \tilde{\lambda}_3^{\dot\beta} \tilde{\lambda}_{3 \dot\alpha} \,,
\\ \left( V_Z^{+} \right)_{\beta}^{~\alpha} &= - \frac{g e}{4 m} x^2 {\lambda}_{3 \beta} {\lambda}_3^{\alpha} \,.
\eld \label{eq:Zeeman1}
\el
For general $s$ in $(\frac{s}{2}, \frac{s}{2})$ representation we have:
\bl
\left( V_{Z,2s}^{+} \right)^{\dot\b_1 \cdots \dot\b_{s} ~~~~~~~ \a_1 \cdots \a_{s}}_{~~\dot\a_1 \cdots \dot\a_{s} ~ \b_1 \cdots \b_s} &= \sum_i \left( V_Z^{+} \right)^{\dot\b_i}_{~\dot\a_i} \bar{\iden}_i + \bar{\iden}_i \left( V_Z^{+} \right)_{\b_i}^{~\a_i} \,. 
\el
To compare with our three-point amplitude we contract the $SL(2,\IC)$ indices with massive spinor helicity variables, yielding
\bl
\bld
\left( V_{Z,2s}^{+} \right)& = s\frac{g e}{4} \left\{ \frac{[ \mathbf{2}3 ][3\mathbf{1}]}{m^2} \left(\frac{[ \bf{2}\bf{1}]}{m}\right)^{s{-}1} \left(\frac{ \la \bf{2}\bf{1} \ra}{m}\right)^{s} - \frac{x^2 \la \mathbf{2}3 \ra \la 3\mathbf{1} \ra}{m} \left(\frac{[ \bf{2}\bf{1}]}{m}\right)^{s} \left(\frac{ \la \bf{2}\bf{1} \ra}{m}\right)^{s -1} \right\}
\\ &= - \frac{s g}{2} e x \frac{x \la \mathbf{2}3 \ra \la 3\mathbf{1} \ra}{m^2} \left(\frac{ \la \bf{2}\bf{1} \ra}{m}\right)^{2s -1} + \cdots \,.
\eld \label{ZeemanForm}
\el

Now let us compare with our minimal coupling amplitude. We add to eq.\eqc{ZeemanForm} the scalar potential $V_\text{S}$ in $(\frac{s}{2}, \frac{s}{2})$ representation, $e x \left(\frac{[ \bf{2}\bf{1}]}{m}\right)^{s} \left(\frac{ \la \bf{2}\bf{1} \ra}{m}\right)^{s}$. The three-point amplitude then takes the form 
\bl
M_3^{+1,s,s} &= e x \frac{\la \bold{2} \bold{1} \ra^{2s}}{m^{2s}} - e x \frac{s (g-2)}{2} \frac{\la \bold{2} \bold{1} \ra^{2s-1} \la \bold{2} 3 \ra \la 3 \bold{1} \ra}{m^{2s+1}} + \cdots  \,.
\el
Compared with the general 3pt ansatz in eq.(\ref{eq:poshel3pt}), we immediately see that our minimal coupling leads to $g=2$ for arbitrary spin. Thus the simplest amplitude with photon coupling is also characterized by the classical magnetic dipole moment being $2$! Note that terms denoted as $\cdots$ corresponds to higher multipole moments as they are of higher order in $\tilde{\lambda}_3$, indicating higher derivative terms on the field strength. Since minimal coupling is also related to good high energy behaviour, this indicates that for an isolated charged spin-$s$ particle with good UV behaviour, the classical magnetic moment must be $2$. By isolated we are referring to the case where there are no other states with similar mass.  Indeed the classical value for $g$ is 2 for massive vectors arising from Higgs mechanism, and it is known that when constrained to operators with two derivative couplings, tree-level unitarity requires $g=2$ for isolated massive spinning particles~\cite{Ferrara:1992yc}. 

Finally, we see from above that the presence of $g_1$ indicates $(g-2)$ contributions. Indeed as one finds that the coupling parameterized by $g_1$ is generated at one loop~\cite{Arkani-Hamed:2017jhn}.

\subsection{Gravitational minimal coupling}
We now turn to gravity. The minimal three-point coupling for a positive helicity graviton is\,,
\eq
 \frac{m^2}{M_{pl}} x^2\, \left(\frac{\langle \bf{1}\bf{2}\rangle}{m}\right)^{2s}\,.
\eqe
Following our photon discussion, we can ask whether there is a gravitational analogue of Zeeman coupling for gravitomagnetic interactions, and the minimum coupling correspond to a particular value for the \emph{gravitomagnetic dipole moment}. Indeed one can consider a Kaluza-Klein decomposition of the metric:
\eq
h_{00}=2\Phi,\quad h_{0i}=-\mathcal{A}_i,\quad h_{ij}=2\Phi\delta_{ij}\,,
\eqe 
where $\Phi$ will be identified with the gravitational potential and $\vec{\mathcal{A}}$ the vector potential for gravitational version of magnetic field. The full gravitational potential then takes the form:
\eq
V=m\Phi+\alpha \vec{S}\cdot\vec{\mathcal{B}}\,.
\eqe 
We can also run a parallel analysis of the argument for Zeeman coupling in the previous section, with $h_{\m\n} = 4 \e^+_\m \e^+_\n$ for normalisation of the scalar potential $V_{\text{S}}^+ = m \Phi = \frac{1}{2m} h_{\m\n} P^\m P^\n = m x^2$. This sets $\CA_\m = \frac{1}{m} h_{\m\n} P^\n = 2x \sqrt{2} \e^+_\m$, therefore computations of the previous section can be reused with the substitution $\frac{ge}{2m} \to -\a$ with overall scaling $2x$.
\bl
\bld
\left( V_{Z,2s}^{+} \right)& = - s (2x) \frac{m \a}{2} \left\{ \frac{[ \mathbf{2}3 ][3\mathbf{1}]}{m^2} \left(\frac{[ \bf{2}\bf{1}]}{m}\right)^{s{-}1} \left(\frac{ \la \bf{2}\bf{1} \ra}{m}\right)^{s} - \frac{x^2 \la \mathbf{2}3 \ra \la 3\mathbf{1} \ra}{m} \left(\frac{[ \bf{2}\bf{1}]}{m}\right)^{s} \left(\frac{ \la \bf{2}\bf{1} \ra}{m}\right)^{s -1} \right\}
\\ &= 2 \a s m x^2 \frac{x \la \mathbf{2}3 \ra \la 3\mathbf{1} \ra}{m^2} \left(\frac{ \la \bf{2}\bf{1} \ra}{m}\right)^{2s -1} + \cdots \,.
\eld
\el
Including the scalar potential contribution in the $(\frac{s}{2}, \frac{s}{2})$ representation yields the 3pt amplitude
\bl
M_3^{+2,s,s} &= m x^2 \frac{\la \bold{2} \bold{1} \ra^{2s}}{m^{2s}} + m x^2 s (1 + 2 \a) \frac{\la \bold{2} \bold{1} \ra^{2s-1} \la \bold{2} 3 \ra \la 3 \bold{1} \ra}{m^{2s+1}} + \cdots  \,.
\el
Note that in contrast to the photon case, here $\alpha$ is fixed by the requirement that the resulting Hamiltonian reproduces the correct evolution of the spin operator $\vec{S}$, which is dictated from general covariance. Thus we seem to have a potential contradiction: since the gravitomagnetic dipole moment is completely fixed from general covariance, minimal coupling is inconsistent if $\a$ determined from general covariance doesn't reproduce the minimal coupling value $\alpha=-\frac{1}{2}$. However, from an on-shell point of view, there is no apparent sickness either in its high energy behaviour or consistent embedding in a four-point amplitude, as we will see in section \ref{sec:Compton_amplitude_for _arbitrary_spin}. Not surprisingly, we will find that minimal coupling \textit{exactly} reproduces the correct result! We leave the details for computing $\a$ by demanding general covariance to appendix~\ref{FermiWalkerTransport}.

The discussion above indicates that the gravitomagnetic dipole moment should be universal on grounds of general covariance\footnote{This is reminiscent of Weinberg's soft theorems~\cite{Weinberg:1965nx}, where photon soft theorems only require charge conservation, and thus allowing any charge for a given state. On the other hand graviton soft theorems leads to unversal coupling constants and hence the equivalence principle. }. We can also consider the same problem from Lagrangians of higher spin massive particles. Since the dipole moment is associated with minimal coupling as well as the coefficient of $\lambda^2_3$,  this implies that one can simply consider an arbitrary diffeomorphism invariant action, and read off the latter coefficient. The result would be universal! Again introducing a scalar like kinetic term for general spin-$s$ field for integer $s$, we start with the on-shell action:
\eq\label{GravS}
S =\frac{1}{M_{pl}} \int \sqrt{-g} \frac{(-1)^{s}}{2} \left( D^{\mu} \phi^{\nu_1 \cdots \nu_s} D_{\mu} \phi_{\nu_1 \cdots \nu_s} - m^2 \phi^{\nu_1 \cdots \nu_s} \phi_{\nu_1 \cdots \nu_s} \right)\,,
\eqe
where the sign factor $(-1)^s$ is there to make sure that the kinetic term for physical degrees of freedom have the right sign. Note that while additional terms are generally needed to impose tracelessness and transversality condition~\cite{PhysRevD.9.898}, such terms cannot generate non-zero $g_1$ and have been neglected. This can be seen by noting that such terms can be recast into linear combinations of $D_\mu \phi^{\mu\nu_2 \cdots \nu_s}D^\rho \phi_{\rho\nu_2 \cdots \nu_s}$ and $\phi R\phi$ via integration by parts identities, where the latter is a schematic representation with index contractions suppressed. The former term vanishes due to transverse tracelessness of the polarisation tensors, while terms involving the Reimann tensor yields $g_i$ terms with $i>1$~\cite{NewPaper}. Expanding around the flat metric, terms linear in graviton can be separated into two terms:
\eqa
\bar{T}_{\m\n} &=& ({-})^s \left[ (\p_\m \phi^{\s_1 \cdots \s_s}) (\p_\n \phi_{\s_1 \cdots \s_s}) {+} s (\p^\l \phi_{\m}^{~\s_2 \cdots \s_s}) (\p_\l \phi_{\n \s_2 \cdots \s_s}){-} s m^2 \phi_{\m}^{~\s_2 \cdots \s_s} \phi_{\n \s_2 \cdots \s_s} \right] {-} \eta_{\m\n} \CL\nonumber
\\ G^{\m\n\l} &=& \frac{({-})^s s}{2} \left( \phi^{\n \s_2 \cdots \s_s} \p^{\m} \phi^{\l}_{~ \s_2 \cdots \s_s} + \phi^{\n \s_2 \cdots \s_s} \p^{\l} \phi^{\m}_{~ \s_2 \cdots \s_s} - \phi^{\l \s_2 \cdots \s_s} \p^{\m} \phi^{\n}_{~ \s_2 \cdots \s_s} {+} (\m \leftrightarrow \n) \right)
\eqae
where we've separated the piece that stems from expanding $\G^{\l}_{\m\n}$ as $G^{\m\n\l}$. 
The stress tensor is then given as $T_{\m\n}=\bar{T}_{\m\n}-\partial^\l G_{\m\n\l}$. These two sources will contribute to the 3pt amplitude as following terms.
\bl
\bld
- \half h_{\m\n} \bar{T}^{\m\n} &\to x^2\frac{m^2}{M_{pl}}\left( \frac{\la \bold{2} \bold{1} \ra}{m}\right)^s \left(\frac{[ \bold{2} \bold{1} ]}{m}\right)^s
\\ - \half (\p_\l h_{\m\n}) G^{\m\n\l} &\to \frac{sx}{M_{pl}}\left( \frac{\la \bold{2} \bold{1} \ra}{m}\right)^s\left(\frac{[ \bold{2} \bold{1} ]}{m}\right)^{s-1}[\bold{2}3][3\bold{1}]
\eld
\el
Using eq.(\ref{eq:3ptkin1}), we convert the expression into pure chiral form:
\eq\label{Gravy}
x^2\frac{m^2}{M_{pl}} \left(\frac{\la \bold{2}\bold{1} \ra}{m^2}\right)^{2s} - \frac{s(s-1)}{2} \frac{1}{M_{pl}m^2 }\left(\frac{\la \bold{2} \bold{1} \ra}{m}\right)^{2s-2} \la \bold{2} 3 \ra^2 \la 3 \bold{1} \ra^2 + \cdots\,.
\eqe 
Note that there are no $\lambda^2_3$ couplings, so $g_1=0$! Thus it appears that general covariance simply tells us that  $\lambda^2_3$ couplings are forbidden. In section \ref{sec:Compton_amplitude_for _arbitrary_spin}, we will present an alternative on-shell view point of why nonzero $g_1$ is prohibited, this time under the constraint of consistent factorisation.

Finally we comment that the action in eq.(\ref{GravS}) yields deviations from minimal coupling that begins at $\lambda_3^4$, with coefficient $- \frac{s(s-1)}{2}$. Indeed as was pointed out in~\cite{Cucchieri:1994tx}, such action leads to violation of tree-level unitarity for longitudinal scattering. For $s<3$ this can be completely resolved by introducing a new coupling to the Reimann tensor 
\eq
h\frac{s(s-1)}{2}\phi^{\mu \rho \mu_3 \cdots \mu_s} R_{\mu\nu\rho\sigma}\phi^{\nu \sigma}\,_{ \mu_3 \cdots \mu_s}
\eqe 
with $h$ set to 1. We see from the above, this is precisely the requisite choice to cancel the  $\lambda_3^4$ term, consistent with the conclusion that terms beyond minimal coupling lead to bad UV behaviours.  Note that string theory in general has $h\neq1$, as discussed in~\cite{Giannakis:1998wi}, where it evades the UV unitarity disaster by introducing an infinite tower of states whose mass scale is the same as the state in question. 
\subsection{Universality of $g$ for perturbative string states}
The requirement that $g_1=0$ for gravitational couplings has important implications for systems in which the three-point coupling to a graviton is given by the square of the coupling to a photon. More precisely, the spin-$2s$ spin-$2s$ coupling to a graviton is given by the square of three-point amplitude of spin-$s$, spin-$s$ and photon:
\eq
M_3({\bf 1}^{2s},{\bf 2}^{2s},3^{+2})=\left[M_3({\bf 1}^{s},{\bf 2}^{s},3^{+1})\right]^2\,,
\eqe
where we've used bolded numbers to indicate the massive legs, and their exponent indicating their spin. An immediate example is perturbative string theories, where type II closed string amplitudes are given by the square of type I, and similarly closed bosonic string is given by the square of open string. In such case, if $g\neq2$ for the charged states, which implies $g_1\neq0$, then the cross terms in the double copy procedure will lead to $g_1\neq0$ in the gravitation sector. Thus we conclude that for systems with double copy relation between gauge and gravity three point amplitudes, the charged states in the gauge theory sector must have $g=2$. This is applicable to not only leading trajectory, but also all daughter trajectory states. Indeed such result was found previously in~\cite{Ferrara:1992yc}.
\section{Black holes as the $s\gg1$ limit of minimal coupling} \label{sec:BHmincoup}
In light of the discussion in the previous section, we see that if we consider the ``simplest" three-point amplitude with $g_i=0$ for $i>0$, we have the bonus simplicity in the UV: it matches minimal coupling in the UV and has the best high energy behavior. For spin-$\frac{1}{2}$, 1, this is precisely the couplings for particles in the standard model. It is then natural to ask the following:  are there particles in nature with $s>1$ that have such minimal couplings ?

Given the good UV behaviour of string theory, one might expect that the higher spin string resonances would be a perfect candidate. Interestingly it is quite the contrary. For example, the three-point coupling between a photon and the leading trajectory states in open bosonic string theory is given by:
\eq
M_3^{+1,s,s} = x \sum_{n=0}^s \sum_{k=0}^n (\alpha')^{2s-n+\frac{k-1}{2}}\left(\begin{array}{c}s \\n\end{array}\right)\left(\begin{array}{c}n \\ k\end{array}\right)  {s{-}n{-}k{+}1 \over 2^{s{-}n}\left( s{-}n{+}1 \right)!} \langle \textbf{12}\rangle^{2n-k} \left(x\langle \textbf{2}3\rangle\langle 3\textbf{1}\rangle\right)^{2s-2n+k}\,.
\eqe   
One sees that the coupling is ``maximally complex" in that all $g_i\neq 0$ except for $g_1$. Note that this does not violate our discussion with regards to the violation of UV unitarity, since at the energy level where the $\frac{1}{m}$ factor becomes singular, we are at the string scale and the infinite string resonances now come into play. 

Instead of looking to the UV, we consider the IR. For a most general approach, we consider the one body effective action of a point particle coupled to gravity, introduced by Goldberger and Rothstein~\cite{Goldberger:2004jt} and generalised to cases involving spin by Porto~\cite{Porto:2005ac}\footnote{The authors would like to thank Rafael Porto for an explanation on the historical development of the subject.}. This is an effective action where the internal degree of freedom in the object is integrated away, and shows up as ``higher dimensional operators" multiple moments. This is given by the following world-line action:
\eq\label{1bdy}
S=\int d\sigma \;\left\{-m\sqrt{u^2}-\frac{1}{2}S_{\mu\nu}\Omega^{\mu\nu}+L_{SI}\left[u^\mu,S_{\mu\nu}, g_{\mu\nu}(y^\mu)\right]\right\}
\eqe
where $u^\mu\equiv \frac{dy^\mu}{d\sigma}$, $S_{\mu\nu}$ correspond to the spin-operator and $\Omega_{\mu\nu}$ is the angular velocity. The first two terms correspond to minimal coupling and are universal, irrespective of the details of the 
point-like particle, while the terms in $L_{SI}$ correspond to spin-interaction terms that are beyond minimal coupling, and depend on the 
inner structure of the particle. The angular velocity $\O^{\m\n}$ is defined as $\O^{\m\n} := e^\m_A \frac{D e^{A\n}}{D\s}$, where $e^{\m}_A (\s)$ is the tetrad attached to the worldline of the particle. The defining relation for this tetrad is $\eta^{AB} e^{\m}_A (\s) e^{\n}_B (\s) = g^{\m\n}$. Generalising the quadrupole moment operator introduced in \cite{Porto:2006bt}, the non-minimal spin-interaction terms can be parameterized as \cite{Levi:2018nxp}:
\bl
\bld
L_{SI} &= \sum_{n=1}^{\infty} \frac{(-1)^n}{(2n)!} \frac{C_{\text{ES}^{2n}} }{m^{2n-1}} D_{\m_{2n}} \cdots D_{\m_{3}} \frac{E_{\m_{1} \m_{2}} }{\sqrt{u^2}} S^{\m_{1}} S^{\m_{2}} \cdots S^{\m_{2n-1}} S^{\m_{2n}}
\\ &\phantom{=} + \sum_{n=1}^{\infty} \frac{(-1)^n}{(2n+1)!} \frac{C_{\text{BS}^{2n+1}} }{m^{2n}} D_{\m_{2n+1}} \cdots D_{\m_{3}} \frac{B_{\m_{1} \m_{2}} }{\sqrt{u^2}} S^{\m_{1}} S^{\m_{2}} \cdots S^{\m_{2n}} S^{\m_{2n+1}}\,.
\eld \label{eq:obEFTLag}
\el
where $E$ and $B$ are the electric and magnetic components of the Weyl tensor defined as:
\bl
E_{\m\n} &:= R_{\m\a\n\b} u^{\a} u^{\b}
\nonumber\\ B_{\m\n} &:= \half \e_{\a\b\g\m} R^{\a\b}_{~~\delta \n} u^{\g} u^{\delta}\,,
\el
Note that here the Riemann tensors are taken to be linear perturbations around flat space, and the information with regards to non-trivial backgrounds is encoded in the Wilson coefficients $C_{\#}$. For generic astrophysical objects the Wilson-coefficients are obtained by matching with the multipole moments used in numerical simulations. For Kerr black-holes the coefficient is 1, which we review in appendix \ref{sec:BHWilson}.

\subsection{Universal part of the one body EFT }
We first consider the terms besides $L_{SI}$ in eq.(\ref{1bdy}) which are universal for all particles. The spin-independent part of minimal coupling is given by $L = - m \sqrt{u^2}$, 
\bl
- m \sqrt{u^2} = - m \sqrt{\eta_{\m\n} u^\m u^\n + \k h_{\m\n} u^\m u^\n} = - \frac{\k m}{2} h_{\m\n} u^\m u^\n +\mathcal{O}(h^2) \label{eq:ptptcaction}\,.
\el
Keeping only linear order in $h_{\m\n}=2\ve_{\m}\ve_{\nu}$\footnote{The extra factor of $2$ was inserted to make equations simpler.}, and identifying $x = \sqrt{2} ( \ve^+ \cdot u)$, this term simply yields the scalar three-point interaction:
\eq
- m \sqrt{u^2}=-\frac{\kappa m x^2}{2}\,.
\eqe
Next we consider the minimal spin coupling $ - \half S_{\m\n} \O^{\m\n}$, given as \cite{Porto:2005ac,Levi:2015msa},
\bl
- \half S_{\m\n} \O^{\m\n} = - \half S_{AB} \omega^{~AB}_\m u^\m\,.
\el
As usual the spin connection  $\omega_{\m~B}^{~A}$ is defined as $\omega_{\m~B}^{~A} =  e_{B \n} \p_\m e^{A \n} + e_{B \n} \G^{\n}_{\m\l} e^{A \l}$.  Since we are only interested in the three point amplitude with one graviton, the derivative on the tetrad will not contribute, and we have:
\bl
- \half S_{AB} \omega^{~AB}_\m u^\m &= - \half S_{\m\n} u^\l \G^{\n}_{\l\s} g^{\s\m} \label{eq:minspinint}
\el
In classical mechanics, the spin $S^{\m\n}$ can be defined as the difference between the angular momentum that the orbital part:
\eq\label{InSSC}
J^{\mu\nu}=X^{[\mu}P^{\nu]}+S^{\m\n}\,.
\eqe 
However, this separation is ambiguous without a reference frame to define the origin and hence $X$. The choice of the origin can be translated into an additional constraint on $S^{\m\n}$ known as \emph{spin supplementary condition} (SSC). Of the various choices for SSC known in the literature, one that can be generalised to curved space without any ambiguity is the \emph{covariant} SSC $S^{\m\n}p_\n = 0$, also known as Tulczyjew SSC or Tulczyjew-Dixon SSC. Adoption of this condition can be met by the following choice of $S^{\m\n}$, where the vector $u^\m$ is defined as $u^\m = \frac{p^\m}{m}$.
\bl
S^{\m\n} &= J^{\m\n} + u^\m J^{\n\l} u_\l - u^\n J^{\m\l} u_\l \,.
\el
Note that this spin operator can be cast in the form $S^{\m\n} = - \frac{1}{m} \e^{\m\n\l\s} p_\l S_\s$, where $S^\m$ is the Pauli-Lubanski psuedo-vector eq.\eqc{eq:PLvecdef}. The 3pt amplitude can be computed by adopting the following definitions and replacement rules:
\bl
\bld
\G^{\m}_{\n\l} &= \frac{\k}{2} \left[ h^{\m}_{~\n,\l} + h^{\m}_{~\l,\n} - h_{\n\l}^{~~,\m} \right]
\\ u^\m &= \frac{1}{m} p_1^\m
\\ \partial_\m &\Rightarrow -i p_{3\m}
\\ h_{\m\n} &\Rightarrow 2 \varepsilon^+_\m \varepsilon^+_\n\,.
\eld
\el
Combined with three particle kinematics, eq.\eqc{eq:minspinint} becomes
\eq
 - \frac{\k}{2} x \left[ - i p_{3\m} ( \sqrt{2} \varepsilon^+_{\n} - x u_\n ) J^{\m\n} \right] =   - \frac{x^2}{2} \ket{3} \bra{3}=- \frac{\k m x^2}{2} \left( - \frac{p_3 \cdot S}{m} \right)^{~\b}_{\a}\label{eq:minspinamplag}
\eqe
where we've used the spin-$\frac{1}{2}$ representation of the Lorentz generator in the chiral representation, and eq.\eqc{PSDef}  to obtain the rightmost expression. Generalisation to higher spin follows from eq.\eqc{eq:spinophs}, which gives $\left( - \frac{p_3 \cdot S}{m} \right)$ with the understanding that,
\bl
\left( \frac{p_3 \cdot S}{m} \right)_{\a_1 \a_2 \cdots}^{~~~\b_1 \b_2 \cdots} &= \frac{xs}{m} \ket{3} \bra{3}
\\ \left( \frac{p_3 \cdot S}{m} \right)^{\dot{\a}_1 \dot{\a}_2 \cdots}_{~~~\dot{\b}_1 \dot{\b}_2 \cdots} &= - \frac{s}{mx} \sket{3} \sbra{3}\,.
\el
A caveat with using this form is that higher degree of this operator must be evaluated from the definition of Lie algebra eq.\eqc{eq:largespinop}. For example,
\bl
\bld
\left[ \left( \frac{p_3 \cdot S}{m} \right)^2 \right]_{\a_1 \cdots \a_{2s}}^{~~~~\b_1 \cdots \b_{2s}} &= 2s(2s-1) \frac{x^2}{(2m)^2} \ket{3}_{\a_1} \ket{3}_{\a_2} \bra{3}^{\b_1} \bra{3}^{\b_2}
\\ \left[ \left( \frac{p_3 \cdot S}{m} \right)^3 \right]_{\a_1 \cdots \a_{2s}}^{~~~~\b_1 \cdots \b_{2s}} &= 2s(2s-1)(2s-2) \frac{x^3}{(2m)^3} \ket{3}_{\a_1} \ket{3}_{\a_2} \ket{3}_{\a_3} \bra{3}^{\b_1} \bra{3}^{\b_2} \bra{3}^{\b_3}
\eld \label{eq:SpinProdNumFacts}
\el
when symmetrization is taken into account.

Thus the universal piece of the 1 body EFT translates into the following three-point interaction:\footnote{{When the graviton is chosen to have negative helicity, the sign of $p_3 \cdot S$ term flips.}}
\eq\label{Universal}
- \frac{\k m x^2}{2} \left(\mathbb{I}- \frac{p_3 \cdot S}{m} \right)\,.
\eqe
where both the operator $\mathbb{I}$ and $p_3 \cdot S$ are defined to act on the Hilbert space of $SL(2,\IC)$ irreps.

\subsection{The three-point amplitude from $L_{SI}$}
We now consider the three-point amplitude arising from the Wilson operators in eq.(\ref{eq:obEFTLag}). The electric and magnetic components of the Weyl tensor are converted to:
\bl
\bld
 E_{\m\n} &\Rightarrow \frac{\k x^2}{2} p_{3\m} p_{3\n}
\\ B_{\m\n} S^{\m} &\Rightarrow \frac{\k x}{2} \left[ p_{3\a} ( \sqrt{2} \varepsilon^+_{\b} - x u_\b ) J^{\a\b} \right] p_{3\n}
\eld
\el
The one-body EFT Lagrangian eq.\eqc{eq:obEFTLag} then translates to the following form for three-particle kinematics:\footnote{{The sign of $k_3 \cdot S$ term in the last line flips when negative helicity is chosen for the graviton, which is consistent with the sign flip in eq.\eqc{Universal}.}}
\bl
\bld
 & - \sum_{n=1}^{\infty} \frac{C_{\text{ES}^{2n}}}{(2n)!} \frac{\k m x^2}{2} \left( \frac{p_3 \cdot S}{m} \right)^{2n} - \sum_{n=1}^{\infty} \frac{C_{\text{BS}^{2n+1}}}{(2n+1)!} \frac{\k x}{2} \left[ - i p_{3\a} ( \sqrt{2} \varepsilon^+_{\b} - x u_\b ) J^{\a\b} \right] \left( \frac{p_3 \cdot S}{m} \right)^{2n}
\\ & \hskip 10pt = -\sum_{n=2}^{\infty} \frac{\k m x^2}{2} \frac{C_{\text{S}^n}}{n!} \left( - \frac{p_3 \cdot S}{m} \right)^{n}
\eld \label{eq:SIspincoup}
\el
The Wilson coefficients $C_{\text{S}^n}$ are defined as $C_{\text{S}^{2m}} = C_{\text{ES}^{2m}}$ for even $n = 2m$ and $C_{\text{S}^{2m+1}} = C_{\text{BS}^{2m+1}}$ for odd $n = 2m + 1$. 
It is possible to add the universal pieces in eq.\eqc{Universal}, so that the sum starts from $n=0$, with the definition $C_{\text{S}^0} = C_{\text{S}^1} = 1$.

We will be interested in the three-point scattering amplitude of a spin-$s$ particle emitting a graviton described by the effective action eq.(\ref{1bdy}). Again the incoming and out going momenta will be $p_1,p_2$, while the graviton being $p_3$. The polarization tensor for a spin-$s$ particle is given by:
\eq
\varepsilon^{(I_1\cdots I_sJ_1\cdots J_s)}_{\alpha_1\dot\alpha_1\cdots \alpha_s\dot\alpha_s}=2^{s/2}\frac{\lambda^{(I_1}_{\alpha_1}\cdots\lambda^{I_s}_{\alpha_s}\tilde\lambda^{J_1}_{\dot\alpha_1}\cdots\tilde\lambda^{J_s)}_{\dot\alpha_s}}{m^s}\,,
\eqe  
where the total symmetrization of the Little group indices ensures the transversality of the polarization tensor. As the polarization tensors are contracted with the operators in the effective action, terms with spin-operator of degree $n$, with $n\leq s$, will contribute. Furthermore, for each fixed $n$, we sum over the all possible distributions of the $n$ spin operators between the chiral and anti-chiral indices of the polarization tensor. This results in the following three-point amplitude:\footnote{The irrelevant overall factor of $2^s$ has been neglected.}  
\eq
M_3^{+2,s,s} =\sum_{a+b\leq s}\; x^2 C_{\text{S}^{a+b}}  \tilde{c}^{s}_{a,b}\lan \bold{2} \bold{1} \ra^{s-a} \left( - \frac{x \lan \bold{2}3 \ra \lan 3\bold{1} \ra}{2m} \right)^a [ \bold{2} \bold{1} ]^{s-b} \left( \frac{[ \bold{2}3 ] [ 3\bold{1} ]}{2mx} \right)^b,\quad \tilde{c}^{s}_{a,b}\equiv\left( \begin{array}{c} s \\ a \end{array} \right) \left( \begin{array}{c} s \\ b \end{array} \right)  \label{eq:WilsonCoeff3ptExp}
\eqe
where the $+$ subscript indicates that this is the plus helicity graviton amplitude, and we denote the combinatoric factors as $\tilde{c}^{s}_{a,b}$ for reasons that will be clear shortly. In a sense this provides an alternative parameterization for the general three-point amplitude, where the information of the specific interaction is encoded in the Wilson coefficients $C_{\text{S}^{a+b}}$. Again for Kerr black holes they are unity.

\subsection{The matching to minimal coupling}
We are now ready to recast our minimal coupling to the above EFT basis. While minimal coupling for the positive helicity graviton is simple in the chiral basis, the EFT basis in eq.(\ref{eq:WilsonCoeff3ptExp}) is in the symmetric basis. To convert the chiral basis into the symmetric basis, we use the following identity, 
\bl
\lan \bold{2} \bold{1} \ra^2 = \lan \bold{2} \bold{1} \ra [ \bold{2} \bold{1} ]  + \lan \bold{2} \bold{1} \ra \frac{[ \bold{2}3 ][ 3\bold{1} ]}{2mx} - \frac{x\lan \bold{2}3 \ra \lan 3\bold{1} \ra}{2m} [ \bold{2} \bold{1} ] - \frac{2 \lan \bold{2}3 \ra \lan 3\bold{1} \ra [ \bold{2}3 ][ 3\bold{1} ]}{(2m)^2} \label{eq:angsymexp}
\el
This relation can be readily generalised to integer higher spin.
\bl
\lan \bold{2} \bold{1} \ra^{2s} &= \left( \lan \bold{2} \bold{1} \ra^2 \right)^s = \sum_{a,b=0}^{s} c^{s}_{a,b} \lan \bold{2} \bold{1} \ra^{s-a} \left( - \frac{x \lan \bold{2}3 \ra \lan 3\bold{1} \ra}{2m} \right)^a [ \bold{2} \bold{1} ]^{s-b} \left( \frac{[ \bold{2}3 ] [ 3\bold{1} ]}{2mx} \right)^b \label{eq:ElemCoupl3ptExp}
\el
Note that the ratio of $c^{s}_{a,b}$ with respect to $\tilde{c}^{s}_{a,b}$ yields the Wilson coefficients for the minimal coupling. The coefficient $c^{s}_{a,b}$ can be readily computed by identifying it as simply the coefficients of $(1 + x + y + 2xy)^s$,
\bl
(1 + x + y + 2xy)^s &= \sum_{a,b=0}^{s} c^{s}_{a,b} ~ x^a y^b
\\ c^{s}_{a,b} &= \sum_{c=0}^{\min (a,b)} \frac{2^c s!}{(s-a-b+c)!(a-c)!(b-c)!c!}
\el
Note that if $2^c$ in $c^{s}_{a,b}$ was substituted by $1$, which is equivalent to using $(1+x+y+xy)^s$ to evaluate $c^{s}_{a,b}$, then we would simply have ${\tilde{c}^{s}_{a,b}} = c^{s}_{a,b}$! This observation can be used to derive the following formula.
\bl
\bld
c^{s}_{a,b} &= \sum_{i=0}^{\min(a,b)} \left( \begin{array}{c} s \\ i \end{array} \right) \tilde{c}^{s-i}_{a-i,b-i}
\\ &= \tilde{c}^{s}_{a,b} + s \tilde{c}^{s-1}_{a-1,b-1} + \frac{s(s-1)}{2} \tilde{c}^{s-2}_{a-2,b-2} + \cdots
\eld
\el
Since $\tilde{c}^{s}_{a,b}$ tends to $\frac{s^{a+b}}{a! b!}$ for asymptotically large $s$, each term in the series is subleading in powers of $\frac{1}{s}$ for fixed set of $a$ and $b$. In other words,
\eq
\framebox[6cm][c]{$\displaystyle c^{s}_{a,b} = \tilde{c}^{s}_{a,b} (1 + \CO(1/s))\,.$}
\eqe
There are no $1/s$ corrections when either $a$ or $b$ of $c^{s}_{a,b}$ is zero; $c^{s}_{a,0} = \tilde{c}^{s}_{a,0}$ and $c^{s}_{0,b} = \tilde{c}^{s}_{0,b}$. It is worthy of note that since $C_{\text{S}^1}$ is fixed to be unity, $c^{s}_{1,0} = \tilde{c}^{s}_{1,0}$ and $c^{s}_{0,1} = \tilde{c}^{s}_{0,1}$; these conditions imply that introduction of $M_3^{2+} \supset x^3 \lan \bold{2} \bold{1} \ra^{2s-1} \lan \bold{2} 3 \ra \lan 3\bold{1} \ra$ term in the graviton 3pt amplitude, or introduction of non-zero $g_1$, is forbidden in this context as well. 

Thus we see that in the $s\gg1$ limit, the minimal coupling reproduces the Wilson coefficient of a Kerr black hole! The fact that one should take the large spin limit is not surprising since the spin of a black hole takes a macroscopic value. The reader might wonder that since the matching is occurring at the large spin limit, it may very well be that deviation from minimal coupling is subleading in $s$ and hence suppressed. In such case, the matching of minimal coupling to black holes is simply a reflection of it being the leading contribution in the limit. We now show this is not the case.

The simplest deformation from minimal coupling is introducing $\l^4$ coupling to the 3pt amplitude. The three-point amplitude then becomes  
\bl
\bld
M _3^{+2,s,s} &= x^2 \lan \bold{2} \bold{1} \ra^{2s} + g_2 x^2 \frac{x^2}{m^2} \lan \bold{2} \bold{1} \ra^{2s-2} \lan \bold{2} 3 \ra^2 \lan 3 \bold{1} \ra^2
\\ &= \sum_{a,b=0}^{s} B^{s}_{a,b} \lan \bold{2} \bold{1} \ra^{s-a} \left( - \frac{x \lan \bold{2}3 \ra \lan 3\bold{1} \ra}{2m} \right)^a [ \bold{2} \bold{1} ]^{s-b} \left( \frac{[ \bold{2}3 ] [ 3\bold{1} ]}{2mx} \right)^b
\eld
\el
$B^{s}_{a,b}$ is determined to be $B^{s}_{a,b} = c^{s}_{a,b} + 4 g_2 c^{s-2}_{a-1,b-1}$. The Wilson coefficients $C_{\text{S}^n}$ for asymptotic $s$ is then given as 
\bl
B^{s}_{a,b} = \tilde{c}^{s}_{a,b} (1 + 4 ab \tilde{g}_2 + \CO(1/s)) \Longrightarrow C_{\text{S}^n} = 1 + \frac{2 n (n - 1)}{3}\tilde{g}_2 + \CO(1/s)
\el
where $\tilde{g}_2=g_2/s^2$. The natural value for $g_2$ can be deduced from eq.(\ref{Gravy}) to be $\sim s^2$ in the large $s$. Thus we see that introducing terms that generate deviations to minimal coupling does indeed modify the Wilson coefficients from the black hole value.

Note that this is consistent with the intuition that the terms beyond minimal couplings represent finite size effects that indicate deviation from point particle. In other words, the fact that black holes are given by minimal coupling is a kinematic way of saying that it has no ``hair".

\section{Compton amplitudes for arbitrary spin}\label{sec:Compton_amplitude_for _arbitrary_spin}
Consistent factorization at four-points often imposes new constraints for the underlying theory that are not visible at three-points. For example, the color algebra associated with non-abelian theories can be recovered by simply enforcing that the residue from one factorization channel can be made consistent with that of another~\cite{Benincasa:2007xk, Arkani-Hamed:2017jhn}. For massive amplitudes, the application of such consistency condition has been initiated in~\cite{Arkani-Hamed:2017jhn}, which led to bounds on the spin of isolated massive particles. Here we will systematically construct the Compton amplitude, as well as its gravitational counterpart, for general massive spin-$s$ particle, utilizing consistent factorizations. The gravitational Compton amplitude will later serve an important ingredient in extracting the spin-dependent piece of the 2 PM potential.

Let us first give an over view of our strategy. We will start from gluing the known 3pt amplitudes on $s$-channel together. Putting the result on an $s$-channel propagator gives a putative ansatz for the four-point amplitude:
\eq
Ansatz=\frac{M_3(1,2,P)\times M_3(3,4,-P)}{s-m^2}\,.
\eqe
Without loss of generality, we take legs $1$ and $4$ to be the massive spin-$s$ state, and legs $2$ and $3$ to be either photons or gravitons. For minimal couplings, the gluing on a specific channel is not local, reflecting the presence of another factorization channel. Thus we will need to check whether the factorization constraint on the other channel is also satisfied. If $s$ channel gluing in our Compton amplitude carries $u$-channel information,\footnote{For the discussion with regards to amplitudes, we follow the notation that $s=(p_1+p_2)^2, t=(p_1+p_4)^2$ and $u=(p_1+p_3)^2$.} correct factorization in the $u$-channel is guaranteed if the $s$-channel residue is given in a form symmetric under $(1 \leftrightarrow 4)$ exchange. In general for photon Compton amplitude, we will find:
\begin{equation}\label{eq:Photon_s_res_ansatz}
\begin{split}
& M_3(\textbf{1},2^{+1}, P)\times M_3(-P,3^{-1},\textbf{4}) \Big|_{s= m^2} = \frac{f_{su}}{u - m^2} + f_s \\
\Rightarrow & Ansatz =  \frac{f_{su}}{\sm \um} + \frac{f_s}{s - m^2} + \frac{f_u}{u - m^2} \,,
\end{split}
\end{equation}
where $f_u$ can be deduced from $f_s$ via $1\leftrightarrow 4$ symmetry.  For graviton Compton amplitudes, we will find:
\begin{equation}\label{eq:Graviton_s_res_ansatz}
\begin{split}
& M_3(\textbf{1},2^{+2}, P)\times M_3(-P,3^{-2},\textbf{4}) \Big|_{s= m^2} = \frac{f_{stu}}{(u - m^2)t} + \frac{ f_{st} }{t} + f_s \\
\Rightarrow & Ansatz =  \frac{f_{stu}}{\sm (u - m^2)t} + \frac{ f_{st} }{\sm t} + \frac{ f_{ut} }{\um t} + \frac{f_s}{s - m^2} +  \frac{f_u}{u - m^2} + \frac{f_t}{t} \,.
\end{split}
\end{equation}
This procedure fixes the four-point amplitude up to polynomial terms, which do not have poles and therefore are not subject to previous constraints. Importantly, for (photon) graviton couplings $s\leq (1) 2$ the possible polynomials must be of higher order in $\frac{1}{m}$ suppressions, which reflects the fact that these are finite size effects. For $s>(1) 2$, the order of $\frac{1}{m}$ for such ambiguity is of the same order as terms in the $ Ansatz $. Thus the result given here are ``correct" only up to polynomial ambiguities for charged spinning particles with $s>1$, and gravitationally coupled spin states with $s>2$.

\subsection{Photon}\label{subsec:Photon}

The minimal coupling 3pt amplitude of a photon with 2 massive spin $s$ particles is given by:
\begin{equation}\label{eq:Minimal_Photon_3pt}
M^{+1,s,s}_3 = x\frac{\AB{\BS{12}}^{2s}}{m^{2s-1}}, \qquad M^{-1,s,s}_3 = \frac{1}{x} \frac{\SB{\BS{12}}^{2s}}{m^{2s-1}}
\end{equation}

\subsubsection{Photon Compton Amplitude with $s \leq 1$}\label{subsubsec:Photon_Lower_Spin}

\noindent
$s$-channel gluing gives:
\begin{equation}\label{eq:Photon_s_Channel_residue}
\begin{split}
M_3(\textbf{1},2^{+1}, P) \times M_3(-P,3^{-1},\textbf{4})
& = \frac{1}{m^{2(2s-1)}} \frac{x_{12}}{x_{34}} ( \MixLeft{\BS{1}}{P}{\BS{4}} )^{2s}\\
& = -\frac{\MixLeft{3}{p_1}{2}^{2-2s}}{t} (\SpDept)^{2s}
\end{split}
\end{equation}
\noindent
with $P$ as the momentum of the $s$-channel propagator and the second equality in eq.\eqref{eq:Photon_s_Channel_residue} comes from solving the conditions :
\begin{equation}\label{eq:P_on-shell_condition}
P_{\alpha \dot{\alpha}}\tilde{\lambda}_2^{\dot{\alpha}} = -m x_{12}\lambda_{2\alpha}, \quad P_{\alpha \dot{\alpha}}\tilde{\lambda}_3^{\dot{\alpha}} = m x_{34}\lambda_{3\alpha}, \quad
P^2 = m^2
\end{equation}
yielding
\begin{equation}\label{eq:x_over_x}
\MixLeft{\BS{1}}{P}{\BS{4}} = m^2 \frac{\SpDept}{\MixLeft{3}{p_1}{2}}
\end{equation}
and by the definition of the $x$-factors:
\begin{equation}
\frac{x_{12}}{x_{34}} = - \frac{\MixLeft{3}{p_1}{2}^2}{m^2 t}
\end{equation}
Since there's no 3-photon interaction to be considered in the $t$ channel, we identify the $t$ in the denominator as $-\um$. Putting back the $\sm$, we obtain an ansatz for photon Compton amplitudes:
\begin{equation}\label{eq:Lower_Spin_Photon_Compton_Ansatz}
Ansatz = \frac{\MixLeft{3}{p_1}{2}^{2 - 2s}}{\sm \um} (\SpDept)^{2s}
\end{equation}
for $s \leq 1$ this is precisely the Compton amplitudes. On the other hand, for $s>1$, there will be spurious poles $\ThreePTwo$ in the denominator and the ansatz ceases to be local. Then we conclude for $s\leq1$,  
\begin{equation}\label{eq:Lower_Spin_Photon_Compton}
M(\boldsymbol{1}^{s},2^{+1},3^{-1},\boldsymbol{4}^{s}) = \frac{\MixLeft{3}{p_1}{2}^{2 - 2s}}{\sm \um} (\SpDept)^{2s}\,.
\end{equation}

\subsubsection{Photon Compton Amplitude with $s > 1$}\label{subsubsec:Photon_Higher_Spin}
For higher spin charged particles, we need to more work to find a completely local ansatz. The assumption that went into eq.\eqref{eq:Lower_Spin_Photon_Compton_Ansatz} was that we used a representation of $P$ such that it matches both the s and u-channel residue. This anticipates the fact that the $s$-channel residue sits on top of a $u$-channel pole as well. However, it is also possible that part of the $s$-channel residue is in fact local, and thus do not need to satisfy any $u$-channel constraint. Thus the task becomes that of separating the residue into local and non-local pieces. Again starting with the $s$-channel residue:
\begin{equation}\label{eq:Photon_Higher_Spin_s_Res}
Res[M(\boldsymbol{1}^{s},2^{+1},3^{-1},\boldsymbol{4}^{s})] \Big|_{s=m^2} = \frac{ \MixLeft{3}{p_1}{2}^2 }{u-m^2} \Bigg( \frac{\AB{\boldsymbol{4} 3}\SB{\boldsymbol{1}2} + \AB{\boldsymbol{1} 3}\SB{\boldsymbol{4}2}}{\MixLeft{3}{p_1}{2}} \Bigg)^{2s}
\end{equation}
\noindent
We rewrite the term in the parenthesis, which is simply $\MixLeft{\BS{1}}{P}{\BS{4}}$, as 
\begin{equation}\label{eq:almost_local_identity_symmetric}
\begin{split}
\MixLeft{\BS{1}}{P}{\BS{4}} 
&= \frac{1}{2} \Bigg( \frac{\SB{ \BS{14} }}{m} + \frac{ \AB{\BS{4}2}\SB{2\BS{1}} - \AB{\BS{1}2}\SB{2\BS{4}} }{2m^2} \Bigg) + \frac{1}{2}\Bigg( \frac{\AB{ \BS{14} }}{m} + \frac{ \AB{\BS{1}3}\SB{3\BS{4}} - \AB{\BS{4}3}\SB{3\BS{1}} }{2m^2} \Bigg)  + \frac{t\AB{3\boldsymbol{4}} \SB{2\boldsymbol{1}}}{2m^2\MixLeft{3}{p_1}{2}} \\
&\equiv \frac{1}{2}(F + \tilde{F}) + B \equiv \mathcal{F} + B
\end{split}
\end{equation}
\noindent 
where $\mathcal{F}$ is written in a way that is symmetric under angle square exchange for the massive legs. Now, the $\mathcal{F}$ term is completely local and satisfies the correct spin-statistics property under $(1 \leftrightarrow 4)$, with the price of introducing extra factors of $m$ in the denominator. In expanding $(\mathcal{F}+B)^{2s}$, one of the $B$ factor in the $B$ dependent terms will cancel the $u$ pole, since $-t=u-m^2$ when $s=m^2$, and its spurious pole will be canceled by the prefactor. Thus these terms will be pure $s$-channel terms. The remaining $B$ factors still contain unphysical pole, but can be removed by imposing $s$-channel kinematics:
\begin{equation}\label{eq:B_identity}
\begin{split}
\frac{t \SB{2 \boldsymbol{1}} \AB{3 \boldsymbol{4}}}{2m^2\MixLeft{3}{p_1}{2}}\Bigg|_{s=m^2} 
&= -\frac{\AB{\boldsymbol{4}3} \SB{32} \AB{2 \boldsymbol{1}}}{4m^3} - \frac{ \SB{ \BS{4}3 } \AB{32} \SB{2 \BS{1} } }{4m^3} \\
\end{split}
\end{equation}
\noindent 
\noindent 
We now see that only the $\mathcal{F}^{2s}$ term carries both $s$ and the $u$ channel poles. The pure $u$-channel term will be fixed by $(1 \leftrightarrow 4)$ symmetry.

Now we conclude that the photon Compton amplitude for $s>1$ to be:
\begin{tcolorbox}[title = Photon Compton Amplitude for $s > 1$, fontupper=\footnotesize]
\begin{equation}\label{eq:Photon_Higher_Spin_Compton_Amplitude}
\begin{split}
M(\boldsymbol{1}^{s},2^{+1},3^{-1},\boldsymbol{4}^{s}) =
&\frac{\MixLeft{3}{p_1}{2}^2}{(s - m^2)(u - m^2)}\mathcal{F}^{2s} \\
& - \Bigg\{\frac{\MixLeft{3}{p_1}{2} \AB{3 \boldsymbol{4}} \SB{2 \boldsymbol{1}}}{2m^2 (s - m^2)} \Bigg[\sum_{r=1}^{2s}\binom{2s}{r} \mathcal{F}^{2s-r}\Big(\frac{ -\AB{\boldsymbol{4}3} \SB{32} \AB{2 \boldsymbol{1}}}{4m^3} - \frac{ \SB{\boldsymbol{4}3} \AB{32} \SB{2 \boldsymbol{1}}}{4m^3} \Big)^{r-1}\Bigg] \\
& +  \frac{\MixLeft{3}{p_1}{2} \AB{3 \boldsymbol{1}} \SB{2 \boldsymbol{4}}}{2m^2 (u - m^2)} \Bigg[\sum_{r=1}^{2s}\binom{2s}{r} (-1)^r\mathcal{F}^{2s-r}\Big(\frac{ -\AB{\boldsymbol{1}3} \SB{32} \AB{2 \boldsymbol{4}}}{4m^3} - \frac{ \SB{\boldsymbol{1}3} \AB{32} \SB{2 \boldsymbol{4}}}{4m^3} \Big)^{r-1}\Bigg] \Bigg\}
\end{split}
\end{equation}
\end{tcolorbox}
\noindent where we dropped the $\sm$ terms in eq.\eqref{eq:B_identity} since it would not contribute to the residue at any poles.

\subsection{Graviton}\label{subsec:Graviton}

The minimal coupling 3pt amplitude of a graviton with 2 massive spin $s$ particles is given by:
\begin{equation}\label{eq:Minimal_Graviton_3pt}
M^{+2,s,s}_3 = x^2 \frac{1}{M_{pl}} \frac{\AB{\BS{12}}^{2s}}{m^{2s-2}}, \qquad M^{-2,s,s}_3 = \frac{1}{x^2} \frac{1}{M_{pl}} \frac{\SB{\BS{12}}^{2s}}{m^{2s-2}}
\end{equation}
\noindent

\subsubsection{Graviton Compton Amplitude for $s \leq 2$}\label{subsubsec:Graviton_Lower_Spin}
We again start out with $s$-channel gluing of the graviton Compton amplitude, 
\begin{equation}\label{eq:Graviton_s_Channel_residue}
\begin{split}
M_3(\textbf{1},2^{+2}, P)\times M_3(-P,3^{-2},\textbf{4})
&= \frac{1}{m^{2(2s-2) M_{pl}^2}} \frac{ x_{12}^2 }{ x_{34}^2 } \MixLeft{\BS{1}}{P}{\BS{4}}^{2s} \\
&= \frac{\MixLeft{3}{p_1}{2}^{4-2s}}{t^2 M_{pl}^2} (\SpDept)^{2s}
\end{split}
\end{equation}
\noindent
where eq.\eqref{eq:P_on-shell_condition} and eq.\eqref{eq:x_over_x} is applied in the second equality in eq.\eqref{eq:Graviton_s_Channel_residue}. The double pole $t^2$ in the denominator can be identified as one massive $u$-channel pole and one massless $t$ channel pole comming from the 3-graviton interaction. So the ansatz for graviton Compton amplitude is 
\begin{equation}\label{eq:Lower_Spin_Graviton_Compton_Ansatz}
Ansatz = -\frac{\MixLeft{3}{p_1}{2}^{4 - 2s}}{\sm \um t M_{Pl}^2} (\SpDept)^{2s}\,.
\end{equation}
Note that now we should also check that this ansatz correctly factorizes in the $t$ channel. Here, we should match both the MHV $(\SB{23} = 0)$ and anti-MHV $(\AB{23} = 0)$ $t$-channel residues:
\begin{equation}\label{eq:MHV_t_res}
M_3(2^{+2}, 3^{-2}, P^{-2}) M_3(-P^{+2}, \textbf{1},\textbf{4}) = -\frac{\MixLeft{3}{p_1}{2}^4}{\sm \um} \frac{ \AB{\BS{14}}^{2s} }{m^{2s}} \equiv -\frac{\MixLeft{3}{p_1}{2}^4}{\sm \um} \tilde{F}_1^{2s}
\end{equation}
\begin{equation}\label{eq:anti_MHV_t_res}
M_3(2^{+2}, 3^{-2}, P^{+2}) M_3(-P^{-2}, \textbf{1},\textbf{4}) = -\frac{\MixLeft{3}{p_1}{2}^4}{\sm \um} \frac{ \SB{\BS{14}}^{2s} }{m^{2s}} \equiv -\frac{\MixLeft{3}{p_1}{2}^4}{\sm \um} {F}_1^{2s}
\end{equation}
which is indeed the case. Now that our ansatz eq.\eqref{eq:Lower_Spin_Graviton_Compton_Ansatz} consistently factorizes in all three channels and that it contains no other poles for $s \leq 2$, it gives us the graviton Compton amplitude:
\begin{equation}\label{eq:Lower_Spin_Graviton_Compton}
M( \BS{1}^{s}, 2^{+2}, 3^{-2}, \BS{4}^{s} ) = -\frac{\MixLeft{3}{p_1}{2}^{4 - 2s}}{\sm \um t M_{Pl}^2} (\SpDept)^{2s} \quad \text{for} \quad s \leq 2
\end{equation}
Again there will be spurious poles when $s > 2$ and thus need to be further taken care of.
\begin{tcolorbox}[colframe=gray, floatplacement=h!, float, title = Definition of variables, fontupper=\footnotesize]
\label{Box:Definition of Variables}
The variables we'll be using are defined as follow:
$$F_1 = \frac{\SB{ \BS{14} }}{m}, \quad F_2 = \frac{ \AB{\BS{4}2}\SB{2\BS{1}} - \AB{\BS{1}2}\SB{2\BS{4}} }{2m^2}$$
$$\tilde{F}_1 = \frac{\AB{ \BS{14} }}{m}, \quad \tilde{F}_2 =  \frac{ \AB{\BS{1}3}\SB{3\BS{4}} - \AB{\BS{4}3}\SB{3\BS{1}} }{2m^2} $$
$$\mathcal{F} = \frac{1}{2}(F + \tilde{F}), \quad \mathcal{F}_1 = \frac{1}{2}(F_1 + \tilde{F}_1), \quad \mathcal{F}_2 = \frac{1}{2}(F_2 + \tilde{F}_2)$$
The $\mathcal{F}$ is defined such that it remains invariant under $(F_1 \leftrightarrow \tilde{F}_1)$ and $(F_2 \leftrightarrow \tilde{F}_2)$. Also, for the pure $t$-channel terms, we'll be needing:
$$
C_{\SB{23}} = \frac{\SB{23}\AB{\boldsymbol{1}3}\AB{3\boldsymbol{4}}}{m}, \quad C_{\AB{23}} = - \frac{\AB{23}\SB{\boldsymbol{1}2}\SB{2\boldsymbol{4}}}{m}
$$
and
$$
K \equiv  \frac{\AB{3 \boldsymbol{4}} \SB{2 \boldsymbol{1}}}{2m^2} - \frac{\AB{3 \boldsymbol{1}} \SB{2 \boldsymbol{4}}}{2m^2} 
$$
in the functions:
\begin{equation}\nonumber
h(n) \equiv \frac{K^2 C^2}{2^{n-1}} \binom{2s}{n+1} \mathcal{F}_1^{2s-n-1} 
\end{equation}
\begin{equation}\nonumber
\begin{split}
g(n)  
& \equiv  - \frac{K^2 C^2}{2^n}  \sum_{r=1}^{2s-n-1}(2r+1) \binom{2s}{r+n+1} \mathcal{F}_1^{2s-r-n-1} \mathcal{F}_2^{r-1}\\
& \qquad \quad + \Big( \frac{s-u}{2} \Big) \frac{K^3 C}{2^{n-1}}  \sum_{r=1}^{2s-n-1}(r+1) \binom{2s}{r+n+1} \mathcal{F}_1^{2s-n-1-r} \mathcal{F}_2^{r-1}  
\end{split}
\end{equation}
We'll be taking $C = C_{\AB{23}}$ for $g_A(n)$ and $h_A(n)$, $C = C_{\SB{23}}$ for $g_S(n)$ and $h_s(n)$, with $g(n)$ satisfying $g(n \geq 2s-1) = 0$.
They will be used in the numerator of the pure-$t$ channel:
\begin{equation}\nonumber
Poly = - \MixLeft{3}{p_1}{2}^2 K^2  \sum_{r=1}^{2s-1} r \binom{2s}{r+1} \mathcal{F}_1^{2s-r-1} \mathcal{F}_2^{r-1}
\end{equation}
\begin{equation}\nonumber
Poly_{\AB{23}} = \sum_{r=0}^{ \lceil s \rceil - 3} h_{A}(4 + 2r) ( F_1 - \tilde{F}_1 )^{2r+1} +  \sum_{r=0}^{ \lceil s \rceil - 2} g_{A}(2 + 2r) ( F_1 - \tilde{F}_1 )^{2r} - \frac{ \MixLeft{3}{p_1}{2} K^2 C_{\AB{23}} }{2} \binom{2s}{3} \mathcal{F}_1^{2s-3} 
\end{equation}
\begin{equation}\nonumber
Poly_{\SB{23}} = \sum_{r=0}^{ \lceil s \rceil - 3} h_{S}(4 + 2r) (\tilde{F}_1 - F_1)^{2r+1} +  \sum_{r=0}^{ \lceil s \rceil - 2} g_{S}(2 + 2r) (\tilde{F}_1 - F_1)^{2r} - \frac{\MixLeft{3}{p_1}{2} K^2 C_{\SB{23}} }{2} \binom{2s}{3} \mathcal{F}_1^{2s-3} 
\end{equation}
\end{tcolorbox}
\subsubsection{Graviton Compton Amplitude for $s>2$}\label{subsubsec:Graviton_Higher_Spin}

Just as in section \ref{subsubsec:Photon_Higher_Spin}, for $s>2$ we will relax the constraint that the $s$-channel residue sits on both the $t$- and $u$- channel poles. Instead the $s$-channel residue will be converted into one that has both $t$ and $u$-channel poles, one that only has either $t$ or $u$- channel poles, and one that is completely local. The $u$-channel image will be fixed by $(1\leftrightarrow4)$ again. Simply doing so still wouldn't give us a consistently factorizing amplitude, since we need to ensure that the ansatz also matches that of the $t$-channel pole. We will find that the $t$-channel residue of our ansatz differs from eq.\eqref{eq:MHV_t_res} and eq.\eqref{eq:anti_MHV_t_res}, by local polynomial terms, and hence the mismatch can be removed by a pure $t$-channel term.

We again go back to the $s$-channel gluing eq.\eqref{eq:Graviton_s_Channel_residue} and apply the identity eq.\eqref{eq:almost_local_identity_symmetric}. Putting back $\sm$ and fixing the $u$-channel by spin statistics, our ansatz become
\begin{equation}\label{eq:M_trial}
\begin{split}
&\quad Ansatz\\
&= -\frac{\MixLeft{3}{p_1}{2}^4}{(s-m^2) (u-m^2) t }\mathcal{F}^{2s} + \frac{2s \MixLeft{3}{p_1}{2}^3}{t (s-m^2) }\frac{\AB{3 \boldsymbol{4}} \SB{2 \boldsymbol{1}}}{2m^2}\mathcal{F}^{2s-1} - \frac{2s \MixLeft{3}{p_4}{2}^3}{t (u-m^2)}\frac{\AB{3 \boldsymbol{1}} \SB{2 \boldsymbol{4}}}{2m^2}\mathcal{F}^{2s-1} \\
&\qquad + \Bigg\{\frac{\MixLeft{3}{p_1}{2}^2 \AB{3 \boldsymbol{4}}^2 \SB{2 \boldsymbol{1}}^2 }{4m^4 (s-m^2)}  \Bigg[\sum_{r=2}^{2s}\binom{2s}{r} \mathcal{F}^{2s-r}\Big( \frac{ -\AB{\boldsymbol{4}3} \SB{32} \AB{2 \boldsymbol{1}}}{4m^3} - \frac{ \SB{\boldsymbol{4}3} \AB{32} \SB{2 \boldsymbol{1}}}{4m^3} \Big)^{r-2}\Bigg]  \\
 & \qquad + (-1)^{2s} \frac{\MixLeft{3}{p_1}{2}^2 \AB{3 \boldsymbol{1}}^2 \SB{2 \boldsymbol{4}}^2 }{4m^4 (u-m^2)}  \Bigg[\sum_{r=2}^{2s} \binom{2s}{r} (-1)^{2s-r} \mathcal{F}^{2s-r}\Big(\frac{ -\AB{\boldsymbol{1}3} \SB{32} \AB{2 \boldsymbol{4}}}{2m^3} - \frac{ \SB{\boldsymbol{1}3} \AB{32} \SB{2 \boldsymbol{4}}}{4m^3}  \Big)^{r-2}\Bigg] \Bigg\}
\end{split}
\end{equation}
\noindent
Taking the $t$-channel residue of eq.\eqref{eq:M_trial} for both MHV and anti-MHV poles, we find that it yields eq.\eqref{eq:MHV_t_res} and eq.\eqref{eq:anti_MHV_t_res} plus additional pure polynomlias. All we need to do is subtracting them off, adding minus the polynomial terms over $t$:

\begin{equation}\label{eq:M_trial_t_residue_anti-MHV}
\begin{split}
Res[Ansatz]\Big|_{\AB{23}=0} 
&= -\frac{\MixLeft{3}{p_1}{2}^4}{(s-m^2) (u-m^2)  M_{pl}^2 } \mathcal{F}^{2s} \\
&\qquad + \frac{2s \MixLeft{3}{p_1}{2}^3}{ (s-m^2) }\frac{\AB{3 \boldsymbol{4}} \SB{2 \boldsymbol{1}}}{2m^2 M_{pl}^2} \mathcal{F}^{2s-1} - \frac{2s \MixLeft{3}{p_4}{2}^3}{ (u-m^2)}\frac{\AB{3 \boldsymbol{1}} \SB{2 \boldsymbol{4}}}{2m^2 M_{pl}^2} \mathcal{F}^{2s-1} \\
&= -\frac{\MixLeft{3}{p_1}{2}^4}{\sm \um} F_1^{2s} + \frac{Poly + Poly_{\SB{23} } }{M_{pl}^2}
\end{split}
\end{equation}

\begin{equation}\label{eq:M_trial_t_residue_MHV}
\begin{split}
Res[Ansatz]\Big|_{\SB{23}=0} 
&= -\frac{\MixLeft{3}{p_1}{2}^4}{(s-m^2) (u-m^2)  M_{pl}^2 } \mathcal{F}^{2s} \\
&\qquad + \frac{2s \MixLeft{3}{p_1}{2}^3}{ (s-m^2) }\frac{\AB{3 \boldsymbol{4}} \SB{2 \boldsymbol{1}}}{2m^2 M_{pl}^2} \mathcal{F}^{2s-1} - \frac{2s \MixLeft{3}{p_4}{2}^3}{ (u-m^2)}\frac{\AB{3 \boldsymbol{1}} \SB{2 \boldsymbol{4}}}{2m^2 M_{pl}^2} \mathcal{F}^{2s-1} \\
&= -\frac{\MixLeft{3}{p_1}{2}^4}{\sm \um} \tilde{F}_1^{2s} + \frac{Poly + Poly_{\AB{23}} }{M_{pl}^2}
\end{split}
\end{equation}

\noindent
We conclude that the graviton Compton amplitude for $s>2$ is:
\begin{tcolorbox} [title = Graviton Compton Amplitude for $s > 2$, fontupper=\footnotesize]
\begin{equation}\label{eq:All_Spin_Answer} 
\begin{split}
(-1)^{2s}M_4(s>2)
&= -\frac{\MixLeft{3}{p_1}{2}^4}{(s-m^2) (u-m^2) t M_{pl}^2 } \mathcal{F}^{2s} \\
&\quad+ \frac{2s \MixLeft{3}{p_1}{2}^3}{t (s-m^2) }\frac{\AB{3 \boldsymbol{4}} \SB{2 \boldsymbol{1}}}{2m^2 M_{pl}^2} \mathcal{F}^{2s-1} - \frac{2s \MixLeft{3}{p_4}{2}^3}{t (u-m^2)}\frac{\AB{3 \boldsymbol{1}} \SB{2 \boldsymbol{4}}}{2m^2 M_{pl}^2} \mathcal{F}^{2s-1} \\
&\quad + \Bigg\{\frac{\MixLeft{3}{p_1}{2}^2 \AB{3 \boldsymbol{4}}^2 \SB{2 \boldsymbol{1}}^2 }{4m^4 (s-m^2) M_{pl}^2}  \Bigg[\sum_{r=2}^{2s}\binom{2s}{r} \mathcal{F}^{2s-r}\Big(\frac{ - \AB{\boldsymbol{4}3} \SB{32} \AB{2 \boldsymbol{1}}}{4 m^3}  -  \frac{\SB{\boldsymbol{4}3} \AB{32} \SB{2 \boldsymbol{1}} }{4 m^3}  \Big)^{r-2}\Bigg]  \\
 & \qquad + \frac{\MixLeft{3}{p_1}{2}^2 \AB{3 \boldsymbol{1}}^2 \SB{2 \boldsymbol{4}}^2 }{4m^4 (u-m^2) M_{pl}^2}  \Bigg[\sum_{r=2}^{2s} \binom{2s}{r} (-1)^{r} \mathcal{F}^{2s-r}\Big( \frac{ -\AB{\boldsymbol{1}3} \SB{32} \AB{2 \boldsymbol{4}}}{4m^3} - \frac{ \SB{\boldsymbol{1}3} \AB{32} \SB{2 \boldsymbol{4}}}{4m^3}  \Big)^{r-2}\Bigg] \Bigg\} \\
& \quad - \frac{Poly + Poly_{\SB{23} } + Poly_{\AB{23}} }{t M_{pl}^2}
\end{split}
\end{equation}
\end{tcolorbox}
\noindent
which is consistent with the ansatz eq.\eqref{eq:Graviton_s_res_ansatz}.\footnote{Note that eq\eqref{eq:All_Spin_Answer} is derived by requiring consistent factorization on all three channels and locality. So it can actually be extrapolated to $s \leq 2$ particles. For $s=0$, $\frac{1}{2}$ and $1$, eq\eqref{eq:All_Spin_Answer} is exactly the same as eq\eqref{eq:Lower_Spin_Graviton_Compton}. For $s=\frac{3}{2}$ and $2$, eq\eqref{eq:All_Spin_Answer} differs with eq\eqref{eq:Lower_Spin_Graviton_Compton} by contact terms that does not contribute on any factorization limit. } 

We can see that $\mathcal{F}$, $\mathcal{F}_1$ and $\mathcal{F}_2$ carry inverse powers of $m$, and do not have a healthy high energy behaviour. So we can again conclude that massive particles with $s > 2$ cannot be elementary. 

\subsection{The Compton Amplitude for Non-Minimal Coupling}\label{subsec:non_minimal_Compton}

In previous sections, we've seen that minimal coupling can always be embedded into a local consistent four-point amplitude, and no constraint other than possible high energy sickness was revealed. In this subsection, we proceed and investigate the case of non-minimal couplings. Recall that we've argued through general covariance, that $\lambda \lambda$ couplings are forbidden for gravitational couplings. We will see this constraint as a consequence of inconsistent factorizations for the four-point amplitude.

\subsubsection{$\lambda \lambda$ deformation}\label{subsubsec:lambda2_deformation}
We again start with the $s$-channel gluing of the three point amplitudes. With the $\lambda \lambda$ deformation, the 3pt amplitudes are 
$$\left( x^{h_2}_{12} + x_{12}^{h_2+1}\lambda\lambda \vphantom{\tilde{\lambda}} \right) \otimes \left( x_{34}^{-h_2} + x_{34}^{-h_2 - 1} \tilde{\lambda}\tilde{\lambda} \right)$$
\noindent
Expanding the expression, we find that need to consider 4 contributions here. One is the pure minimal contribution which is already known from previous sections, which will not be repeated in this section. The other three are 
$$
(a): \hphantom{a} x^{h+1} (\lambda \lambda) \otimes \frac{1}{x^h} \qquad 
(b): \hphantom{a} x^h \otimes  \frac{\tilde{\lambda} \tilde{\lambda} }{x^{h+1}}  \qquad
(c): \hphantom{a} x^{h+1}(\lambda \lambda) \otimes  \frac{\tilde{\lambda} \tilde{\lambda} }{x^{h+1}} 
$$
\noindent
The $x$-factors of the both sides' non-minimal gluing cannot be completely absorbed in the graviton case and is the reason causing inconsistent factorization.

Let's start with photons, where the minimal coupling 3pt amplitude is given by eq.\eqref{eq:Minimal_Photon_3pt} and the $(\lambda \lambda)$ defomation 3pt amplitude is given by:
\begin{equation}\label{eq:photon_lambda2_deformation_3pt_amplitude}
M^{+1, s, s}_{3} = x_{12}^2 \frac{ \AB{ \BS{1I} }^{2s-1} \AB{ \BS{1}2 } \AB{ 2 \BS{I} } }{m^{2s}}, 
\qquad 
M^{-1, s, s}_{3} = \frac{1}{x_{34}^2} \frac{ \SB{ \BS{I4} }^{2s-1} \SB{ \BS{I}3 } \SB{ 3 \BS{4} } }{m^{2s}}
\end{equation}
$s$-channel gluing of case (i) yields:
\begin{equation}\label{eq:M4_min_nml_s_res}
M^{+1,s,s}_{3} \times M^{-1,s,s}_{3} = - \frac{\MixLeft{3}{p_1}{2}^{2}}{ t } \Bigg(\frac{ \AB{23} \SB{2 \boldsymbol{4}} \SB{2 \boldsymbol{1}} }{m \MixLeft{3}{p_1}{2}}\Bigg)  \Bigg( \frac{\AB{\boldsymbol{4}3} \SB{\boldsymbol{1} 2} + \AB{\boldsymbol{1}3} \SB{\boldsymbol{4} 2}}{\MixLeft{3}{p_1}{2}} \Bigg)^{2s-1}
\end{equation}
\noindent 
where we can see that we'll confront spurious poles again when $s > 1$. For $0 < s \leq 1$, we have
\begin{equation}\label{eq:M4_min_nml_convenient}
\widetilde{M}^{(a)}_4(0 < s \leq 1) = \frac{\MixLeft{3}{p_1}{2}}{ \sm \um } \frac{ \AB{23} \SB{2 \boldsymbol{4}} \SB{2 \boldsymbol{1}} }{m} \Bigg( \frac{\AB{\boldsymbol{4}3} \SB{\boldsymbol{1} 2} + \AB{\boldsymbol{1}3} \SB{\boldsymbol{4} 2}}{\MixLeft{3}{p_1}{2}} \Bigg)^{2s-1}
\end{equation}
where the tilde denotes that it is a partial contribution to the full non-minimal coupling amplitude. For higher spin, we 
follow the procedure of dealing with spurious poles demonstrated in section \ref{subsubsec:Photon_Higher_Spin}, leading to the following result
\begin{equation}\label{eq:l_is_1_photon_amplitude_higher_spin}
\begin{split}
\widetilde{M}_4^{(a)} 
& = -\frac{\MixLeft{3}{p_1}{2} }{(u-m^2)(s-m^2)} \frac{\AB{23} \SB{2 \boldsymbol{4}} \SB{2 \boldsymbol{1}}}{m} \mathcal{F}^{2s-1} \\
& \qquad + \Bigg\{\frac{ \SB{2 \boldsymbol{1}} \AB{3 \boldsymbol{4}} }{2m^2 (s-m^2)} \Big( \frac{  \AB{23} \SB{2 \boldsymbol{4}} \SB{2 \boldsymbol{1}} }{m }\Big) \Bigg[ \sum_{r=1}^{2s-1} \binom{2s-1}{r} \mathcal{F}^{2s-1-r} \Bigg( -\frac{\AB{\boldsymbol{4}3} \SB{32} \AB{2 \boldsymbol{1}}}{2m^3} \Bigg)^{r-1} \Bigg] \\
& \qquad \qquad + (1 \leftrightarrow 4) \Bigg\}
\end{split}
\end{equation}
Finally, the completely local contribution $(c)$ is:
\begin{equation}\label{eq:NM_NM_l_is_1_M4}
\begin{split}
\widetilde{M}_4^{(c)}  =&\frac{\SB{\BS{1}2} \AB{3 \BS{4}}}{(s-m^2)2s m^{2s}} \left( \frac{\AB{\BS{14}}}{2} + \frac{\SB{\BS{14}}}{2} + \frac{ \AB{\BS{1}3} \SB{3\BS{4}}}{2m} - \frac{ \AB{\BS{1}2} \SB{2\BS{4}}}{2m} \right)^{2s-2}\\
&\times\Bigg\{
2s \frac{\MixLeft{3}{p_1}{2}}{m}  \left(\frac{\AB{\BS{14}}}{2} + \frac{\SB{\BS{14}}}{2} + \frac{ \AB{\BS{1}3} \SB{3\BS{4}}}{2m} - \frac{ \AB{\BS{1}2} \SB{2\BS{4}}}{2m} \right) + (2s-1) \SB{\BS{1}2} \AB{3 \BS{4}} \Bigg\}\\
& \qquad + (-1)^{2s} (1 \leftrightarrow 4)
\end{split}
\end{equation}
So, we have obtained a $\lambda \lambda$ deformed photon Compton amplitude that consistently factorizes in all channels:
\eq
\widetilde{M}_4^{(Min)}+\widetilde{M}_4^{(a)}+\widetilde{M}_4^{(b)}+\widetilde{M}_4^{(c)}\,,
\eqe 
where $\widetilde{M}_4^{(Min)}$ is the Compton amplitude derived in the previous section. One thing worth mentioning is that for the mixed contribution, $x$-factors in the $s$ channel gluing that cannot be absorbed by $\lambda$ or $\tilde{\lambda}$ via:
\begin{equation}\label{eq:x_factor_absorbing_identity}
x \lambda^{\alpha} = \frac{p^{\dot{\alpha} \alpha}}{m} \tilde{\lambda}_{\dot{\alpha}}, \quad \frac{\tilde{\lambda}_{\dot{\alpha}}}{x} = \frac{p_{\alpha \dot{\alpha}}}{m}\lambda^{\alpha}
\end{equation}
becomes a $\um$ pole by using the identity eq.\eqref{eq:x_over_x}. On the other hand, $x$ factors in the $(c)$ contribution can be completely absorbed because we have enough $\lambda$ and $\tilde{\lambda}$ to use eq.\eqref{eq:x_factor_absorbing_identity} so that it is completely local. This discussion will be important to see the inconsistent factorization of $\lambda \lambda$ deformed Compton amplitude.

Let's now turn to the non-minimal graviton Compton scattering. The minimal coupling 3pt amplitude is given by eq.\eqref{eq:Minimal_Graviton_3pt} and the $(\lambda \lambda)$ deformation is:
\begin{equation}\label{eq:graviton_lambda2_deformation_3pt_amplitude}
M^{+2, s, s}_3 = x_{12}^3 \frac{ \AB{ \BS{1I} }^{2s-1} \AB{ \BS{1}2 } \AB{ 2 \BS{I} } }{m^{2s-1} M_{pl}}, 
\qquad
M^{-2, s, s}_3 = \frac{1}{x_{34}^3} \frac{ \SB{ \BS{I4} }^{2s-1} \SB{ \BS{I}3 } \SB{ 3 \BS{4} } }{m^{2s-1} M_{pl}}
\end{equation}
The mixed coupling $(a) + (b)$ contribution for spin $\frac{1}{2}$ is
\begin{equation}\label{eq:spin_one_half_Min-NM}
\begin{split}
\widetilde{M}_4^{(Mix)} 
&= \widetilde{M}^{(a)} + \widetilde{M}^{(b)} \\
&= \frac{\MixLeft{3}{p_1}{2}^3}{(s-m^2)(u-m^2)t} \frac{ \SB{\boldsymbol{1} 2} \AB{23} \SB{2 \boldsymbol{4}} }{m M_{pl}^2} - \frac{\MixLeft{3}{p_1}{2}^3}{(s-m^2)(u-m^2)t} \frac{ \AB{\boldsymbol{1} 3} \SB{23} \AB{3 \boldsymbol{4}} }{m M_{pl}^2}\,.
\end{split}
\end{equation}
\noindent
Importantly, the $t$-channel residue is already correctly reproduced by that of minimal coupling and eq.\eqref{eq:spin_one_half_Min-NM}! This poses a problem because when we include the all non-minimal couplings contribution:
\begin{equation}\label{eq:spin_one_half_NMNM}
\begin{split}
\widetilde{M}_4^{(c)} &= -\frac{\MixLeft{3}{p_1}{2}^3}{2(s-m^2)(u-m^2)} \frac{ \SB{\boldsymbol{1}2} \AB{3 \boldsymbol{4}} + \SB{\boldsymbol{4}2} \AB{3 \boldsymbol{1}} }{m^2 M_{pl}^2} \\
& \qquad + \frac{\MixLeft{3}{p_1}{2}^3}{2(s-m^2)t} \frac{ \SB{\boldsymbol{1}2} \AB{3 \boldsymbol{4}} -  \SB{\boldsymbol{4}2} \AB{3 \boldsymbol{1}} }{m^2 M_{pl}^2} - \frac{\MixLeft{3}{p_4}{2}^3}{2(u-m^2)t} \frac{ \SB{\boldsymbol{4}2} \AB{3 \boldsymbol{1}} -  \SB{\boldsymbol{1}2} \AB{3 \boldsymbol{4}} }{m^2 M_{pl}^2}
\end{split}
\end{equation}
we find that there is further $t$-channel singularity. Note that this mismatch is not local, and thus cannot be removed by modifying the expression by pure $t$-channel contributions. Thus we have failed to obtain a local amplitude that correctly factorizes in all channels. Note that the source of this can be traced back to the excess $x$-factors. There is a factor of $\frac{x_{12}}{x_{34}}$ left in the gluing procedure that gives the extra $t$ channel in the $(c)$ contribution due to eq.\eqref{eq:x_over_x}. Thus, we conclude that $\lambda \lambda$ coupling is forbidden for gravity. This is consistent with the previous results shown in the 3pt amplitude.

Finally, a side note on the high energy behaviour of the $\lambda \lambda$ deformed photon Compton amplitude. The $(a)$ contribution eq.\eqref{eq:M4_min_nml_convenient} scales at least as $O(\frac{1}{m})$ in HE. The $(b)$ contribution is not given because it should behave the same as $(a)$ contribution in HE due to symmetry. For higher spins, dealing with the spurious poles introduces higher power of $\frac{1}{m}$. In other words, the counting of the factors of $\frac{1}{m}$ is no more just the multiplication of the ones in the 3pt amplitudes. The  $(a)$ contribution for $ s > 1$ eq.\eqref{eq:l_is_1_photon_amplitude_higher_spin} scales at least as $O(\frac{1}{m^{6s-1}})$ at HE. And the $(c)$ contribution  scales at least as $O(\frac{1}{m^{4s}})$ at HE for all spins. That is, worse than both $(a)$ and $(b)$ contributions when $0 < s \leq 1$, but not for higher spin charged particles.



\subsubsection{$(\lambda \lambda)^2$ deformation}\label{subsubsec:lambda4_deformation}
For $(\lambda\lambda)^2$ deformations, we should in general consider
$$\left( g_0 x^{h_2}_{12} + g_1 x_{12}^{h_2+1}\lambda\lambda + g_2 x_{12}^{h_2+2}(\lambda\lambda)^2 \vphantom{\tilde{\lambda}} \right) \otimes \left( g_0 x_{34}^{-h_2} + g_1 x_{34}^{-h_2 - 1} \tilde{\lambda}\tilde{\lambda} + g_2 x_{34}^{-h_2 - 2} (\tilde{\lambda}\tilde{\lambda})^2 \right)$$
where $g_1 = 0$ for gravitons. In this section, we'll be mostly ineterested in the contributions 
$$
(a'): \hphantom{a} x^{h+2} (\lambda \lambda)^2 \otimes \frac{1}{x^h}, \qquad
(b'): \hphantom{a} x^h \otimes  \frac{( \tilde{\lambda} \tilde{\lambda} )^2 }{x^{h+2}}, \qquad
(c'): \hphantom{a} x^{h+2}(\lambda \lambda)^2 \otimes  \frac{(\tilde{\lambda} \tilde{\lambda})^2 }{x^{h+2}} $$

We again start with photons. From our experience above, we will only be interested in $(a')$ and $(b')$ contributions to the deformed Compton amplitude since this is the only structure that causes the $\frac{1}{m}$ counting differ from that of 3pt counting in higher spin. The $(\lambda \lambda)^2$ coupling of photon is given by:
\begin{equation}\label{eq:photon_lambda4_deformation_3pt_amplitude}
M^{+1, s, s}_3 = x_{12}^3 \frac{ \AB{ \BS{1I} }^{2s-2} \AB{ \BS{1}2 }^2 \AB{ 2 \BS{I} }^2 }{ m^{2s + 1}}, 
\qquad 
M^{-1, s, s}_3 = \frac{1}{x_{34}^3} \frac{ \SB{ \BS{I4} }^{2s-2} \SB{ \BS{I}3 }^2 \SB{ 3 \BS{4} }^2 }{ m^{2s+1}}
\end{equation}
This $(a')$ contribution for $s=1$ photon Compton scattering is:

\begin{equation}\label{eq:l_is_2_photon_Compton}
\widetilde{M}^{(a')}_4(s=1) = \frac{ \AB{23}^2 \SB{2 \boldsymbol{4}}^2 \SB{2 \boldsymbol{1}}^2 }{m^2 \sm \um } 
\end{equation}
with $O(m^{-2})$ in HE. And for $s > 1$, we'll need to deal with spurious poles. The $(a')$ contribution to the non-minimal photon Compton amplitudes for $S>1$ is:
\begin{equation}\label{eq:l_is_2_higher_spin_photon_amplitude}
\begin{split}
\widetilde{M}^{(a')}_4(s>1)
&= -\frac{\MixLeft{3}{p_1}{2}^{2}}{ (u-m^2)(s-m^2) } \mathcal{F}^{2s-2} F_2^2\\ 
&\qquad+\Bigg\{ \frac{\MixLeft{3}{p_1}{2} \SB{2 \boldsymbol{1}} \AB{3 \boldsymbol{4}} }{2m^2(s-m^2)} \Bigg[ \mathcal{F}^{2s-2} \sum_{i=1}^2 \binom{2}{i} F_2^{2-i}\Big(  -\frac{\AB{\boldsymbol{4}3} \SB{32} \AB{2 \boldsymbol{1}} }{2m^3}\Big)^{i-1} \\
&\qquad \qquad +  F_2^2 \sum_{j}^{2s-2} \binom{2s-2}{j} \mathcal{F}^{2s-2-j} \Big(-\frac{\AB{\boldsymbol{4}3} \SB{32} \AB{2 \boldsymbol{1}} }{2m^3}\Big)^{j-1}\\
&\qquad \qquad +  \sum_{i=1}^2 \sum_{j=1}^{2s-2} \binom{2}{i} \binom{2s-2}{j} F_2^{2-i} \mathcal{F}^{2s-2-j} \Big(-\frac{\AB{\boldsymbol{4}3} \SB{32} \AB{2 \boldsymbol{1}} }{2m^3}\Big)^{i+j-1} \Bigg] \\
&\qquad \qquad + (1 \leftrightarrow 4)\Bigg\}
\end{split}
\end{equation}
which at least scales as $O(m^{-6s+3})$ in HE.

Now we apply the same analysis for gravitons. Since $(\lambda \lambda)$ coupling is forbidden for gravitons, we will elaborate more on the $(\lambda \lambda)^2$ coupling. The $(\lambda \lambda)^2$ graviton 3pt amplitude is:
\begin{equation}\label{eq:NM2-3pt}
M^{+2, s, s}_3 = x_{12}^4 \frac{1}{M_{pl}} \frac{ \AB{\BS{1I}}^{2s-2} \AB{\BS{1}2}^2 \AB{2 \BS{I}}^2 }{m^{2s}}, \quad M^{-2, s, s}_3 = \frac{1}{x_{34}^4} \frac{1}{M_{pl}} \frac{ \SB{\BS{I4}}^{2s-2} \SB{\BS{I}3}^2 \SB{3 \BS{4}}^2 }{m^{2s}}
\end{equation}
\noindent
The 4pt mixed coupling $(a') + (b')$  contribution to the amplitude  for $1 \leq s \leq 2$ is given by:
\begin{equation}\label{eq:4pts_NM_Min_No_Spurious}
\widetilde{M}_4^{(Mix)}(s\leq2) = \frac{-\MixLeft{3}{p_1}{2}^{4-2s} \big( \AB{\BS{4}3}\SB{\BS{1}2} + \AB{\BS{1}3}\SB{\BS{4}2} \big)^{2s-2} }{(s-m^2)(u-m^2) t } \frac{ \SB{\BS{1}2}^2\AB{23}^2\SB{2\BS{4}}^2 + \AB{\BS{1}3}^2\SB{23}^2\AB{3\BS{4}}^2 }{m^2 M_{pl}^2} 
\end{equation}
\noindent
And for  $s > 2$ particles
\begin{equation}\label{eq:4pts_AB}
\begin{split}
\widetilde{M}_{4}^{(a')}(s>2) = \frac{ C_{\AB{23}}^2 }{M_{pl}^2} &\Big\{ -\frac{\MixLeft{3}{p_1}{2}^2 \mathcal{F}^{2s-2} }{(s-m^2)(u-m^2) t }  \\
& \quad + (2s-2) \Big[\frac{ \MixLeft{3}{p_1}{2} \SB{2 \BS{1}} \AB{3 \BS{4}} }{2m^2 t (s-m^2)} - \frac{ \MixLeft{3}{p_4}{2} \SB{2 \BS{4}} \AB{3 \BS{1}} }{2m^2 t (u-m^2)}  \Big] \mathcal{F}^{2s-3} \\
& \quad + \frac{\SB{\BS{1}2}^2 \AB{3 \BS{4}}^2}{4m^4 (s-m^2)} \Big( \sum_{r=2}^{2s-2} \binom{2s-2}{r} \mathcal{F}^{2s-2-r} \Big(\frac{ -\AB{\boldsymbol{4}3} \SB{32} \AB{2 \boldsymbol{1}}}{2m^3} \Big)^{r-2}\Big) \\
& \quad +  \frac{\SB{\BS{4}2}^2 \AB{3 \BS{1}}^2}{4m^4 (u-m^2)} \Big( \sum_{r=2}^{2s-2} \binom{2s-2}{r} (-1)^{r} \mathcal{F}^{2s-2-r} \Big(\frac{ -\AB{\boldsymbol{4}3} \SB{32} \AB{2 \boldsymbol{1}}}{2m^3} \Big)^{r-2}\Big) \\
& \quad + \frac{K^2}{t} \Big[ \sum_{r=1}^{2s-3} r\binom{2s-2}{r+1} \tilde{F}_1^{2s-3-r}\tilde{F}_2^{r-1} \Big]  \Big\}
\end{split}
\end{equation}
\begin{equation}\label{eq:4pts_SB}
\begin{split}
\widetilde{M}_{4}^{(b')}(s>2) = \frac{ C_{\SB{23}}^2 }{M_{pl}^2} &\Big\{ -\frac{\MixLeft{3}{p_1}{2}^2 \mathcal{F}^{2s-2} }{(s-m^2)(u-m^2) t }  \\
& \quad + (2s-2) \Big[\frac{ \MixLeft{3}{p_1}{2} \SB{2 \BS{1}} \AB{3 \BS{4}} }{2m^2 t (s-m^2)} - \frac{ \MixLeft{3}{p_4}{2} \SB{2 \BS{4}} \AB{3 \BS{1}} }{2m^2 t (u-m^2)}  \Big] \mathcal{F}^{2s-3} \\
& \quad + \frac{\SB{\BS{1}2}^2 \AB{3 \BS{4}}^2}{4m^4 (s-m^2)} \Big( \sum_{r=2}^{2s-2} \binom{2s-2}{r} \mathcal{F}^{2s-2-r} \Big(\frac{ -\AB{\boldsymbol{4}3} \SB{32} \AB{2 \boldsymbol{1}}}{2m^3} \Big)^{r-2}\Big) \\
& \quad +  \frac{\SB{\BS{4}2}^2 \AB{3 \BS{1}}^2}{4m^4 (u-m^2)} \Big( \sum_{r=2}^{2s-2} \binom{2s-2}{r} (-1)^{r} \mathcal{F}^{2s-2-r} \Big(\frac{ -\AB{\boldsymbol{4}3} \SB{32} \AB{2 \boldsymbol{1}}}{2m^3} \Big)^{r-2}\Big) \\
& \quad + \frac{K^2}{t} \Big[ \sum_{r=1}^{2s-3} r \binom{2s-2}{r+1} F_1^{2s-3-r} F_2^{r-1} \Big]  \Big\}
\end{split}
\end{equation}
\noindent 
So the mixed coupling contribution to higher spin Compton amplitude is:
\begin{equation}\label{eq:mix_lambda4}
\widetilde{M}_4^{Mixed}(s>2) = \widetilde{M}_{4}^{(a')}(s>2) + \widetilde{M}_{4}^{(b')}(s>2)
\end{equation}
Finally, for the $(\lambda\lambda)^2 \otimes (\lambda\lambda)^2$ contribution to the  amplitude, the $x$-factors can be completely absorbed by the $\lambda$'s, so the gluing is going to be completely local. For spin $1$, the amplitude is:
\begin{equation}\label{eq:NMNM_S=1}
\widetilde{M}_4^{(c')}({s=1}) = \frac{ \MixLeft{3}{p_1}{2}^2 \SB{\BS{1}2}^2 \AB{3 \BS{4}}^2 }{(s-m^2) m^4 M_{pl}^2} + \frac{ \MixLeft{3}{p_4}{2}^2 \SB{\BS{4}2}^2 \AB{3 \BS{1}}^2 }{(u-m^2) m^4 M_{pl}^2} 
\end{equation}
\noindent 
For $s=\frac{3}{2}$, the amplitude is:
\begin{equation}\label{eq:NMNM_S=3half}
\begin{split}
\widetilde{M}^{(c^\prime)}\left(s = \frac{3}{2} \right) =
& 
\frac{\SB{\BS{1}2}^2 \AB{3\BS{4}}^2}{(s-m^2)m^3 M_{pl}^2} \Bigg\{ \frac{\MixLeft{3}{p_1}{2}^2  }{m^2} \left( \frac{\AB{\BS{14}}}{2} + \frac{\SB{\BS{14}}}{2} + \frac{ \AB{\BS{1}3} \SB{3\BS{4}}}{2m} - \frac{ \AB{\BS{1}2} \SB{2\BS{4}}}{2m}\right)  \\
&\quad + \frac{2}{3}  \frac{\MixLeft{3}{p_1}{2}\AB{3 \BS{4}} \SB{\BS{1}2}}{m} \Bigg\} + (-1)^{2s}(1 \leftrightarrow 4)
\end{split}
\end{equation}
\noindent
And for $s \geq 2$:
\begin{equation}\label{eq:NMNM_S_req_2}
\begin{split}
\widetilde{M}_4^{(c^\prime)} (s \geq 2) =&\frac{\SB{\BS{1}2}^2 \AB{3 \BS{4}}^2}{(s-m^2)2s(2s-1)m^{2s} M_{pl}^2} \left(  \frac{\AB{\BS{14}}}{2} + \frac{\SB{\BS{14}}}{2} + \frac{ \AB{\BS{1}3} \SB{3\BS{4}}}{2m} - \frac{ \AB{\BS{1}2} \SB{2\BS{4}}}{2m}\right)^{2s-4}\\
&\times\Bigg\{
2s(2s-1)\frac{\MixLeft{3}{p_1}{2}^2 }{m^2}  \left(  \frac{\AB{\BS{14}}}{2} + \frac{\SB{\BS{14}}}{2} + \frac{ \AB{\BS{1}3} \SB{3\BS{4}}}{2m} - \frac{ \AB{\BS{1}2} \SB{2\BS{4}}}{2m} \right)^2 \\
&\qquad+ 2(2s-1)(2s-2) \frac{\MixLeft{3}{p_1}{2} }{m}  \left( \frac{\AB{\BS{14}}}{2} + \frac{\SB{\BS{14}}}{2} + \frac{ \AB{\BS{1}3} \SB{3\BS{4}}}{2m} - \frac{ \AB{\BS{1}2} \SB{2\BS{4}}}{2m}\right) \SB{\BS{1}2} \AB{3 \BS{4}}\\
&\qquad + (2s-2)(2s-3)\SB{\BS{1}2} ^2\AB{3 \BS{4}}^2
\Bigg\} + (-1)^{2s} (1 \leftrightarrow 4)
\end{split}
\end{equation}
\noindent
The full amplitude containing the $(\lambda \lambda)^2$ non-minimal coupling is the sum of the mixed one and the pure non-minimal one. The mixed one has already matched the $t$-channel residue and there are no $t$ channel poles to be considered in the pure non-minimal piece because the $x$-factors are completely absorbed.

\subsubsection{UV behaviour of the 4pt Amplitudes}
It is natural to ask the following question. Given that the minimal coupling 3pt amplitude has the best high energy behavior among all possible structures, does the Compton amplitude obtained from gluing the minimal 3pt amplitudes together automatically has better high energy behavior than the non-minimal contributions? The naive answer to this question is yes, because the $\frac{1}{m}$ counting in the minimal coupling 3pt amplitude for a given spin is always lower than the non-minimal couplings. This is true for $s<1$ photon Compton amplitudes and $s<2$ graviton Compton amplitudes, but it no longer holds for higher spin Comptons. The procedure of removing the spurious poles introduces various powers of $\frac{1}{m}$. This can make the HE behaviour of the minimal Compton amplitude worse than the non-minimal ones.

Now, we summarize the HE behaviour of all the Compton amplitudes we have obtained until now.
\begin{enumerate}[leftmargin=*]
\item Photon
\begin{itemize}[leftmargin=*]
\item The pure minimal Compton amplitude has no $\frac{1}{m}$ factors for $0 \leq s \leq 1$ and thus has a good HE behaviour and scales at least as $O(m^{-6s+1})$ at HE for $s > 1$.
\item The $(a)$ and $(b)$ contribution from $(\lambda \lambda)$ coupling deformation scales at least as $O(m^{-1})$ for $s \leq 1$ and $O(m^{-6s+3})$ for $s>1$.
\item The $(c)$ contribution from $(\lambda \lambda)$ coupling deformation scales at least as $O(m^{-2})$ for $s \leq 2$ and at least $O(m^{-4s})$ for $s>2$.
\item The $(a)$ and $(b)$ contribution from $(\lambda \lambda)^2$ coupling deformation scales at least as $O(m^{-6s+2})$.
\end{itemize}
\item Graviton
\begin{itemize}[leftmargin=*]
\item The pure minimal Compton amplitude has no $\frac{1}{m}$ factors for $0 \leq s \leq 2$ and thus has a good HE behaviour and scales at least as $O(m^{-6s+2})$ at HE for $s > 2$.
\item The $(a')$ and $(b')$ contribution from $(\lambda \lambda)^2$ coupling deformation scales at least as $O(m^{-6s+6})$.
\item The $(c')$  contribution from $(\lambda \lambda)^2$ coupling deformation scales at least as $O(m^{-4s})$ for all spins.
\end{itemize}

\end{enumerate}
Finally we conclude that the HE behaviour predicted by 3pt amplitudes only holds for lower spins. For higher spins, the powers of $\frac{1}{m}$ are determined by the number of the spurious poles cancelled. One factor of $\frac{1}{\MixLeft{3}{p_1}{2}}$ cancelled raises the inverse mass factor counting by $\frac{1}{m^2}$ more.

\section{Computing the classical potential (1 PM)}\label{sec:GenMatchProc}
Now that we have identified the $s \gg 1$ limit of minimal coupling three-point amplitudes as that describing Kerr black holes from the one body EFT framework, we can utilize this fact to compute the spin-dependent part of the classical potential between two black holes, which is defined as Fourier transform of the non-relativistic centre of mass amplitude as in \cite{Holstein:2008sx}. When we talk about the classical potential, we are referring to a long range effect which in momentum space corresponds to the zero momentum limit. The standard textbook setup is then to consider the amplitude for a massless exchange between two massive states, 
$$\includegraphics[scale=0.5]{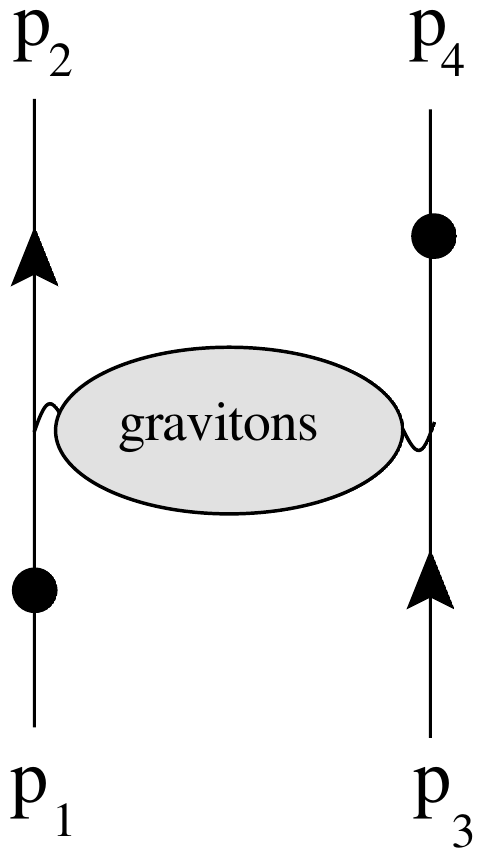}\,,$$
As we will be interested in long range effects, this requires taking the $q^2=(\vec{q})^2\rightarrow0$ limit of the above process, where $\vec{q}$ is the momentum transferred in the center of mass frame. Thus the classical potential can be extracted by taking the four-point amplitude and extracting the $t=(P_1-P_2)^2=-q^2$ channel massless pole and discontinuity, schematically given as:
\eq
\left.\frac{M_{4}}{4E_aE_b}\right|_{q^2\rightarrow0}\,.
\eqe 
The computation can be organized in terms of powers of Newton constant $G$, which corresponds to post Minkowskian (PM) expansion. At leading order (1 PM), the classical potential is given by the residue of the massless pole in the tree-amplitude. At (2 PM), the contribution to the classical potential arrises from the $t$-channel triangle integral in the scalar integral basis. There the relevant factor would then be the scalar triangle integral coefficient.

Note that these contributions are on-shell in nature: the residues in a tree-level amplitude are simply products of lower point amplitudes, and at one-loop the integral basis coefficients are computable by generalized unitarity cut methods~\cite{Bern:1994zx, Bern:1994cg},  which again are given by products of tree-amplitude. Thus in principle, one should be able to bypass the need for the full four-amplitude, and compute the potential using these on-shell building blocks. There is one obstruction, however, in that in the centre of mass (COM) frame the transfer momentum is space-like, and the $t\rightarrow 0$ limit is only reachable by taking the limit of vanishing transfer momentum. On the other hand, to fully utilize the on-shell nature, we need to have light-like exchange momentum.

Cachazo and Guevara~\cite{Cachazo:2017jef, Guevara:2017csg} demonstrated how this can be circumvented by utilizing the ``holomorphic classical limit" (HCL), which keeps the total exchanged (complex) momenta to be light-like and then perform a non-relativistic expansion. This expansion is in $\sqrt{r^2 - 1}$, with 
\eq
r=\frac{P_1 \cdot P_3}{m_a m_b}\,.
\eqe
The general procedure for obtaining the classical potential then proceeds as follows:  
\begin{itemize}
  \item (A) One first computes the Leading Singularity (LS) for the wanted order in $G$ in the HCL. At tree-level the LS is just the $t$-channel residue which is the product of three-point amplitudes on both sides of the pole. For one-loop it is the pole at infinity for a specific parametrisation of the triangle cut, which yields the triangle coefficient. 

Since for the HCL the kinematic setup is essentially that of two three-point on-shell kinematics, the LS can be expanded as: 
\eq\label{eq:LSExp}
\text{LS} = \sum_{i=0}^{2s_a} \sum_{j=0}^{2s_b} N^{s_a,s_b}_{i,j} \tilde{A}_{i,j} (Sp_a)^i (Sp_b)^j,\quad  N^{s_a,s_b}_{i,j} = \frac{(2s_a)!}{(2s_a - i)!} \frac{(2s_b)!}{(2s_b-j)!} \, ,
\eqe
where 
\eq
Sp_a = \frac{\sket{\hat\l}\sbra{\hat\l}}{m_a}, \quad Sp_b = \frac{\sket{\l}\sbra{\l}}{m_b} \label{eq:SpaSpbDef}
\eqe
and the spinors $|\hat{\lambda}]$ and $|\lambda]$ are defined from the null exchanged momenta in the HCL. Here the LS is written with free $SL(2,\IC)$ indices, as $2s_a$ and $2s_b$ external massive spinors have been stripped.  The coefficients $\tilde{A}_{i,j} $ will be functions of the external momenta, and have an expansion in  $\sqrt{r^2 - 1}$.   

  \item (B) Next one considers a set of local Lorentz invariant operators also evaluated in the HCL. Although the number of operators are possibly infinite, the number of combinations actually used to form the basis is not high due to the observation that number of spin operators and number of momentum transfer vector appearing in the classical potential are closely related~\cite{Guevara:2017csg}. 
   These operators will in general have different powers of $\sqrt{r^2 - 1}$ in the HCL, and thus be unambiguously mapped to the LS. We thus have a representation of the LS in terms of local operators.
  \item (C) Divide the LS obtained above by the additional factor of $\frac{1}{4 E_a E_b}$. This additional factor changes the normalisation of density of particles from one particle per volume $\propto\frac{m}{\abs{E}}$, relevant for relativistic scattering, to one particle per unit volume, relevant for non-relativistic scattering. The classical potential is then simply the non-relativistic expansion of this result.
\bl
&\phantom{asdf} \frac{\text{LS}}{4E_a E_b} \stackrel{NR}{\longrightarrow} V_{Cl}
\\ &\bld
\varepsilon^\ast (P_2) \varepsilon(P_1) &\stackrel{NR}{\longrightarrow} \varepsilon^\ast(p_a) \left[ \iden - \frac{i}{2m_a^2} \left( \vec{p_a} \times \vec{q} \right) \cdot \vec{S_a} + \cdots \right] \varepsilon(p_a)
\\ \varepsilon^\ast (P_4) \varepsilon(P_3) &\stackrel{NR}{\longrightarrow} \varepsilon^\ast(p_b) \left[ \iden + \frac{i}{2m_b^2} \left( \vec{p_b} \times \vec{q} \right) \cdot \vec{S_b} + \cdots \right] \varepsilon(p_b) \,.
\eld \label{eq:SOcorrection}
\el
where the last two lines indicate the extra contributions that arises from the effect of putting back the polarization tensors. Appendix \ref{sec:SOcorrections} outlines how the non-relativistic results for the last two lines were worked out.

\end{itemize}

Note that from eq.(\ref{eq:LSExp}), we see that for a given scattering of particle with spins $s_a$ and $s_b$, we can compute terms in the potential that is up to degree $2s_a$ in $\vec{S_a}$ and $2s_b$ in $\vec{S_b}$. For example for spin-$\{1,\frac{1}{2}\}$ we can compute terms with
\eq
\vec{S_a}^2, \quad\vec{S_a}, \quad\vec{S_b},\quad \vec{S_a}\vec{S_b}, \quad\vec{S_a}^2\vec{S_b}\,.
\eqe
Importantly, a given operator in the potential may appear in many different choice of $\{s_a,s_b\}$, and they all must give identical results. For example the operator $\vec{S}_a\cdot \vec{S}_a$, should emerge from the LS of spin-$\{\frac{1}{2},\frac{1}{2}\}$, $\{1, 1\}$, $\{2, 2\}$ e.t.c. For consistency, they should all agree.

Before starting with explicit examples, as we will be interested in the spin-dependent part of the potential, some comments for the \emph{spin supplementary condition} (SSC), discussed around eq.(\ref{InSSC}), is in order. The covariant condition $S^{\m\n} p_{\n} = 0$ used in former sections was used in \cite{Porto:2005ac} and \cite{Levi:2010zu} to compute leading order (LO) gravitational spin-orbit interactions, which also have been reproduced in the following sections. However, this choice of SSC is in conflict with canonical Poisson bracket relations~\cite{Steinhoff:2010zz}. To get canonical variables, another choice called Newton-Wigner (NW) SSC is needed~\cite{Levi:2010zu,Steinhoff:2010zz}, the choice referred to as baryonic condition in \cite{Porto:2005ac}. In curved space NW SSC can be formulated as $S^{\m\n} (p_{\n} + m e^{0}_{\n}) = 0$ where $e^0$ is the time-like vielbein, and following sections will be mostly concerned with this choice. In the following we will use the tree-level LS (1 PM) as detailed examples to illustrate the details of this procedure, while reproducing known results in the literature. In the next section, we will present the corresponding 2PM results, which will be mostly new.

\subsection{Kinematic variables and basis for operators} \label{sec:LScompdefs}
\subsubsection{Kinematic variables} \label{sec:LScompmisc}
Kinematic variables will be taken to have the following parametrisation in the centre of mass frame, which is the parametrisation adopted by Guevara in \cite{Guevara:2017csg}\footnote{While it was implicitly assumed that $\b = \b'$ in \cite{Guevara:2017csg}, unless $m_a = m_b$ this does not hold true in general.};
\bl
\bld
P_1 &= ( E_a, \vec{p} + \vec{q} / 2 ) = \sket{\hat\eta} \bra{\hat\l} + \sket{\hat\l}\bra{\hat\eta}
\\ P_2 &= ( E_a, \vec{p} - \vec{q} / 2 ) = \b' \sket{\hat\eta} \bra{\hat\l} + \frac{1}{\b'} \sket{\hat\l}\bra{\hat\eta} + \sket{\hat\l} \bra{\hat\l}
\\ P_3 &= ( E_b, - \vec{p} - \vec{q} / 2 ) = \sket{\eta} \bra{\l} + \sket{\l}\bra{\eta}
\\ P_4 &= ( E_b, - \vec{p} + \vec{q} / 2 ) = \b \sket{\eta} \bra{\l} + \frac{1}{\b} \sket{\l}\bra{\eta} + \sket{\l} \bra{\l}
\\ K &= P_1 - P_2 = (0,\vec{q}) = - \sket{\hat\l}\bra{\hat\l} + \CO(\b-1) = \sket{\l}\bra{\l} + \CO(\b-1) \,.
\eld
\el
The HCL corresponds to the limit $\b \to 1$, which is equivalent to the limit $\b' \to 1$. Since the ``approaching speed'' of the limit is the same for both cases, the limit $\b \to 1$ will be used to denote the HCL. Note that this on-shell limit has been reached by complex momenta $K$.

The usual definitions for the Mandelstam variables, $s=(P_1 + P_3)^2$ and $t = (P_1 - P_2)^2$, has been adopted. In this frame $t = -q^2$, where $q^2 = (\vec{q})^2$. All external momenta are taken to be on-shell; $P_1^2 = P_2^2 = m_a^2$ and $P_3^2 = P_4^2 = m_b^2$. The spinor brackets are taken to be constrained by the conditions $\la \hat\l \hat\eta \ra = [\hat\l \hat\eta] = m_a$ and $\la \l \eta \ra = [\l \eta] = m_b$. The variables $u$, $v$, and $r$ are defined as follows;
\bl
\bld
u &= \sbra{\l} P_1 \ket{\eta}
\\ v &= \sbra{\eta} P_1 \ket{\l}
\\ r &= \frac{P_1 \cdot P_3}{m_a m_b}
\eld
\el
In the HCL, the variables $u$ and $v$ tend to the values $u \to m_a m_b x_1 \bar{x}_3$ and $v \to m_a m_b \bar{x}_1 x_3$. Following relations can be derived from kinematic constraints.
\bg
\bgd
\sbra{\eta} P_1 \ket{\eta} \sbra{\l} P_1 \ket{\l} = u v - m_a^2 m_b^2
\\ \sbra{\l} P_1 \ket{\l} = - \frac{(\b-1)^2}{\b} m_b^2 + (1-\b) v + \frac{\b-1}{\b} u \,.
\egd
\eg
To compute the classical potential, an expansion in $r$ or $\e = \sqrt{r^2 - 1}$ is needed. This expansion is obtained by utilising the following relations that hold in the HCL.
\bl
\bld
u &= m_a m_b (r + \sqrt{r^2 - 1})
\\ v &= m_a m_b (r - \sqrt{r^2 - 1}) \,.
\eld
\el

\subsubsection{Lorentz-invariant combination of operators}  \label{sec:matchingopbasis}
The independent four-vector kinematical variables are $P_1$, $P_3$, $S_a$, $S_b$, and $K$. Some examples of non-trivial invariants (in the COM) that can be constructed from these variables are;
\bl
\bld
K \cdot S_i &= \vec{q} \cdot \vec{S_i}
\\ \e_{\m\n\l\s} P_1^\m P_3^\n K^\l S_a^\s &= (E_a + E_b) ( \vec{S_a} \cdot \vec{p_a} \times \vec{q} )
\\ P_1 \cdot S_b &= \left( 1 + \frac{E_a}{E_b} \right) \left( \vec{p_b} \cdot \vec{S} \right)
\\ \e_{\m\n\l\s} P_1^\m P_3^\n S_a^\l S_b^\s &= (E_a + E_b) ( \vec{p_a} \cdot \vec{S_a} \times \vec{S_b} ) \,.
\eld \label{eq:LorInvOpValues}
\el
However, not all such invariants are of interest. The invariants relevant for computing the classical potential should reduce to powers of $Sp_a$ and $Sp_b$ defined in eq.\eqc{eq:SpaSpbDef} when HCL is taken. This removes the candidacy of the last two terms in \eqc{eq:LorInvOpValues}. Indeed the first two terms in \eqc{eq:LorInvOpValues} in the HCL takes the form:
\bl
\bld
K \cdot S_a &\to \left( + \half \sket{\hat\l} \sbra{\hat\l} ~\text{or}~ - \half \ket{\hat\l} \bra{\hat\l} \right) + \CO(\b-1)
\\ K \cdot S_b &\to \left( - \half \sket{\l} \sbra{\l} ~\text{or}~ + \half \ket{\l} \bra{\l} \right) + \CO(\b-1)
\\ \e_{\m\n\l\s} P_1^\m P_3^\n K^\l S_a^\s &\to \frac{i}{2} ( v - u ) ( K \cdot S_a ) + \CO(\b-1) = - i m_a^2 m_b \sqrt{r^2 - 1} \left( \frac{K \cdot S_a}{m_a} \right) + \CO (\b - 1)
\\ \e_{\m\n\l\s} P_1^\m P_3^\n K^\l S_b^\s &\to \frac{i}{2} ( v - u ) ( K \cdot S_b ) + \CO(\b-1) = - i m_a m_b^2 \sqrt{r^2 - 1} \left( \frac{K \cdot S_b}{m_b} \right) + \CO (\b - 1) \,.
\eld \label{eq:LorInvOpBasisList}
\el
The last two results are obtained from the definitions of $u$ and $v$, which are consistent with eq.(B.4) of \cite{Guevara:2017csg} up to phase
. These Lorentz-invariant combination of operators constitute the basis on which the computed LS is expanded, so that classical potential can be read out from the results.

\subsubsection{Symmetric basis for polarisation tensors}
As undotted spinor index and and dotted spinor index can be interchanged freely by the ``Dirac equation relations'' eq.\eqc{eq:SHvarDiracEq}, all LS computations can be simplified by expressing all operators to act on dotted indices. The end result can be expressed as eq.\eqc{eq:LSExp}
\bl
\text{LS} = \sum_{i=0}^{2s_a} \sum_{j=0}^{2s_b} N^{s_a,s_b}_{i,j} \tilde{A}_{i,j} \epsilon^{2s_a{-}i}(Sp_a)^i \epsilon^{2s_b{-}j}(Sp_b)^j,\quad  N^{s_a,s_b}_{i,j} = \frac{(2s_a)!}{(2s_a - i)!} \frac{(2s_b)!}{(2s_b-j)!} \, , \label{eq:LSExp2}
\el
where definitions for $Sp_a$ and $Sp_b$ are the same as in eq.\eqc{eq:SpaSpbDef}.
\bl
Sp_a = \frac{\sket{\hat\l}\sbra{\hat\l}}{m_a}, \quad Sp_b = \frac{\sket{\l}\sbra{\l}}{m_b} \,. \label{eq:SpaSpbDef2}
\el
As the above notation indicates, we've expressed the LS in purely dotted basis, i.e. the external massive spinors (wave functions) have been stripped off. Note that eq.\eqc{eq:LSExp2} tells us that the $SL(2,\IC)$ indices of the external wave functions are contracted either with $Sp_{a,b}$s or the Levi-Cevita tensors $\epsilon_{\dot{\alpha}\dot{\beta}}$. We stress that this simplification is a consequence of the HCL. The normalisation factors $N^{s_a,s_b}_{i,j}$ are the combinatoric factors that appear due to Lie algebra properties as in eq.\eqc{eq:SpinProdNumFacts}.

While it is tempting to match $Sp_a$ and $Sp_b$ in the above expansion to Lorentz-invariant combination of operators discussed in section \ref{sec:matchingopbasis}, the basis used to compute $\tilde{A}_{i,j}$ does not treat dotted and undotted indices democratically at all. It is natural to relate the operators considered in section \ref{sec:matchingopbasis} to one-body effective action operators introduced in section \ref{sec:BHmincoup}, and the natural basis on which these operators act should treat dotted and undotted indices equivalently. The expansion eq.\eqc{eq:LSExp2} needs to be massaged so that LS is expanded in one-body effective action operators.

An educated guess that could be made from the identity eq.\eqc{eq:3ptkin1} for three-point kinematics is that half of $Sp_a$ and $Sp_b$ in eq.\eqc{eq:SpaSpbDef2} do not encode the dynamics but come from kinematics, 
therefore this kinematical contribution which is irrelevant to the dynamics should be factored out from the computed coefficients $\tilde{A}_{i,j}$. A trick\footnote{The authors would like to thank Justin Vines for helpful discussions.} that can be used is to repackage the coefficients by constructing power series in variables $x_a$ and $x_b$ in the following way.
\bl
\sum_{i,j} A_{i,j} (x_a)^i (x_b)^j &= e^{-x_a/2} e^{-x_b/2} \sum_{i,j} \tilde{A}_{i,j} (x_a)^i (x_b)^j \,. \label{eq:AtildeToARelation}
\el
The multiplication by the factor $e^{-x_a/2} e^{-x_b/2}$ has the effect of factoring out the kinematical (or non-dynamical) contributions from $\tilde{A}_{i,j}$ coming from eq.\eqc{eq:3ptkin1} when expressing the LS in the anti-chiral (or purely dotted index) basis. Thus, the data of dynamics is encoded by the series expansion
\bl
\sum_{i,j} A_{i,j} (Sp_a)^i (Sp_b)^j
\el
which can be viewed as the expression for LS in the vector (or the symmetric) basis. The $Sp_a$ and $Sp_b$ in the above expression are operators which act on all indices, where their action on anti-chiral indices is given by eq.\eqc{eq:SpaSpbDef2}. While their action on chiral indices is not specified, eq.\eqc{eq:SpaSpbDef2} and eq.\eqc{eq:LorInvOpBasisList} can be compared to match $Sp_a$ and $Sp_b$ with HCL operators $(K \cdot S_a)$ and $(K \cdot S_b)$;
\bl
\bld
Sp_a &\to 2 \left( \frac{K \cdot S_a}{m_a} \right)
\\ Sp_b &\to - 2 \left( \frac{K \cdot S_b}{m_b} \right) \,.
\eld \label{eq:SpaSpbReadingRules}
\el
Reduction to the form $K \cdot S$ in the HCL does not mean that the operator is necessarily the operator $K \cdot S$. The exact matching onto Lorentz-invariant combination depends on the power of $\sqrt{r^2 - 1}$ that may appear; e.g. the following matching rules can be devised from inspecting the list eq.\eqc{eq:LorInvOpBasisList} closely.
\bl
\bld
\sqrt{r^2 - 1} Sp_a &\to \frac{2i}{m_a^2 m_b} \e_{\m\n\l\s} P_1^\m P_3^\n K^\l S_a^\s
\\ \sqrt{r^2 - 1} Sp_b &\to - \frac{2i}{m_a m_b^2} \e_{\m\n\l\s} P_1^\m P_3^\n K^\l S_b^\s \,.
\eld \label{eq:RelSpaSpbReadingRules}
\el


\subsection{Matching at leading order (1 PM)}
\bfig
\centering
\includegraphics[width=0.4\textwidth]{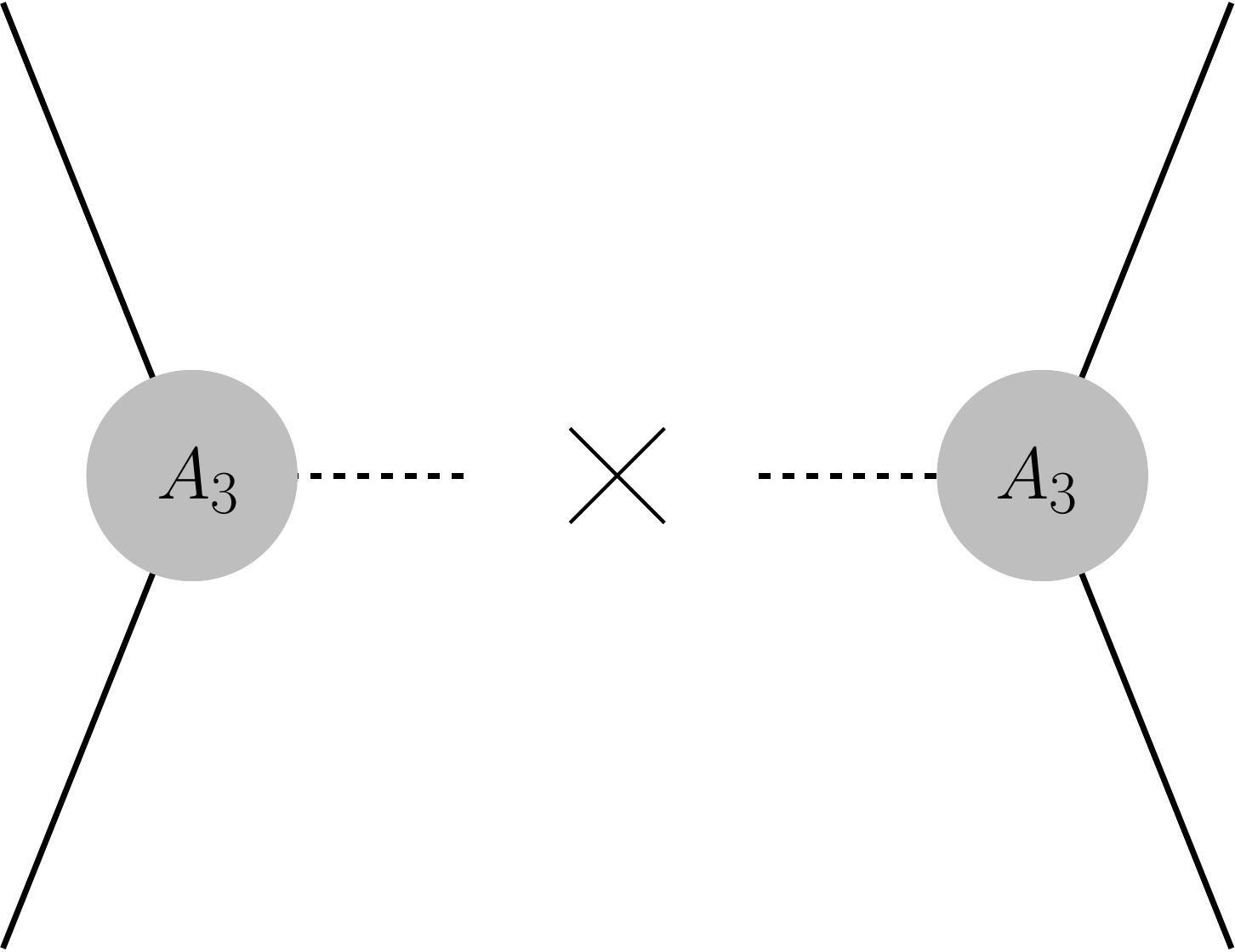}
\caption{A graphical representation of gluing two 3pt amplitudes.} \label{fig:glue3pt}
\efig
The leading order contribution corresponds to the tree-level amplitude. The LS at this level is just a product of two 3pt amplitudes, multiplied by the massless graviton propagator $\frac{1}{t}$ as represented in \fig{fig:glue3pt};
\bl
\text{LS} &= \frac{A_{3a}^+ A_{3b}^- + A_{3a}^- A_{3b}^+}{t} = \a^{2} m_a^2 m_b^2 \frac{( x_1 \bar{x}_3)^2 \la \bold{1} \bold{2} \ra^{2s_a} [ \bold{3} \bold{4} ]^{2s_b} + ( \bar{x}_1 {x}_3)^2 [\bold{1}  \bold{2} ]^{2s_a} \la \bold{3} \bold{4} \ra^{2s_b}}{t} \,.
\el
Following Guevara~\cite{Guevara:2017csg}, this LS can be cast into a purely anti-chiral form. Adopting the definitions in \ref{sec:LScompdefs}, this expression simplifies to \cite{Guevara:2017csg}
\bl
\text{LS} &= - \frac{\a^2}{q^2} \left[ u^2 (1 - Sp_a)^{2s_a} + v^2 (1 - Sp_b)^{2s_b} \right]
\el
where $Sp_a = \frac{\sket{\hat\l}\sbra{\hat\l}}{m_a}$ and $Sp_b = \frac{\sket{\l}\sbra{\l}}{m_b}$, as defined above.

A note of caution is that all $\tilde{A}_{i,j}$ coefficients of LSs were computed as if LSs were operators acting from left to right to match the sign choices in \cite{Guevara:2017csg}, i.e. the polarisation tensors for the incoming particles are put to the left while that for the outgoing particles are put to the right. The factor $e^{-x_a/2}e^{-x_b/2}$ in eq.\eqc{eq:AtildeToARelation} must be changed to $e^{x_a/2}e^{x_b/2}$ in this convention to eliminate the non-dynamical kinematical factors, since the sign appearing in the middle of eq.\eqc{eq:3ptkin1} will be flipped. 
Also, the rule eq.\eqc{eq:SOcorrection} will have a sign flip due to this convention. The rules (modified) eq.\eqc{eq:SOcorrection}, eq.\eqc{eq:SpaSpbReadingRules}, and eq.\eqc{eq:RelSpaSpbReadingRules} are then applied to yield the expression for the classical potential.

\subsubsection{Spin-independent Newtonian gravity}
Tree-level computation of LS yields the following result for $A_{0,0}$, the term responsible for $V = -\frac{GMm}{r}$ of Newtonian gravity.
\bl
A_{0,0} &= \tilde{A}_{0,0} = -\frac{16 \pi  G m_a^2 m_b^2 }{q^2}-\frac{32 \pi  G m_a^2 m_b^2}{q^2} \sqrt{r^2-1}^2 \,.
\el
Since only this term can contribute to the spin-independent part of the LS, this part determines the spin-independent part of the classical potential. The classical potential to leading and subleading orders in PN can be read out by multiplying a factor of $\frac{1}{4E_a E_b}$, which is consistent with the results of \cite{Holstein:2008sx}.
\bl
\left. \frac{\text{LS}}{4E_a E_b} \right|_{S_a^0 S_b^0} &= -\frac{4 \pi  G m_a m_b}{q^2} -\frac{2 \pi  G p^2 \left(8 m_a m_b+3 m_a^2+3 m_b^2\right)}{q^2 m_a m_b} \,.
\el

\subsubsection{Spin-orbit contributions} \label{sec:LOSO}
The result for $A_{1,0}$ up to subleading order in $\sqrt{r^2 - 1}$ is the following.
\bl
\bld
\tilde{A}_{1,0} &= \frac{8 \pi G m_a^2 m_b^2}{q^2}+\frac{16 \pi G m_a^2 m_b^2 \sqrt{r^2-1}}{q^2}
\\ A_{1,0} &= \frac{16 \pi G  m_a^2 m_b^2 \sqrt{r^2-1}}{q^2}+\frac{8 \pi G m_a^2 m_b^2 \sqrt{r^2-1}^3}{q^2} \,.
\eld
\el
Of the two set of rules eq.\eqc{eq:SpaSpbReadingRules} and eq.\eqc{eq:RelSpaSpbReadingRules}, only the second rule matches the order in $\sqrt{r^2 - 1}$. Thus it can be concluded that $\left. \text{LS} \right|_{S_a^1 S_b^0} = \frac{32 i \pi G m_b}{q^2} \e_{\m\n\l\s} P_1^\m P_3^\n K^\l S_a^\s$ by matching the orders in $\sqrt{r^2 - 1}$, or
\bl
\bld
\left. \frac{\text{LS}}{4E_a E_b} \right|_{S_a^1 S_b^0} &= \frac{8 \pi i G \left(m_a+m_b\right)}{q^2 m_a} ( \vec{S_a} \cdot \vec{p_a} \times \vec{q} )
\\ &= - \frac{2G}{r^2} \frac{m_a + m_b}{m_a} ( \vec{S_a} \cdot \vec{p_a} \times \hat{n} )
\eld
\el
to leading PN order where eq.\eqc{eq:LorInvOpValues} was used to evaluate non-relativistic expression which is consistent with the known results; eq.(48) of \cite{Porto:2005ac} and eq.(71) of \cite{Levi:2010zu}. This result corresponds to the choice of covariant SSC $S^{\m\n} p_\n = 0$.

Taking eq.\eqc{eq:SOcorrection} into account, spin-independent term contributes an additional factor which matches other results known in the literature; eq.(51) of \cite{Porto:2005ac}, eq.(53) of \cite{Holstein:2008sx}, and eq.(70) of \cite{Levi:2010zu}; this result corresponds to the choice of NW SSC $S^{\m\n} (p_\n + m e^0_\n) = 0$.
\bl
\left. \frac{\text{LS}}{4E_a E_b} \right|_{S_a^1 S_b^0}^{\text{NW SSC}} &= - \frac{G}{r^2} \frac{4 m_a + 3 m_b}{2 m_a} ( \vec{S_a} \cdot \vec{p_a} \times \hat{n} ) \,.
\el

\subsubsection{Spin-spin interactions}
The computation result 
up to subleading order in $\sqrt{r^2 - 1}$ is the following:
\bg
A_{1,1} = \frac{4 \pi G m_a^2 m_b^2}{q^2} + \frac{8 \pi G m_a^2 m_b^2 \sqrt{r^2-1}^2}{q^2}
\eg
Thus, applying the rules eq.\eqc{eq:SpaSpbReadingRules} to the LS makes the LS to take the form
\bl
\left. \text{LS} \right|_{S_a^1 S_b^1} &= - \frac{16 \pi G m_a m_b}{q^2} (K \cdot S_a)(K \cdot S_b)
\el
and the classical potential to leading PN order is
\bl
\bld
\left. \frac{\text{LS}}{4E_a E_b} \right|_{S_a^1 S_b^1} &= - \frac{4 \pi G }{q^2} (\vec{S_a} \cdot \vec{q})(\vec{S_b} \cdot \vec{q})
\\ &= - \frac{G}{r^3} \left( \vec{S_a} \cdot \vec{S_b} - 3 (\vec{S_a} \cdot \vec{n}) (\vec{S_b} \cdot \vec{n}) \right)
\eld
\el
consistent with the results eq.(6.9) of \cite{Levi:2015msa} and eq.(90) of \cite{Holstein:2008sx}. Note that contributions due to eq.\eqc{eq:SOcorrection} do not change the potential in the leading order in PN, since additional dependence on $\vec{p_i}$ makes the potential subleading in powers of $\frac{v^2}{c^2}$.

\subsubsection{Quadratic in spin effects}
$A_{2,0}$ is relevant for this computation.
\bl
A_{2,0} &= -\frac{2 \pi  G m_a^2 m_b^2}{q^2} - \frac{4 \pi G m_a^2 m_b^2 \sqrt{r^2-1}^2}{q^2} \,.
\el
This is the first two leading terms in expansion over $\sqrt{r^2 - 1}$. The LS and the classical potential up to leading PN order then takes the form
\bl
\left. \text{LS} \right|_{S_a^2 S_b^0} &= - \frac{8 \pi G m_b^2}{q^2} (K \cdot S_a)^2
\\ \left. \frac{\text{LS}}{4E_a E_b} \right|_{S_a^2 S_b^0} &= - \frac{2 \pi  G m_b}{q^2 m_a} (\vec{S_a} \cdot \vec{q})^2 \label{eq:spinquadLO}
\\ &= - \frac{G m_b}{2 m_a r^3} \left( \vec{S_a} \cdot \vec{S_a} - 3 (\vec{S_a} \cdot \vec{n})^2 \right)
\el
also consistent with the results eq.(6.10) of \cite{Levi:2015msa}, provided that $C_{1(\text{ES}^2)} = 1$. Similar to spin-spin interaction term, the application of eq.\eqc{eq:SOcorrection} does not change the potential at this order.

\subsubsection{Cubic in spin effects} \label{sec:LOcubicspin}
There are two terms to consider; $A_{3,0}$ and $A_{2,1}$. Their first two leading terms in the $\sqrt{r^2 - 1}$ expansion are given below.
\bl
A_{3,0} &= \frac{2 \pi G m_a^2 m_b^2 \sqrt{r^2-1} }{3 q^2}+\frac{\pi G m_a^2 m_b^2 \sqrt{r^2-1}^3 }{3 q^2}
\\ A_{2,1} &= -\frac{2 \pi  G m_a^2 m_b^2 \sqrt{r^2-1}}{q^2}-\frac{\pi G m_a^2 m_b^2 \sqrt{r^2-1}^3 }{q^2} \,.
\el
Computation of $S_a^3$-term is straightforward, since there are no ambiguities.
\bl
\left. \text{LS} \right|_{S_a^3 S_b^0} &= \frac{16 i \pi G m_b}{3 m_a^2 q^2} (K \cdot S_a)^2 \e_{\m\n\l\s} P_1^\m P_3^\n K^\l S_a^\s
\\ \left. \frac{\text{LS}}{4E_a E_b} \right|_{S_a^3 S_b^0} &= \frac{4 i \pi  G \left(m_a+m_b\right)}{3 q^2 m_a^3} (\vec{S_a} \cdot \vec{q})^2 ( \vec{S_a} \cdot \vec{p_a} \times \vec{q} ) \nn
\\ &= - \frac{G (m_a + m_b)}{r^4 m_a^3} (\vec{S_a} \cdot \vec{p_a} \times \vec{n}) \left( \vec{S_a} \cdot \vec{S_a} - 5 (\vec{S_a} \cdot \vec{n})^2 \right) \,.
\el
This leading PN order expression matches the terms proportional to $C_{1(\text{BS}^3)}$ in eq.(3.10) of \cite{Levi:2014gsa} when it is set to unity. When eq.\eqc{eq:SOcorrection} is taken into account, there is an additional term generated from eq.\eqc{eq:spinquadLO} that contributes to this potential which matches the terms proportional to $C_{1(\text{ES}^2)}$ when it is set to unity.
\bl
\left. \frac{\text{LS}}{4E_a E_b} \right|_{S_a^3 S_b^0}^{\text{NW SSC}} &= \frac{i \pi  G \left(4 m_a+m_b\right)}{3 q^2 m_a^3} (\vec{S_a} \cdot \vec{q})^2 ( \vec{S_a} \cdot \vec{p_a} \times \vec{q} ) \nn
\\ &= - \frac{G (4 m_a + m_b)}{4 r^4 m_a^3} (\vec{S_a} \cdot \vec{p_a} \times \vec{n}) \left( \vec{S_a} \cdot \vec{S_a} - 5 (\vec{S_a} \cdot \vec{n})^2 \right) \,.
\el
At first sight, computing the contribution from $A_{2,1}$ seems complicated by the fact that there are two combinations that reduce to the same factor ($\frac{i m_a^3 m_b^2}{8} \sqrt{r^2 - 1} Sp_a^2 Sp_b$) in the HCL. Nevertheless, it is possible to write this LS as a linear combination of the two by introducing an arbitrary real parameter $\a$.
\bl
\bld
\left. \text{LS} \right|_{S_a^2 S_b^1} &= 
\frac{16 i \pi G}{m_a q^2} \left[ \a (K \cdot S_a)^2 \e_{\m\n\l\s} P_1^\m P_3^\n K^\l S_b^\s \right.
\\ & \phantom{\frac{16 i \pi G}{m_a q^2}\frac{16 i \pi G}{m_a q^2}} \left. + (1-\a) (K \cdot S_a)(K \cdot S_b) \e_{\m\n\l\s} P_1^\m P_3^\n K^\l S_a^\s \right] \,.
\eld
\el
Computing the classical potential up to leading PN order requires taking the non-relativistic limit.
\bl
\bld
\left. \frac{\text{LS}}{4E_a E_b} \right|_{S_a^2 S_b^1} &= 
\frac{4 i \pi  G \left(m_a+m_b\right)}{q^2 m_a^2 m_b} \left[ \a (\vec{S_a} \cdot \vec{q})^2 ( \vec{S_b} \cdot \vec{p_b} \times \vec{q} ) \right.
\\ & \phantom{asdfasdfasdfasdf} \left. - (1 - \a) (\vec{S_a} \cdot \vec{q})(\vec{S_b} \cdot \vec{q}) ( \vec{S_a} \cdot \vec{p_a} \times \vec{q} ) \right] \,.
\eld
\el
However, the following vector identity\cite{Levi:2014gsa} can be used to relate the different combinations.
\bl
\vec{A_1} (\vec{A_2} \cdot \vec{A_3} \times \vec{A_4}) &= \vec{A_2} (\vec{A_1} \cdot \vec{A_3} \times \vec{A_4}) + \vec{A_3} (\vec{A_1} \cdot \vec{A_4} \times \vec{A_2}) + \vec{A_4} (\vec{A_1} \cdot \vec{A_2} \times \vec{A_3}) \,.
\el
Setting $\vec{A_1} = \vec{S_a}$, $\vec{A_2} = \vec{S_b}$, $\vec{A_3} = \vec{p}$, and $\vec{A_4} = \vec{q}$, it can be shown that
\bl
(\vec{q} \cdot \vec{S_a}) (\vec{S_b} \cdot \vec{p} \times \vec{q}) &= (\vec{q} \cdot \vec{S_b}) (\vec{S_a} \cdot \vec{p} \times \vec{q}) + q^2 (\vec{p} \cdot \vec{S_a} \times \vec{S_b})
\el
therefore
\bl
(\vec{q} \cdot \vec{S_a}) (\vec{S_b} \cdot \vec{p_b} \times \vec{q}) &= - (\vec{q} \cdot \vec{S_b}) (\vec{S_a} \cdot \vec{p_a} \times \vec{q}) - q^2 (\vec{p} \cdot \vec{S_a} \times \vec{S_b}) \label{eq:cubicspinequivclass}
\el
and since there is an overall factor of $q^{-2}$ in the amplitude already, changes in $\a$ is reflected in the classical potential as derivative delta-like interaction which does not affect the long-distance behaviour; $\a$ is a free parameter that can be tuned arbitrarily without affecting the long-distance behaviour. The non-relativistic limit takes the following form in position space.
\bl
\bld
\left. \frac{\text{LS}}{4E_a E_b} \right|_{S_a^2 S_b^1} &= - \frac{3 G (m_a + m_b)}{r^4 m_a^2 m_b} \left[ \a \left\{ (\vec{S_b} \cdot \vec{p_b} \times \vec{n}) \left( \vec{S_a} \cdot \vec{S_a} - 5 (\vec{S_a} \cdot \vec{n})^2 \right) + 2 (\vec{p_b} \cdot \vec{S_a} \times \vec{S_b}) ( \vec{S_a} \cdot \vec{n} ) \right\} \right.
\\ & \left. - (1-\a) \left\{ - (\vec{S_a} \cdot \vec{n})(\vec{p_a} \cdot \vec{S_a} \times \vec{S_b}) + (\vec{S_a} \cdot \vec{p_a} \times \vec{n}) \left( (\vec{S_a} \cdot \vec{S_b}) - 5 (\vec{S_a} \cdot \vec{n})(\vec{S_b} \cdot \vec{S_a}) \right) \right\} \right] \,.
\eld \label{eq:LOSa2Sb1}
\el
When $\a$ is set to unity, this expression matches the sum of first two terms proportional to $C_{1(\text{ES}^2)}$ in eq.(3.10) of \cite{Levi:2014gsa} provided $C_{1(\text{ES}^2)}$ is set to unity. Note that there are two sources that can contribute to this potential through eq.\eqc{eq:SOcorrection}; the first is the contribution from $A_{2,0}$ which is
\bl
- \frac{\pi i  G}{q^2 m_a m_b} (\vec{S_a} \cdot \vec{q})^2 ( \vec{S_b} \cdot \vec{p} \times \vec{q} ) = \frac{\pi i  G}{q^2 m_a m_b} (\vec{S_a} \cdot \vec{q})^2 ( \vec{S_b} \cdot \vec{p_b} \times \vec{q} )
\el
and the other is the contribution from $A_{1,1}$ which is
\bl
- \frac{2 i \pi G }{q^2 m_a^2} (\vec{S_a} \cdot \vec{q})(\vec{S_b} \cdot \vec{q}) ( \vec{S_a} \cdot \vec{p_a} \times \vec{q} ) = - \frac{2 i \pi G }{q^2 m_a^2} (\vec{S_a} \cdot \vec{q})(\vec{S_b} \cdot \vec{q}) ( \vec{S_a} \cdot \vec{p} \times \vec{q} ) \,.
\el
In position space, these two contributions take the following form.
\bg
- \frac{3 G}{4 r^4 m_a m_b} \left[ (\vec{S_b} \cdot \vec{p_b} \times \vec{n}) \left( \vec{S_a} \cdot \vec{S_a} - 5 (\vec{S_a} \cdot \vec{n})^2 \right) + 2 (\vec{p_b} \cdot \vec{S_a} \times \vec{S_b}) ( \vec{S_a} \cdot \vec{n} ) \right] \label{eq:contributio2 PMA20}
\\ \frac{3 G}{2 m_a^2 r^4} \left[ \vec{S_a} \cdot \vec{p_a} \times \vec{n} \left( \vec{S_a} \cdot \vec{S_b} - 5 (\vec{S_a} \cdot \vec{n})(\vec{S_b} \cdot \vec{n}) \right) + \vec{S_a} \cdot \vec{n} \vec{S_a} \cdot \vec{p_a} \times \vec{S_b} \right]\label{eq:contributio2 PMA11}
\eg
Adding up eq.\eqc{eq:LOSa2Sb1}, eq.\eqc{eq:contributio2 PMA20}, and eq.\eqc{eq:contributio2 PMA11} gives an expression that matches with eq.(3.10) of \cite{Levi:2014gsa} when $C_{1(\text{ES}^2)}$ is set to unity. Note that using eq.\eqc{eq:cubicspinequivclass}, the final potential can be written in the following form in momentum space.
\bl
-\frac{i \pi  G \left(3 m_a+2 m_b\right)}{q^2 m_a^2 m_b} (\vec{S_a} \cdot \vec{q})^2 ( \vec{S_b} \cdot \vec{p_b} \times \vec{q} ) \,.
\el

\subsubsection{Quartic in spin effects}
Relevant terms are $A_{4,0}$, $A_{3,1}$, and $A_{2,2}$; their first two leading terms in $\sqrt{r^2 - 1}$ expansion are
\bl
A_{4,0} &= -\frac{\pi G m_a^2 m_b^2}{24 q^2} - \frac{\pi G m_a^2 m_b^2 \sqrt{r^2-1}^2}{12 q^2}
\\ A_{3,1} &= \frac{\pi G m_a^2 m_b^2}{6 q^2}+\frac{\pi G m_a^2 m_b^2 \sqrt{r^2-1}^2}{3 q^2}
\\ A_{2,2} &= -\frac{\pi G m_a^2 m_b^2}{4 q^2}-\frac{\pi G m_a^2 m_b^2 \sqrt{r^2-1}^2}{2 q^2} \,.
\el
There are no ambiguities for matching the LS from these results.
\bl
\left. \text{LS} \right|_{S_a^4 S_b^0} &= - \frac{2 \pi G m_b^2}{3 m_a^2 q^2} (K \cdot S_a)^4
\\ \left. \text{LS} \right|_{S_a^3 S_b^1} &= - \frac{8 \pi G m_b}{3 m_a q^2} (K \cdot S_a)^3 (K \cdot S_b)
\\ \left. \text{LS} \right|_{S_a^2 S_b^2} &= - \frac{4 \pi G}{q^2} (K \cdot S_a)^2 (K \cdot S_b)^2 \,.
\el
Computing the classical potential to leading PN order is straightforward, which is consistent with the results eq.(4.4) of \cite{Levi:2014gsa} when $C_{1(\text{ES}^2)}$, $C_{2(\text{ES}^2)}$, $C_{1(\text{BS}^3)}$, and $C_{1(\text{ES}^4)}$ are all set to unity. Note that eq.\eqc{eq:SOcorrection} does not induce any corrections at this order.
\bl
&\bld
\left. \frac{\text{LS}}{4E_a E_b} \right|_{S_a^4 S_b^0} &= - \frac{\pi G m_b}{6 m_a^3 q^2} (\vec{q} \cdot \vec{S_a})^4
\\ &= - \frac{3 G m_b}{8 m_a^3 r^5} \left[ (\vec{S_a} \cdot \vec{S_a})^2 - 10 (\vec{S_a} \cdot \vec{S_a}) (\vec{S_a} \cdot \vec{n})^2 + \frac{35}{3} (\vec{S_a} \cdot \vec{n})^4 \right]
\eld
\\ &\bld \left. \frac{\text{LS}}{4E_a E_b} \right|_{S_a^3 S_b^1} &= - \frac{2 \pi G}{3 m_a^2 q^2} (\vec{q} \cdot \vec{S_a})^3 (\vec{q} \cdot \vec{S_b})
\\ &= - \frac{3 G}{2 m_a^2 r^5} \left[ \vec{S_a}^2 (\vec{S_a} \cdot \vec{S_b}) - 5 \left\{ (\vec{S_a} \cdot \vec{S_b}) (\vec{S_a} \cdot \vec{n})^2 + \vec{S_a}^2 (\vec{S_a} \cdot \vec{n})(\vec{S_b} \cdot \vec{n}) \right\} \right.
\\ & \left. \phantom{\left\{ (\vec{S_a} \cdot \vec{S_b}) (\vec{S_a} \cdot \vec{n})^2 + \vec{S_a}^2 (\vec{S_a} \cdot \vec{n})(\vec{S_b} \cdot \vec{n}) \right\}} + \frac{35}{3} (\vec{S_a} \cdot \vec{n})^3(\vec{S_b} \cdot \vec{n}) \right]
\eld
\\ &\bld \left. \frac{\text{LS}}{4E_a E_b} \right|_{S_a^2 S_b^2} &= - \frac{\pi G}{m_a m_b q^2} (\vec{q} \cdot \vec{S_a})^2 (\vec{q} \cdot \vec{S_b})^2
\\ &= - \frac{3 G}{4 m_a m_b r^5} \left[ \vec{S_a}^2 \vec{S_b}^2 + 2 (\vec{S_a} \cdot \vec{S_b})^2 - 5 \left\{ (\vec{S_a} \cdot \vec{n})^2 \vec{S_b}^2 + \vec{S_a}^2 (\vec{S_b} \cdot \vec{n})^2 \right. \right.
\\ & \left. \left. \phantom{ \vec{S_a}^2 \vec{S_a}^2 \vec{S_a}^2 } + 4 (\vec{S_a} \cdot \vec{S_b})(\vec{S_a} \cdot \vec{n})(\vec{S_b} \cdot \vec{n}) - 7 (\vec{S_a} \cdot \vec{n})^2(\vec{S_b} \cdot \vec{n})^2 \right\} \right] \,.
\eld
\el
\section{Results for the classical potential (2 PM)}\label{subsec:2PM_result}
Based on dimensional analysis, it can be argued that two particle irreducible (2PI) diagrams with only one massive internal leg per loop contribute to the classical potential for the case of non-spinning particles\cite{Neill:2013wsa}. The only topology that meets this criteria at one loop is the triangle topology. More precisely, since in four-dimensions one-loop amplitudes can be cast into a scalar integral basis involving box, triangle and bubble integrals~\cite{tHooft:1978jhc, vanNeerven:1983vr}, the statement is that only triangle scalar integrals are relevant for contributions to the classical potential.

To understand why note that the one-loop integrals that are relevant to our problem always contains two massless graviton propagators:
$$\includegraphics[scale=0.5]{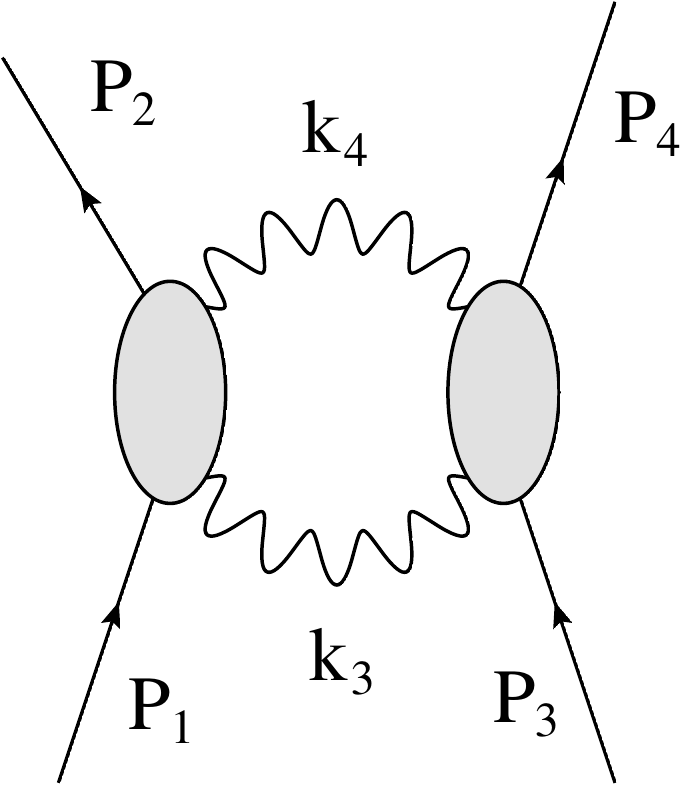}\,.$$
This implies that the result will have non-analyticity in $q^2 = - t$, reflecting the presence of the massless cut. There are two types of such non-analyticity, 
\eq
\sqrt{q^2},\quad \log q^2  \,.
\eqe
The first corresponds to classical contribution and the second quantum~\cite{Holstein:2004dn}. It is then straightforward to march through the scalar integrals, and find that only scalar triangle yields the desired non-analyticity~\cite{Holstein:2008sw}:
\eq\label{TriClass}
\int \frac{d^4\ell}{(2\pi)^4}\frac{1}{\ell^2(\ell+q)^2((\ell+p)^2-m^2)}=\frac{i}{16\pi^2}\frac{1}{m^2}\left[-\frac{\log q^2}{2}-\frac{m \pi^2}{2\sqrt{q^2}}\right]+\mathcal{O}(q) \,,
\eqe
where $p$ will be the momenta of one of the external lines. Thus to extract the classical result at 2 PM amounts to computing the integral coefficient for the scalar triangle. 
\bfig
\centering
\includegraphics[width=0.4\textwidth]{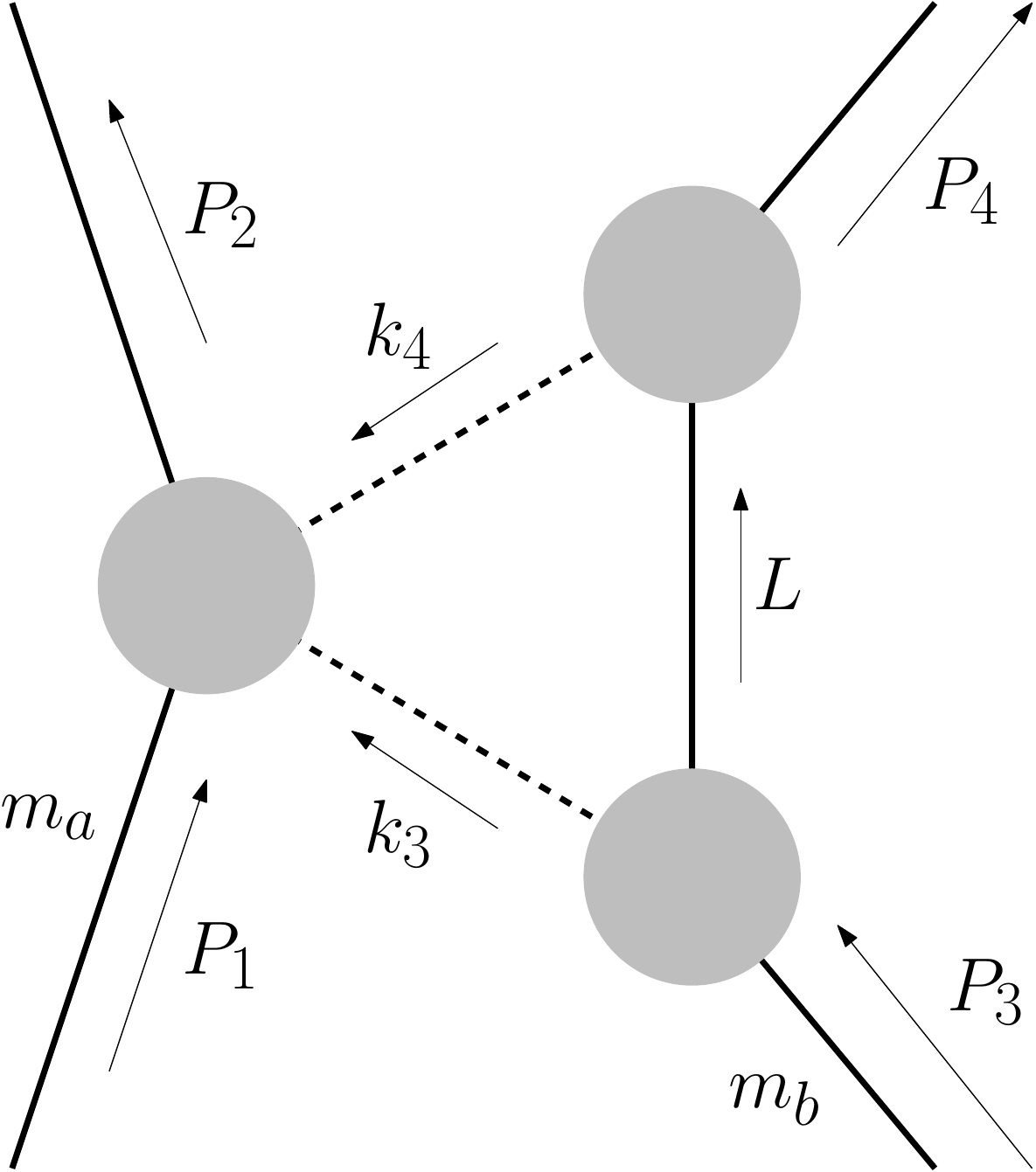}
\caption{The triple-cut diagram for $b$-topology triangle cut. The $a$-topology diagram is obtained by exchanging the labels of particles $a$ and $b$.} \label{fig:guevaratriangle}
\efig

The integral coefficients can be readily computed using generalized unitarity methods~\cite{Bern:1994zx, Bern:1994cg}. As the triangle integral has three propagators, one can explore the kinematic regime of the loop momenta where all three propagators become on-shell, and the ``residue" simply becomes the product of two three-point and a four-point on-shell amplitude, as shown in \fig{fig:guevaratriangle}. This is termed the triangle cut. Note that the triangle integral is not the only basis integral that contributes to the triangle cut. Box integrals with one extra propagator can contribute as well. The challenge is then to separate these two contributions. 

This problem was beautifully solved by Forde~\cite{Forde:2007mi}, which parameterize the loop momenta in terms of four complex variables, and can be fixed as propagators  go on-shell. For the triangle cut, the loop momenta has only one complex variable left, and the cut can be viewed as a function of this variable with poles at finite values as well as infinity. The finite poles represents extra propagator becoming on shell, and hence the presence of box integrals. Thus the contribution from the scalar triangle simply corresponds to the pole at infinity. 

We can again simplify things by evaluating the triangle coefficients in the HCL limit, and match with a preferred local operator basis, after which one performs the non-relativistic expansion to recover the classical potential just as what was done in the 1 PM case. In summary the 2PM result are obtained as follows:

\begin{itemize}
  \item  Compute the following Leading Singularity in the triangle cut
  \eq
  \text{LS}=\int_{\infty} d^4\ell \delta(D_1)\delta(D_2)\delta(D_3)M_3\times M_3\times M_4
  \eqe
  where $D_i = L_i^2 - m_i^2$ represents the three propagators that were put on-shell, and $\int_\infty$ indicates that we are picking the contribution at infinity for the remaining integration variable.   
  \item Due to solving the delta functions, the above generates a Jacobian factor $J$. Thus to get the triangle coefficient, we need to multiply the LS by $J^{-1}$.
  \item Finally we multiply the resulting triangle coefficient to the loop integral and perform the $q^2\rightarrow0$ expansion, and picking out the relevant classical piece, which from eq.(\ref{TriClass}), is given by 
  \eq
  -\frac{i}{32 m\sqrt{q^2}}\,.
  \eqe
  \end{itemize} 
Thus the final 2PM result is given by:
\eq
(2PM)=J^{{-}1}\times \text{LS} \times \left( {-}\frac{i}{32 m\sqrt{ q^2}} \right)\,.
\eqe
The Jacobian factor can be computed explicitly and in the HCL limit, yielding $J={-}\frac{1}{32 m\sqrt{{}q^2}}$, which cancels the last term in the above product! Thus the 2PM classical potential is simply reproduced from the LS along, as pointed out by Cachazo and Guevara \cite{Cachazo:2017jef, Guevara:2017csg}.  

From the previous sections it is clear that the three-point amplitudes that should enter the cut would be that of minimal coupling. For the four-point amplitude, we use the Compton amplitude that was constructed from matching the three-point minimal coupling on the residue. This however, leaves us with polynomial ambiguities as discussed previously. For $s\leq2$, the polynomial ambiguities come with additional $\frac{1}{m}$ factors, which was absent from the answer constructed from residues, and thus can be argued as finite size effects. Such separation is no-longer true for $s>2$. Thus for now, we will constrain ourselves to using the Compton amplitude of $s=2$, which in practice, means we will be limited to terms in the potential that are at most degree 4 in the each particle's spin operator. 

When computing the 1-loop scattering amplitude in the non-relativistic limit, there are terms that diverge as COM average momentum vanishes. These terms have an interpretation as second order perturbation theory effects from the 1 PM potential, or second Born approximation terms. Such terms are artifacts of iterating the 1 PM potential, and they must be subtracted to compute the correct 2 PM contributions to the potential; when these terms are not subtracted, the amplitude computed from the classical potential will double-count such contributions and lead to a wrong answer. It can be shown that such iteration terms only consist of singular terms in COM momentum $p_0:=\abs{\vec{p_a}}$ when non-relativistic propagator is used~\cite{Holstein:2008sx}\footnote{While it is also noted in \cite{Holstein:2008sx} that using non-relativistic propagators to separate iteration terms do not lead to a potential which is useful for computing equations of motion, this prescription will be adopted for its simplicity. A method is provided in the appendix of \cite{Holstein:2008sx} which computes the iteration terms from propagators with relativistic energy-momentum dispersion relations.}, so the following simple prescription for subtracting the iteration terms will be adopted; when there is a divergent term in the expression for LS, all poles in $p_0$ will be interpreted as coming from iteration and will be subtracted. The remaining finite pieces will be interpreted as the 2 PM potential\footnote{The conclusion depends on the order of $p_0$ pole subtraction and flux normalisation; taking the non-relativistic flux normalisation first and then subtracting $p_0$ poles gives the result which matches that of \cite{Vaidya:2014kza}, while subtracting $p_0$ poles first and then taking the non-relativistic flux normalisation gives the result which matches that of \cite{Holstein:2008sx}. The latter is adopted in this manuscript.}.

In this section we compute the spin dependent pieces of the 2 PM classical potential. The analysis is similar to 1 PM case, but 2 PM computations require separation of iterated 1 PM contributions, which is usually referred to as the second Born approximation term~\cite{Holstein:2008sx}. 

\subsection{Parametrisation and computation of the LS}

The parameterisations used in \cite{Guevara:2017csg} was used to compute the LS in this manuscript. The details of the parameterisation apart from the ones given in section \ref{sec:LScompmisc} will be presented here. Consider the triple-cut diagram in \fig{fig:guevaratriangle}, which is referred to as the $b$-topology. The loop momentum $L$ runs through the massive internal leg, and massless internal legs are parameterised as $k_3 = - L + P_3$ and $k_4 = L - P_4$. The parameterisation for the loop momentum $L=L(z)$ is;
\bl
L(z) &= zl + \omega K
\\ l &= \left( \sket{\eta} + B \sket{\l} \right) \left( \bra{\eta} + A \bra{\l} \right) \,.
\el
Imposing the triple-cut conditions $k_3^2 = k_4^2 = L^2 - m_b^2 = 0$ fixes $\omega = - \frac{1}{z}$, and $A(z)$ and $B(z)$ as rational functions of $z$ and $\b$. Defining $y = - \frac{z}{(\b - 1)^2}$ as in \cite{Guevara:2017csg},  the LS from this topology is computed to be
\bl
&\bld
\text{LS} &= \frac{1}{4} \sum_{h_3,h_4 = \pm \abs{h}} \int_{\text{LS}} d^4 L \delta(L^2 - m_b^2) \delta(k_3^2) \delta(k_4^2)
\\ & \phantom{asdfasdfasdf} \times M_4(P_1, -P_2, k_3^{h_3}, k_4^{h_4}) \times M_3(P_3, -L, -k_3^{-h_3}) \times M_3(-P_4, L, -k_4^{-h_4})
\eld
\\ &\bld
\phantom{\text{LS}}&{=} \sum_{h_3,h_4} \frac{\b}{16 (\b^2-1) m_b^2} \int_{\G_{\text{LS}}} \frac{dy}{y} M_4(P_1, -P_2, k_3^{h_3}, k_4^{h_4})
\\ & \phantom{asdfasdfasdfasdfasdfasdf} \times M_3(P_3, -L, -k_3^{-h_3}) \times M_3(-P_4, L, -k_4^{-h_4})
\eld \label{eq:2PMLS}
\el
where $\G_{\text{LS}}$ is taken to be the contour enclosing the pole at $y = \infty$. The product of on-shell amplitudes that constitute the integrand need to be interpreted as operator products, detailed procedure being given in \cite{Guevara:2017csg}. The choice for internal momenta spinor-helicity variables are given below.
\bl
\bld
\ket{k_3} = \frac{1}{\b + 1} \left( (\b^2 -1) \ket{\eta} - \frac{1 + \b y}{y} \ket{\l} \right) &\,,\,\sket{k_3} = \frac{1}{\b + 1} \left( (\b^2 -1) y \sket{\eta} + (1 + \b y) \sket{\l} \right)
\\ \ket{k_4} = \frac{1}{\b + 1} \left( \frac{\b^2 - 1}{\b} \ket{\eta} + \frac{1-y}{y} \ket{\l} \right) &\,,\, \sket{k_4} = \frac{1}{\b + 1} \left( - \b(\b^2 - 1) y \sket{\eta} + (1 - \b^2 y) \sket{\l} \right) \,.
\eld
\el
After having computed the $b$-topology LS, the result is added to the computed result for $a$-topology LS which can be evaluated from the $b$-topology LS by reflection, e.g. $u \leftrightarrow v$, $m_a \leftrightarrow m_b$, etc.

\subsection{Results for the classical potential}\label{subsec:2PM_result}
 In our computations, we are interested in the classical potential up to quartic order in spin, so all the results presented in this section are calculated from spin-2 particle scattering with $M_4(P_1, -P_2, k_3^{h_3}, k_4^{h_4} )$ in eq.\eqref{eq:2PMLS} given by the lower spin Compton amplitude eq.\eqref{eq:Lower_Spin_Graviton_Compton}\footnote{Here we only take the mixed helicity Compton amplitude contribution into account because the same helicity Compton amplitude has zero residue at $y \rightarrow \infty$. Take the $(++)$ channel Compton amplitude for example, the integrand $\frac{1}{y}M_4(P_1, -P_2, k_3^+, k_4^+)M_3(P_3, -L, -k_3^-)M_3(-P_4, L, -k_4^-)$ is of order $O(y)^1$, so that the residue at infinity is zero.}. Recall that from eq.(\ref{eq:LSExp}), the LS for minimally coupled $\{s_a,s_b\}$ particles can capture terms in the potential that is up to degree $2s_a$ in $S^\mu_a$, and $2s_b$ in $S^\mu_b$. Here we have verified that the overlapping results of $\{\frac{1}{2},\frac{1}{2}\}$, $\{1,1\}$, and $\{2,2\}$ are in agreement with each other. At the end of the current section we will discuss how to utilize this fact to fix ambiguities associated with higher spin scattering.

\subsubsection{Spin-independent}
$A_{0,0}^{\text{2 PM}}$ is needed to compute 2 PM contributions to the spin-independent 2PM contribution to the classical potential.
\bl
A_{0,0}^{\text{2 PM}} &= \frac{24 \pi ^2 G^2 m_a^2 m_b^2 \left(m_a+m_b\right)}{q}+\frac{30 \pi ^2 G^2 m_a^2 m_b^2 \left(m_a+m_b\right)\sqrt{r^2-1}^2 }{q} \,.
\el
The resulting leading PN order classical potential is consistent with the results given in \cite{Holstein:2008sx}.
\bl
\left. \frac{\text{LS}^{\text{2 PM}}}{4E_a E_b} \right|_{S_a^0 S_b^0} &= \frac{6 \pi ^2 G^2 m_a m_b \left(m_a+m_b\right)}{q} \,.
\el

\subsubsection{Spin-orbit interaction}
The coefficient $A_{1,0}^{\text{2 PM}}$, which is the 2 PM counterpart to $A_{1,0}$ computed in section \ref{sec:LOSO}, up to first three terms in the $\sqrt{r^2 - 1}$ expansion takes the following form.
\bl
\bld
A_{1,0}^{\text{2 PM}} &= -\frac{2 \pi ^2 G^2 m_a^2 m_b^2 \left(4 m_a+3 m_b\right)}{q \sqrt{r^2-1}}-\frac{6 \pi ^2 G^2 m_a^2 m_b^2 \left(4 m_a+3 m_b\right) \sqrt{r^2-1} }{q}
\\ & \phantom{==} -\frac{9 \pi ^2 G^2 m_a^2 m_b^2 \left(4 m_a+3 m_b\right) \sqrt{r^2-1}^3 }{4 q} \,.
\eld
\el
This is the first of numerous terms that include 1 PM potential iteration pieces; in the stationary limit $r \to 1$ this expression diverges due to the factor $\frac{1}{\sqrt{r^2 - 1}}$ in the first term. Using the following formula
\bl
\sqrt{r^2 - 1} &= \frac{p_0 \sqrt{2 \left(\sqrt{\left(m_a^2+p_0^2\right) \left(m_b^2+p_0^2\right)}+p_0^2\right)+m_a^2+m_b^2}}{m_a m_b}
\el
this expression can be converted to Laurent series in $p_0$, and dropping poles in $p_0$ gives the following expression.
\bl
\bld
\left. A_{1,0}^{\text{2 PM}} \right|_{\text{reg}} &= -\frac{\pi ^2 G^2 \sqrt{r^2-1} m_a^2 m_b^2 \left(62 m_a^2 m_b+57 m_a m_b^2+24 m_a^3+18 m_b^3\right)}{q \left(m_a+m_b\right){}^2} \,.
\eld
\el
All subleading $\sqrt{r^2 - 1}$ expansion pieces were dropped in the above expression. Combined with the contributions from $A_{0,0}^{\text{2 PM}}$ due to eq.\eqc{eq:SOcorrection}, the following expression is obtained for leading PN 2 PM spin-orbit coupling. This is consistent with the results eq.(57) in \cite{Holstein:2008sx}.
\bl
\left. \frac{\text{LS}^{\text{2 PM}}_{\text{reg}}}{4E_a E_b} \right|_{S_a^1 S_b^0}^{\text{final}} &= -\frac{i \pi ^2 G^2 \left(56 m_a^2 m_b+45 m_a m_b^2+24 m_a^3+12 m_b^3\right)}{2 q m_a \left(m_a+m_b\right)} (\vec{S_a} \cdot \vec{p} \times \vec{q}) \,.
\el

\subsubsection{Quadratic order in spin}
Up to first three terms in $\sqrt{r^2 - 1}$ expansion, the terms relevant for quadratic order in spin are;
\bl
&\bld
A_{2,0}^{\text{2 PM}} &= \frac{\pi ^2 G^2 m_a^2 m_b^2 (m_a + m_b)}{2 q \sqrt{r^2-1}^2}+\frac{\pi ^2 G^2 m_a^2 m_b^2 \left(22 m_a + 15 m_b \right)}{4 q}
\\ & \hskip120pt + \frac{5 \pi ^2 G^2 m_a^2 m_b^2\left(19 m_a + 12 m_b\right) \sqrt{r^2-1}^2 }{16 q}
\eld
\\ &\bld
A_{1,1}^{\text{2 PM}} &= -\frac{\pi ^2 G^2 m_a^2 m_b^2 \left(m_a+m_b\right)}{q \sqrt{r^2-1}^2}-\frac{19 \pi ^2 G^2 m_a^2 m_b^2 \left(m_a+m_b\right) }{2 q}
\\ & \hskip120pt -\frac{10 \pi ^2 G^2 m_a^2 m_b^2 \left(m_a+m_b\right) \sqrt{r^2-1}^2 }{q} \,.
\eld
\el
The leading terms after $p_0$ poles are subtracted out are
\bl
\left. A_{2,0}^{\text{2 PM}} \right|_{\text{reg}} &= \frac{\pi ^2 G^2 m_a^2 m_b^2 \left(35 m_a m_b+22 m_a^2+15 m_b^2\right)}{4 q \left(m_a+m_b\right)}
\\ \left. A_{1,1}^{\text{2 PM}} \right|_{\text{reg}} &= -\frac{\pi ^2 G^2 m_a^2 m_b^2 \left(36 m_a m_b+19 m_a^2+19 m_b^2\right)}{2 q \left(m_a+m_b\right)}
\el
which translate into
\bl
\left. \text{LS}^{\text{2 PM}}_{\text{reg}} \right|_{S_a^2 S_b^0} &= \frac{\pi ^2 G^2 m_b^2 \left(35 m_a m_b+22 m_a^2+15 m_b^2\right)}{q \left(m_a+m_b\right)} (K \cdot S_a)^2
\\ \left. \text{LS}^{\text{2 PM}}_{\text{reg}} \right|_{S_a^1 S_b^1} &= \frac{2 \pi ^2 G^2 m_a m_b \left(36 m_a m_b+19 m_a^2+19 m_b^2\right)}{q \left(m_a+m_b\right)} (K \cdot S_a)(K \cdot S_b)
\el
and becomes in the non-relativistic limit
\bl
&\bld
\left. \frac{\text{LS}^{\text{2 PM}}_{\text{reg}}}{4E_a E_b} \right|_{S_a^2 S_b^0} &= \frac{\pi ^2 G^2 m_b \left(35 m_a m_b+22 m_a^2+15 m_b^2\right)}{4 q m_a \left(m_a+m_b\right)} (\vec{q} \cdot \vec{S_a})^2
\eld
\\ &\bld
\left. \frac{\text{LS}^{\text{2 PM}}_{\text{reg}}}{4E_a E_b} \right|_{S_a^1 S_b^1} &= \frac{\pi ^2 G^2 \left(19 m_a^2+36 m_a m_b+19 m_b^2\right)}{2 q \left(m_a+m_b\right)} (\vec{q} \cdot \vec{S_a})(\vec{q} \cdot \vec{S_b}) \,.
\eld
\el
The latter can be compared with eq.(95) of \cite{Holstein:2008sx};
\bl
G^2 \frac{\pi^2}{\sqrt{-t}} \frac{19 m_a^2 + 36 m_a m_b + 19 m_b^2}{2(m_a + m_b)} \left[ (\vec{S_a} \cdot \vec{q}) (\vec{S_b} \cdot \vec{q}) - q^2 \vec{S_a} \cdot \vec{S_b} \right] \,.
\el
The two expression match up to terms proportional to ${q}^2 \vec{S_a} \cdot \vec{S_b}$, which are subleading in the HCL. While this subleading HCL contributions did not affect the long-distance behaviour for LO, this is no longer true for 2 PM; $q^1$ in momentum space is roughly equivalent to $r^{-4}$ in position space. It is not possible at the moment to compute subleading HCL contributions so the answers provided above cannot be complete, but the directional dependence on relative orientation of the bodies $(\vec{S_a} \cdot \vec{r})(\vec{S_b} \cdot \vec{r})$ can solely be attributed to non-vanishing HCL contributions and they can be computed by the methods provided in this manuscript.

Taking such HCL equivalence classes into account, the potential at this order will have the following form in momentum space.
\bl
&\bld
\left. \frac{\text{LS}^{\text{2 PM}}_{\text{reg}}}{4E_a E_b} \right|_{S_a^2 S_b^0} &= \frac{\pi ^2 G^2 m_b \left(35 m_a m_b+22 m_a^2+15 m_b^2\right)}{4 q m_a \left(m_a+m_b\right)} \left[ (\vec{q} \cdot \vec{S_a})^2 + q^2 \CO \right]
\eld
\\ &\bld
\left. \frac{\text{LS}^{\text{2 PM}}_{\text{reg}}}{4E_a E_b} \right|_{S_a^1 S_b^1} &= \frac{\pi ^2 G^2 \left(19 m_a^2+36 m_a m_b+19 m_b^2\right)}{2 q \left(m_a+m_b\right)} \left[ (\vec{q} \cdot \vec{S_a})(\vec{q} \cdot \vec{S_b}) + q^2 \CO \right] \,.
\eld
\el
$\CO$ refers to an unknown operator that is vanishing in the HCL. The corrections induced by eq.\eqc{eq:SOcorrection} does not affect the potential at this order.

\subsubsection{Cubic order in spin}
The leading PN corrections at 2PM order for cubic order spin interactions are formally classified as $4.5$PN corrections, and according to \cite{Vines:2016qwa} they were not known in the literature. The coefficients up to first three terms in $\sqrt{r^2 - 1}$ expansion that will be relevant are the following.
\bl
&\bld
A_{3,0}^{\text{2 PM}} &= -\frac{\pi ^2 G^2 m_a^2 m_b^2 \left(4 m_a+3 m_b\right)}{8 q \sqrt{r^2-1}}-\frac{\sqrt{r^2-1} \left(\pi ^2 G^2 m_a^2 m_b^2 \left(22 m_a+13 m_b\right)\right)}{16 q}
\\ & \hskip120pt -\frac{\sqrt{r^2-1}^3 \left(\pi ^2 G^2 m_a^2 m_b^2 \left(32 m_a+17 m_b\right)\right)}{64 q}
\eld
\\ &\bld
A_{2,1}^{\text{2 PM}} &= \frac{\pi ^2 G^2 m_a^2 m_b^2 \left(11 m_a+10 m_b\right)}{8 q \sqrt{r^2-1}}+\frac{\pi ^2 G^2 \sqrt{r^2-1} m_a^2 m_b^2 \left(117 m_a+100 m_b\right)}{32 q}
\\ & \hskip120pt +\frac{7 \pi ^2 G^2 \sqrt{r^2-1}^3 m_a^2 m_b^2 \left(6 m_a+5 m_b\right)}{32 q} \,.
\eld
\el
After subtraction of $p_0$ poles, leading PN terms take the following form.
\bl
\left. A_{3,0}^{\text{2 PM}} \right|_{\text{reg}} &= -\frac{\pi ^2 G^2 \sqrt{r^2-1} m_a^2 m_b^2 \left(53 m_a^2 m_b+45 m_a m_b^2+22 m_a^3+13
   m_b^3\right)}{16 q \left(m_a+m_b\right){}^2}
\\ \left. A_{2,1}^{\text{2 PM}} \right|_{\text{reg}} &= \frac{\pi ^2 G^2 \sqrt{r^2-1} m_a^2 m_b^2 \left(312 m_a^2 m_b+297 m_a m_b^2+117 m_a^3+100
   m_b^3\right)}{32 q \left(m_a+m_b\right){}^2} \,.
\el
The $S_a^3$-term has no ambiguities, apart from HCL-vanishing contributions.
\bl
\bld
\left. \text{LS}^{\text{2 PM}}_{\text{reg}} \right|_{S_a^3 S_b^0} &= -\frac{i \pi ^2 G^2 m_b \left(53 m_a^2 m_b+45 m_a m_b^2+22 m_a^3+13 m_b^3\right)}{2 q m_a^2 \left(m_a+m_b\right){}^2}
\\ & \phantom{asdfasdfasdfasdf} \times \left[ (K \cdot S_a)^2 \e_{\m\n\l\s} P_1^\m P_3^\n K^\l S_a^\s + K^2 \CO\right] \,.
\eld
\el
Taking the non-relativistic limit gives
\bl
\bld
\left. \frac{\text{LS}^{\text{2 PM}}_{\text{reg}}}{4E_a E_b} \right|_{S_a^3 S_b^0} &= -\frac{i \pi ^2 G^2 \left(53 m_a^2 m_b+45 m_a m_b^2+22 m_a^3+13 m_b^3\right)}{8 q m_a^3 \left(m_a+m_b\right)}
\\ & \phantom{asdfasdfasdfasdfasdfasdf} \times \left[ (\vec{S_a} \cdot \vec{q})^2 ( \vec{S_a} \cdot \vec{p_a} \times \vec{q} ) + q^2 \CO \right]
\eld
\el
and adding contributions due to eq.\eqc{eq:SOcorrection} gives the final answer.
\begin{tcolorbox}
\bl
\bld
\left. \frac{\text{LS}^{\text{2 PM}}_{\text{reg}}}{4E_a E_b} \right|_{S_a^3 S_b^0}^{\text{final}} &= -\frac{i \pi ^2 G^2 \left(31 m_a^2 m_b+10 m_a m_b^2+22 m_a^3-2 m_b^3\right)}{8 q m_a^3 \left(m_a+m_b\right)}
\\ & \phantom{asdfasdfasdfasdfasdfasdf} \times \left[ (\vec{S_a} \cdot \vec{q})^2 ( \vec{S_a} \cdot \vec{p_a} \times \vec{q} ) + q^2 \CO \right] \,.
\eld
\el
\end{tcolorbox}
The other cubic spin interaction term suffers from an ambiguity that was elaborated in section \ref{sec:LOcubicspin}. Since this ambiguity can be absorbed into the unknown HCL-vanishing contributions, this ambiguity will be ignored in this section. The interpretation for $\left. A_{2,1}^{\text{2 PM}} \right|_{\text{reg}}$ is then
\bl
&\bld
\left. \text{LS}^{\text{2 PM}}_{\text{reg}} \right|_{S_a^2 S_b^1} &= -\frac{i \pi ^2 G^2 \left(312 m_a^2 m_b+297 m_a m_b^2+117 m_a^3+100 m_b^3\right)}{4 q m_a \left(m_a+m_b\right){}^2}
\\ & \phantom{asdfasdfasdfasdf} \times \left[ (K \cdot S_a)^2 \e_{\m\n\l\s} P_1^\m P_3^\n K^\l S_b^\s + K^2 \CO\right] \,,
\eld
\el
and in the non-relativistic limit it takes the form
\bl
\\ &\bld
\left. \frac{\text{LS}^{\text{2 PM}}_{\text{reg}}}{4E_a E_b} \right|_{S_a^2 S_b^1} &= \frac{i \pi ^2 G^2 \left(312 m_a^2 m_b+297 m_a m_b^2+117 m_a^3+100 m_b^3\right)}{16 q m_a^2 m_b \left(m_a+m_b\right)}
\\ & \phantom{asdfasdfasdfasdfasdfasdf} \times \left[ (\vec{S_a} \cdot \vec{q})^2 ( \vec{S_b} \cdot \vec{p_b} \times \vec{q} ) + q^2 \CO \right] \,.
\eld
\el
Taking effects from eq.\eqc{eq:SOcorrection} into account, 2 PM $S_a^2 S_b^1$ potential takes the following form.
\begin{tcolorbox}
\bl
\bld
\left. \frac{\text{LS}^{\text{2 PM}}_{\text{reg}}}{4E_a E_b} \right|_{S_a^2 S_b^1}^{\text{final}} &= \frac{i \pi ^2 G^2 \left(166 m_a^2 m_b+123 m_a m_b^2+73 m_a^3+24 m_b^3\right)}{16 q m_a^2
   m_b \left(m_a+m_b\right)}
\\ & \phantom{asdfasdfasdfasdfasdfasdf} \times \left[ (\vec{S_a} \cdot \vec{q})^2 ( \vec{S_b} \cdot \vec{p_b} \times \vec{q} ) + q^2 \CO \right] \,.
\eld
\el
\end{tcolorbox}
\subsubsection{Quartic order in spin}\label{subsubsec:2PM_Quartic_Order_in_Spin}
Formally, leading PN corrections at 2PM order at quartic order in spin is classified as 5PN corrections, which were also not known in the literature according to \cite{Vines:2016qwa}. At quartic order in spin, the following coefficients computed up to first three terms in $\sqrt{r^2 - 1}$ expansion are relevant.
\bl
&\bld
A_{4,0}^{\text{2 PM}} &= \frac{\pi ^2 G^2 m_a^2 m_b^2 \left(m_a+m_b\right)}{64 q \sqrt{r^2-1}^2}+\frac{\pi ^2 G^2 m_a^2 m_b^2 \left(19 m_a+12 m_b\right)}{128 q}
\\ & \hskip120pt +\frac{\pi ^2 G^2 \sqrt{r^2-1}^2 m_a^2 m_b^2 \left(239 m_a+120 m_b\right)}{1536 q}
\eld
\\ &\bld
A_{3,1}^{\text{2 PM}} &= -\frac{\pi ^2 G^2 m_a^2 m_b^2 \left(m_a+m_b\right)}{16 q \sqrt{r^2-1}^2}-\frac{7 \left(\pi ^2 G^2 m_a^2 m_b^2 \left(5 m_a+4 m_b\right)\right)}{64 q}
\\ & \hskip120pt -\frac{\sqrt{r^2-1}^2 \left(\pi ^2 G^2 m_a^2 m_b^2 \left(27 m_a+20 m_b\right)\right)}{48 q}
\eld
\\ &\bld
A_{2,2}^{\text{2 PM}} &= \frac{3 \pi ^2 G^2 m_a^2 m_b^2 \left(m_a+m_b\right)}{32 q \sqrt{r^2-1}^2}+\frac{95 \pi ^2 G^2 m_a^2 m_b^2 \left(m_a+m_b\right)}{128 q}
\\ & \hskip120pt +\frac{95 \pi ^2 G^2 \sqrt{r^2-1}^2 m_a^2 m_b^2 \left(m_a+m_b\right)}{128 q} \,.
\eld
\el
Subtraction of poles in $p_0$ yields the following leading term expression.
\bl
\left. A_{4,0}^{\text{2 PM}} \right|_{\text{reg}} &= \frac{\pi ^2 G^2 m_a^2 m_b^2 \left(29 m_a m_b+19 m_a^2+12 m_b^2\right)}{128 q \left(m_a+m_b\right)}
\\ \left. A_{3,1}^{\text{2 PM}} \right|_{\text{reg}} &= -\frac{\pi ^2 G^2 m_a^2 m_b^2 \left(59 m_a m_b+35 m_a^2+28 m_b^2\right)}{64 q \left(m_a+m_b\right)}
\\ \left. A_{2,2}^{\text{2 PM}} \right|_{\text{reg}} &= \frac{\pi ^2 G^2 m_a^2 m_b^2 \left(178 m_a m_b+95 m_a^2+95 m_b^2\right)}{128 q \left(m_a+m_b\right)} \,.
\el
Proceeding as in former examples, the relativistic LS takes the following form
\bl
\left. \text{LS}^{\text{2 PM}}_{\text{reg}} \right|_{S_a^4 S_b^0} &= \frac{\pi ^2 G^2 m_b^2 \left(29 m_a m_b+19 m_a^2+12 m_b^2\right)}{8 q m_a^2 \left(m_a+m_b\right)} \left[ (K \cdot S_a)^4 + K^2 \CO \right]
\\ \left. \text{LS}^{\text{2 PM}}_{\text{reg}} \right|_{S_a^3 S_b^1} &= \frac{\pi ^2 G^2 m_b \left(59 m_a m_b+35 m_a^2+28 m_b^2\right)}{4 q m_a \left(m_a+m_b\right)} \left[ (K \cdot S_a)^3(K \cdot S_b) + K^2 \CO \right]
\\ \left. \text{LS}^{\text{2 PM}}_{\text{reg}} \right|_{S_a^2 S_b^2} &= \frac{\pi ^2 G^2 \left(178 m_a m_b+95 m_a^2+95 m_b^2\right)}{8 q \left(m_a+m_b\right)} \left[ (K \cdot S_a)^2(K \cdot S_b)^2 + K^2 \CO \right]
\el
which, with non-relativistic flux normalisation, yields the following expression for the potentials.
\begin{tcolorbox}
\bl
&\bld
\left. \frac{\text{LS}^{\text{2 PM}}}{4E_a E_b} \right|_{S_a^4 S_b^0}^{\text{final}} &= \frac{\pi ^2 G^2 m_b \left(29 m_a m_b+19 m_a^2+12 m_b^2\right)}{32 q m_a^3 \left(m_a+m_b\right)} \left[ (\vec{q} \cdot \vec{S_a})^4 + q^2 \CO \right]
\eld
\\ &\bld \left. \frac{\text{LS}^{\text{2 PM}}}{4E_a E_b} \right|_{S_a^3 S_b^1}^{\text{final}} &= \frac{\pi ^2 G^2 \left(59 m_a m_b+35 m_a^2+28 m_b^2\right)}{16 q m_a^2 \left(m_a+m_b\right)} \left[ (\vec{q} \cdot \vec{S_a})^3 (\vec{q} \cdot \vec{S_b}) + q^2 \CO \right]
\eld
\\ &\bld \left. \frac{\text{LS}^{\text{2 PM}}}{4E_a E_b} \right|_{S_a^2 S_b^2}^{\text{final}} &= \frac{\pi ^2 G^2 \left(178 m_a m_b+95 m_a^2+95 m_b^2\right)}{32 q m_a m_b \left(m_a+m_b\right)} \left[ (\vec{q} \cdot \vec{S_a})^2 (\vec{q} \cdot \vec{S_b})^2 + q^2 \CO \right] \,.
\eld
\el
\end{tcolorbox}
Since the terms introduced by eq.\eqc{eq:SOcorrection} are subleading in PN expansion, they do not need to be considered.

\subsubsection{Partial results for higher order in spin}
Though we cannot obtain the full results for higher spin effects due to the polynomial ambiguities of the higher spin Compton amplitude eq.\eqref{eq:All_Spin_Answer}, we can still obtain partial results from LS computations of spin 2 particle scattering. We present them in the following list:

\begin{itemize}[leftmargin=*]
\item Fifth order in spin: 
\begin{tcolorbox}
\begin{align}
&\bld
\left. \frac{\text{LS}^{\text{2 PM}}}{4E_a E_b} \right|^{\text{final}}_{S_a^4 S_b^1} &= \frac{i \pi ^2 G^2 \left(149 m_a^2 m_b-116 m_a m_b^2+182 m_a^3-128 m_b^3\right)}{384 q m_a^4 m_b \left(m_a+m_b\right)}
\\ &\phantom{asdfasdfasdfasdfasdfasdf} \times \left[ (\vec{S_a} \cdot \vec{q})^4 ( \vec{S_b} \cdot \vec{p_b} \times \vec{q} ) + q^2 \CO \right] \eld \\ &\bld
\left. \frac{\text{LS}^{\text{2 PM}}}{4E_a E_b} \right|^{\text{final}}_{S_a^3 S_b^2} &= \frac{i \pi ^2 G^2 \left(88 m_a^2 m_b-m_a m_b^2+60 m_a^3-44 m_b^3\right)}{192 q m_a^3 m_b^2 \left(m_a+m_b\right)}
\\ &\phantom{asdfasdfasdfasdfas} \times \left[ (\vec{S_a} \cdot \vec{q})^3 (\vec{S_b} \cdot \vec{q}) ( \vec{S_b} \cdot \vec{p_b} \times \vec{q} ) + q^2 \CO \right] \eld
\end{align}
\end{tcolorbox}
\item Sixth order in spin: 
\begin{tcolorbox}
\begin{align}
\left. \frac{\text{LS}^{\text{2 PM}}}{4E_a E_b} \right|^{\text{final}}_{S_a^3 S_b^3} &=  \frac{\pi^2 G^2 (69 m_a m_b + 37 m_a^2+37 m_b^2)}{96 q m_a^2 m_b^2 (m_a+m_b)} \Big[(q \cdot S_a)^3 (q \cdot S_b)^3 + q^2 \mathcal{O}\Big] \\
\left. \frac{\text{LS}^{\text{2 PM}}}{4E_a E_b} \right|^{\text{final}}_{S_a^4 S_b^2} &= \frac{\pi ^2 G^2 \left(413 m_a m_b+239 m_a^2+204 m_b^2\right)}{768 q m_a^3 m_b \left(m_a+m_b\right)} \Big[(q \cdot S_a)^4 (q \cdot S_b)^2 + q^2 \mathcal{O}\Big]
\end{align}
\end{tcolorbox}
\item Seventh order in spin: 
\begin{tcolorbox}
\begin{align}
\bld
\left. \frac{\text{LS}^{\text{2 PM}}}{4E_a E_b} \right|^{\text{final}}_{S_a^4 S_b^3} &= -\frac{i \pi ^2 G^2 \left(494 m_a^2 m_b+681 m_a m_b^2+114 m_a^3+336 m_b^3\right)}{4608 q m_a^4 m_b^3 \left(m_a+m_b\right)}
\\ &\phantom{asdfasdfasdfasdf} \times \left[ (\vec{S_a} \cdot \vec{q})^4 (\vec{S_b} \cdot \vec{q})^2 ( \vec{S_b} \cdot \vec{p_b} \times \vec{q} ) + q^2 \CO \right]
\eld
\end{align}
\end{tcolorbox}
\item Eighth order in spin:
\begin{tcolorbox}
\begin{align}
\left. \frac{\text{LS}^{\text{2 PM}}}{4E_a E_b} \right|^{\text{final}}_{S_a^4 S_b^4} &=  \frac{ 5 \pi^2 G^2 (93 m_a m_b + 50 m_a^2 + 50 m_b^2)}{9216 m_a^2 m_b^2 q (m_a+m_b)} \Big[(q \cdot S_a)^4 (q \cdot S_b)^4 + q^2 \mathcal{O}\Big]
\end{align}
\end{tcolorbox}
\end{itemize}

\subsection{Fixing the local polynomial term at $S^4$ from consistency condition}\label{subsec:fix_quartic_order_in_spin}

In the beginning of this section, we mentioned that certain terms in the potential can be computed with more than one way of assigning the spin to the two particles, and the result from each assignment should be identical. More precisely, for a specific order of spin operator, the coefficient should be independent of which external spins were chosen to extract the potential.  Indeed in derivation of the potential, we have verified that the same result has been reached with different choices of spin assignment. In this subsection we will explore the possibility of using this consistency condition to fix the polynomial ambiguity. Recall that the Compton amplitudes were derived from matching the factorization poles, which leaves us open to polynomial ambiguities. As we will show the polynomial ambiguities can only enter at the quartic order in spin operator in the HCL limit.

In general, a candidate polynomial term needs to satisfy the little group weights and spin-statistics relations. Since we are talking about the one contributing to the classical potential, the contact terms should also survive the HCL. So we require the following conditions:
\begin{enumerate}
\item Correct little group weights: $M(\BS{1}^s, 2^{+2}, 3^{-2}, \BS{4}^{s})$
\item Correct spin-statistics property: $M(\BS{1}^s, 2^{+2}, 3^{-2}, \BS{4}^{s}) = (-1)^{2s} M(\BS{4}^s, 2^{+2}, 3^{-2}, \BS{1}^{s})$
\item Survives the HCL: $M_{\text{contact}} \sim O(\beta-1)^0$
\end{enumerate}
The following list exhausts all possible spinor combinations that survive the HCL limit:
\begin{align}
F_1 = \frac{ \SB{\boldsymbol{14}} }{ m } &\rightarrow -1 \nonumber \\
\tilde{F_1} = \frac{ \AB{\boldsymbol{14}} }{ m } &\rightarrow - 1 + Spa \nonumber \\
\mathcal{F}_1  = \frac{1}{2}( F_1 + \tilde{F}_1 ) &\rightarrow -1 + \frac{Spa}{2} \nonumber \\
F_2 = \frac{ \AB{ \BS{4}2 } \SB{2 \BS{1}} - \AB{ \BS{1}2 } \SB{2 \BS{4}} }{2m^2} &\rightarrow - \frac{(1-y)^2}{4y} Spa \nonumber \\
\tilde{F}_2 = \frac{ \AB{ \BS{1}3 } \SB{3 \BS{4}} - \AB{ \BS{4}3 } \SB{3 \BS{1}} }{2m^2}  &\rightarrow - \frac{(1+y)^2}{4y} Spa \nonumber \\
\mathcal{F}_2 = \frac{1}{2}( F_2 + \tilde{F}_2 ) &\rightarrow -\frac{1 + y^2}{4y}\frac{Spa}{2} \nonumber \\
\mathcal{F} = \mathcal{F}_1 + \mathcal{F}_2 &\rightarrow - 1 - \frac{  (1 - y)^2 }{4 y}Spa \nonumber \\
K \equiv  \frac{\AB{3 \boldsymbol{4}} \SB{2 \boldsymbol{1}}}{2m^2} - \frac{\AB{3 \boldsymbol{1}} \SB{2 \boldsymbol{4}}}{2m^2} &\rightarrow -\frac{1-y^2}{4y} \frac{Spa}{m_a} \nonumber
\end{align}
We can see that the only combination that provides the correct helicity weight of the gravitons is $K^4$. Since this term contains 4 $SU(2)$ indices for both particle 1 and particle 4, contact terms starts to affect the classical potential at quartic order in spin.

Universality of spin effects demands that all $S^4$ potential extracted from particles with $s>2$ should also take the same form. However, one would find that using  eq.\eqref{eq:All_Spin_Answer} in the LS calculation to extract quartic order spin effects for $s>2$ particles will yield a result different from the one we presented in section \ref{subsubsec:2PM_Quartic_Order_in_Spin}. This difference exactly comes from the polynomial ambiguities. Now we propose the following ansatz for the contact terms:
\begin{equation}\label{eq:polynomial_ansatz}
\text{Polynomial}(s) = K^4 \sum_{r=4}^{2s} a_{r}^{(s)} \mathcal{F}_1^{2s-r} \mathcal{F}_2^{r-4}
\end{equation}
where $\mathcal{F}_1$, $\mathcal{F}_2$ provide $2s-4$ $SU(2)$ indicies. This ansatz is motivated from following properties:
\begin{enumerate}
\item The inconsistency begins at quartic order in spin, so we'll need $K^4$ as an overall coefficient of the ansatz to supply the correct helicity weights.
\item We need $\mathcal{F}_1$ and $\mathcal{F}_2$ to supply correct little group weights for the massive particles and the correct  spin statistics property under $(1 \leftrightarrow 4)$ exchange. We use the curly symbols $\mathcal{F}_1$ and $\mathcal{F}_2$ because they are the most symmetric under angle square brackets exchange.
\item To make the correction from polynomial terms begin at $S^4$ for all higher spins, we'll need the combinations of $\mathcal{F}_1$ and $\mathcal{F}_2$ in the summation of the ansatz eq.\eqref{eq:polynomial_ansatz} starting from $Spa^0$ in the spin operator basis. Only $\mathcal{F}$ and $\mathcal{F}_1$ can supply such $Spa^0$ terms. But since $\mathcal{F}$ can be expanded in terms of $\mathcal{F}_1$ and $\mathcal{F}_2$, we choose to write the ansatz in the form of eq.\eqref{eq:polynomial_ansatz}. When $r=4$, this will correspond to the correction to $S^4$ from polynomial terms.
\end{enumerate}
Expanding the ansatz eq.\eqref{eq:polynomial_ansatz} in the spin operator basis, it looks like 
\begin{equation}\label{eq:ansatz_expand}
\begin{split}
\text{Polynomial} \sim a_4^{(s)} Spa^4 + (a_4^{(s)} + a_5^{(s)} ) Spa^5 + (a_4^{(s)} + a_5^{(s)} + a_6^{(s)}) Spa^6  + \cdots
\end{split}
\end{equation}
The interest in this section is fixing $a_4^{(s)}$, so one can write down other ansatz that will not alter $a_4^{(s)}$. On the other hand, we also observed that though eq.\eqref{eq:Lower_Spin_Graviton_Compton} becomes spurious when $s>2$, it still yields a consistent $S^4$ potential for higher spins. So,  we can fix this coefficient by directly comparing $M_4^{\text{lower spin}}$  and $M_4^{\text{higher spin} }$ in the HCL for higher spin particles. Then we require the coefficient of $Spa^4$ for higher spin particles to take the analogous form for the spin 2 particles. We found
\begin{equation}\label{eq:Contact_for_quartic_in_spin}
a_4^{(s)} = -(-1)^{2s} \frac{3\alpha^2}{2}\binom{2s}{4} m_a^2
\end{equation}
with $\alpha=\sqrt{8 \pi G}$. So we conclude that adding eq.\eqref{eq:polynomial_ansatz} with the coefficient $a_4^{(s)}$ given by  eq.\eqref{eq:Contact_for_quartic_in_spin} to the higher spin Compton amplitude eq.\eqref{eq:All_Spin_Answer} will yield the correct $S^4$ classical potential for all spins. That is, we need
\begin{equation}
\begin{split}
M^{\text{1-loop}}(s > 2) 
= \frac{1}{4} \sum_{h_3, h_4} \int_{\Gamma_{LS}} &d^4L \delta(L^2 - m_b^2)\delta(k_3^2)\delta(k_4^2) \\
&\times (M_4^{\text{higher spin}}(s) +\text{ Polynomial }(s)) M_3 M_3
\end{split}
\end{equation}
to obtain a consistent $S^4$ potential. The coeffcients $\{ a_5^{(s)}, \cdots, a_{2s}^{(s)}\}$ in the ansatz eq.\eqref{eq:polynomial_ansatz} cannot be fixed at this level since we do not have another representation of the higher spin Compton amplitude to compare with.

Finally, a remark on the spurious Compton amplitude for higher spins, which is just an extrapolation of eq.\eqref{eq:Lower_Spin_Graviton_Compton} to $s>2$. Though we found that using this spurious Compton amplitude in the LS can give us consistent $S^4$ effects, it still cannot be applied to extract $S^5$ effects and higher. As an example, one can compare the $A_{5,0}$ coefficient of the LS calculated from eq.\eqref{eq:All_Spin_Answer} and eq.\eqref{eq:Lower_Spin_Graviton_Compton}. We found that they differ by:
\begin{equation}\label{eq:Weird_difference}
\begin{split}
A_{5,0}^{\text{higher spin} } 
&- A_{5,0}^{\text{spurious}} \\
&= \frac{\pi ^2 G^2 \epsilon  m_a^2 m_b^3 \left(m_a^6+ 5 m_a^5 m_b+10 m_a^4 m_b^2+11 m_a^3 m_b^3+10 m_a^2 m_b^4+5 m_a m_b^5+m_b^6\right)}{3072 q \left(m_a+m_b\right){}^6}
\end{split}
\end{equation}
One might think that they just differ by a polynomial term. But a polynomial term of the Compton amplitude of particle $a$ should not carry any information of particle $b$. That is, the difference must only carry powers of $m_a$ and no powers of $b$ in the denominator while eq.\eqref{eq:Weird_difference} carry $(m_a + m_b)^6$ in the denominator. So, such difference is definitely not from the polynomial ambiguity of the Compton amplitude. Thus we conclude that the spurious Compton amplitude is not applicable for higher spin effects.
\section{Conclusion and Outlook}
In this paper we systematically study charged and gravitationally coupled higher spin particles. We focus on ``minimal couplings", where the UV limit matches to the minimal derivative coupling. We identify that these interactions can also be characterised in the IR through various physical properties such as $g=2$, and the absence of finite size effects. For spins-1/2 and 1, this corresponds the usual minimal couplings for Dirac fermions as well as $W$ bosons. We also derive the (gravitational) Compton amplitude, up to polynomial ambiguities, for the minimal coupling with arbitrary spin. These are derived through the requirement of consistent factorisation. Applying the same criteria for non-minimal couplings, we find that $\lambda^2$ interactions are forbidden for gravitational coupling. We argue that the absence of $\lambda^2$ deformations, or anomalous gravitomagnetic dipole moment, is a reflection of general covariance. For theories whose gravitational coupling is the square of gauge couplings, this implies that the charged states must have $g=2$, consistent with string theory. 

Having equipped with the Compton amplitude, we proceed to utilize it to compute the spin dependent piece of the classical gravitational potential. We follow the work of Cachazo and Guevara, where the 2PM potential is computed by evaluating the one-loop triangle leading singularity in the holomorphic classical limit, and matched to local Lorentz invariant operators. Using the spin-2 Compton amplitude, we derive the spin-dependent parts of the potential up to degree four in the spin operator of either black holes. We also discuss to which extent the polynomial ambiguities of the higher spin Compton amplitudes can be fixed by requiring that the resulting classical potential yields the same coefficient for the spin operators as that for the lower spins. 

As alluded to in the paper, the leading trajectory states in string theory do not yield the simplest coupling. In fact, it is the most complex, as all allowed deformations except for the (gravito)magnetic dipole moments  are turned on. It would be interesting to see if the couplings for the subleading trajectories are simpler. This would be in line with the expectation that  the large degeneracy for subleading trajectory states become the dominant contribution for black hole microstates, which we know are simple. It would be fascinating if this is the case, as vertex operators for subleading trajectories are generally much more complicated than the leading one, and yet it yields a simpler amplitude, providing further evidence that the worldsheet point of view can often be misleading.

An immediate task is to identify what is the theory that gives minimal coupling for spins $\geq2$. To construct the corresponding Lagrangian, one starts with the quadratic term in eq.(\ref{GravS}), and successively adds terms linear in the Reimann tensor to remove the spin-operator pieces that are induced by eq.(\ref{GravS}), characterizing the deviation from minimal coupling. This is not only of theoretical interest, but it will resolve the polynomial ambiguity in the gravitational Compton amplitude, allowing one to extract spin effects beyond quartic order.

The fact that the infinite number of Wilson coefficients of the one body effective action is reproduced by the minimal coupling which is simply an $x^2$ in our on-shell parameterization, lends one to wonder if further simplification can be achieved by reformulating all computations in the one body EFT approach in terms of computations involving $x^2$. A tantalising example would be the fact for charged black holes, one also has $g=2$~\cite{Carter:1968rr}, and one can conjecture that $x$ gives the correct Wilson coefficient for the electricmagnetic couplings. This would be the simplest example of double copy for classical objects. 

Finally, an important question is whether the relation between minimal coupling and black holes persists through quantum corrections. It is well known that quantum effects generate $(g{-}2)$ for charged particles. On the other hand the gravito-magnetic moment is argued to be universal and thus should be protected. It would be interesting to see why $\lambda^2$ terms are not generated by loop corrections. Furthermore, whether minimal coupling states in gravity stays minimally coupled quantum mechanically, in that all deformations are never turned on. 
\section{Acknowledgements}
We thank Alfredo Guevara, Alexander Ochirov and Justin Vines for their generosity in sharing their on-going work, as well as discussions with regards to polynomial ambiguities in the Compton scattering amplitude. We would also like to thank Radu Roiban and Julio Parra-Martinez for discussions on high energy behaviours of higher spin states, as well as Jan Steinhoff for discussion on spin supplementary conditions. MZC and YTH is supported by MoST Grant No. 106-2628-M-002-012-MY3. The work of JWK and SL was supported in part by the National Research Foundation of Korea (NRF) Grant 2016R1D1A1B03935179. JWK is also supported by Kwanjeong Educational Foundation.

\appendix
\section{Spinor-helicity variables} \label{sec:spin-helconventions}

\subsection{Lorentz algebra}

We work with the metric $\eta_{\m\n} = \text{diag}(+1, -1, -1, -1)$, so that $p^2 = \eta^{\mu\nu} p_\mu p_\nu = (p_0)^2 - (\vec{p})^2$. 
Our convention for the Lorentz generators are fixed by the algebra, 
\bl
\bld
\left[ J^{\m\n} , J^{\l\s} \right] &= -i \left( \eta^{\m\l} J^{\n\s} + \eta^{\n\s} J^{\m\l} - \eta^{\m\s} J^{\n\l} - \eta^{\n\l} J^{\m\s} \right) \,.
\eld \label{eq:LorentzGenDef}
\el

To relate the (connected part) of the Lorentz group $SO(1,3)$ and its double cover $SL(2,\mathbb{C})$, we follow a widely adopted convention for spinors and gamma matrices, 
\begin{align}
\gamma^\mu 
=
\begin{pmatrix}
0 & (\sigma^\mu)_{\a\dot{\b}}
\\
(\bar{\sigma}^\mu)^{\dot{\a}\b} & 0 
\end{pmatrix}
\,, 
\quad
\sigma^\mu = (\iden, \vec{\sigma}) \,, 
\quad
\bar{\sigma}^\mu = (\iden, -\vec{\sigma}) \,, 
\end{align}
where $\vec{\sigma}$ denote Pauli matrices in the standard convention. 
Complex conjugation exchanges undotted and dotted indices. 
It is easy to check that 
\bl
(J^{\mu\nu})_\text{spinor} 
= \frac{i}{4} [\g^\m , \g^\n]  
= \frac{i}{2}
\begin{pmatrix}
(\s^{[\m} \bar{\s}^{\n]})_\a{}^\b & 0 
\\
0 & (\bar{\s}^{[\m} \s^{\n]})^{\dot{\a}}{}_{\dot{\b}} 
\end{pmatrix}
= 
\begin{pmatrix}
(J^{\m\n})_\a{}^\b & 0 
\\
0 & (J^{\m\n})^{\dot{\a}}{}_{\dot{\b}} 
\end{pmatrix}
\label{eq:SL2Celemdef}
\el
forms a representation of the algebra eq.\eqref{eq:LorentzGenDef}. 
Spinor indices are raised and lowered by the invariant tensor of $SL(2,\mathbb{C})$ satisfying 
\bl
\e_{\a\b} = - \e_{\b\a} \,,
\quad 
\e_{\a\b} \e^{\b\g} = \delta_\a{}^\g \,,
\quad 
\e^{12} = + 1\,, 
\quad
\e_{\dot{\a}\dot{\b}} = (\e_{\a\b})^*  \,. 
\label{SL2C-epsilon}
\el
For example, $\l^\a = \e^{\a\b} \l_\b$ and $\tilde{\l}_{\dot{\a}} = \e_{\dot{\a} \dot{\b}} \tilde{\l}^{\dot{\b}}$.

For any (momentum) 4-vector, the bi-spinor notation is defined by 
\bl 
p_{\a\dot{\a}} = p_\mu (\s^\m)_{\a\dot{\a}} 
\,,
\quad 
p^2 = \det(p_{\a\dot{\a}}) 
= \frac{1}{2} \e^{\a\b} \e^{\dot{\a}\dot{\b}} p_{\a\dot{\a}} p_{\b\dot{\b}} \,.
\el

\subsection{Massless momenta} \label{sec:masslessmom}

For massless momenta, $p_{\a\dot{\a}}$ as a $(2\times 2)$ matrix has rank 1, so it can be written as 
\bl
p_{\a\dot{\a}} = \l_\a \tilde{\l}_{\dot{\a}} \,.
\el
For a real momentum, the spinors satisfy the reality condition, 
\bl
(\l_\a)^* = \text{sign}(p_0) \tilde{\l}_{\dot{\a}} \,. 
\el
The Little group $U(1)$ acts on the spinors as 
\bl
\l \rightarrow e^{-i \frac{\theta}{2}} \l 
\,, \quad 
\tilde{\l} \rightarrow e^{i \frac{\theta}{2}} \tilde{\l} \,. 
\el
The spinors for $p$ and those for $(-p)$ must be proportional. 
We fix the relation by setting 
\bl
\l(-p) = \l(p) \,,
\quad 
\tilde{\l}(-p) = - \tilde{\l}(p) \,.
\el

It is customary to introduce a bra-ket notation, 
\begin{align}
|p \rangle \leftrightarrow \l_\a \,,
\quad 
\langle p | \leftrightarrow \l^\a \,,
\quad
|p ] \leftrightarrow \tilde{\l}^{\dot{\a}} \,,
\quad 
[p | \leftrightarrow \tilde{\l}_{\dot{\a}} \,, 
\end{align}
which leads to the Lorentz invariant, Little group covariant brackets, 
\bl
\langle i j \rangle = \l_i^\a \l_j^\b \e_{\a\b} = \l_i^\a \l_{j\b}
\,, \quad
[i j] = \tilde{\l}_{i \dot{\a}} \tilde{\l}_j^{\dot{\a}} \,.
\el
The massless Mandelstam variables, which are both Lorentz invariant and Little group invariant, 
can be expressed as 
\begin{align}
2 p_i \cdot p_j =  \e^{\a\b} \e^{\dot{\a}\dot{\b}} 
(p_i)_{\a\dot{\a}} (p_j)_{\b\dot{\b}} = \langle i j \rangle [j i] \,.
\end{align}

\subsection{Massive momenta} \label{sec:massivemom}

For massive momenta, the on-shell condition in the bi-spinor notation is given by
\bl
\det(p_{\a\dot{\a}}) =  m^2 \,.
\el
The massive helicity spinor variables are defined by
\bl
p_{\a\dot{\a}} = \l_\a{}^I \tilde{\l}_{I\dot{\a}} \,, 
\quad 
\det(\l_\a{}^I) = m = \det(\tilde{\l}_{I\dot{\a}})\,. 
\el
The index $I$ indicates a doublet of the $SU(2)$ Little group. 
The reality condition reads 
\bl
\bld
(\l_\a{}^I)^* &= \text{sign}(p_0) \tilde{\l}_{I\dot{\a}} 
\\ (\l_\a{}_I)^* &= -\text{sign}(p_0) \tilde{\l}^{I}_{\dot{\a}} \,. 
\eld
\el

\paragraph{SU(2)-invariant tensor} 
Given a matrix representation of the doublet of $SU(2)$, 
\bl
\psi^I \rightarrow U^I{}_J \psi^J \,, 
\el
the two defining properties of $SU(2)$ can be written as
\bl
\e_{IK} U^I{}_J U^K{}_L = \e_{JL} \,,
\quad 
U^I{}_J (U^\dagger)^J{}_K 
= \delta^I{}_K \,, 
\label{SU2-def}
\el
where the $SU(2)$-invariant tensor $\e_{IJ}$ shares, by convention, the first three properties in eq.\eqref{SL2C-epsilon}. Just like spinor indices, the Little group indices are raised and lowered by $\e_{IJ}$ and $\e^{IJ}$.
It follows from eq.\eqref{SU2-def} that the two variables below transform in the same way. 
\bl
\psi_I := \e_{IJ} \psi^J \quad \mbox{and} \quad 
\bar{\psi}_I := (\psi^I)^* \,. 
\el
Then, 
\bl
p_{\a\dot{\a}} 
= \l_\a{}^I \tilde{\l}_{I\dot{\a}} 
= - \l_{\a I} \tilde{\l}^I{}_{\dot{\a}} \,,
\quad 
\bar{p}^{\dot{\a}\a} 
= p_\m (\bar{\s}^\m)^{\dot{\a}\a} 
= \e^{\dot{\a}\dot{\b}} \e^{\a\b} p_{\b \dot{\b}} 
= \l^{\a I} \tilde{\l}_I{}^{\dot{\a}} 
= - \l^\a{}_I \tilde{\l}^{I\dot{\a}} \,.
\el
It is also useful to note that
\bl
\e^{\a\b} \l_\a{}^I \l_\b{}^J = \det(\l) \e^{IJ} = m \e^{IJ} \,,
\quad 
\e^{\dot{\a}\dot{\b}} \tilde{\l}_{I\dot{\a}} \tilde{\l}_{J\dot{\b}} = -\det(\tilde{\l}) {\e}_{IJ} = -m {\e}_{IJ} \,.
\el

\paragraph{Dirac spinors}

By definition, the massive spinor helicity variables satisfy 
\bl
p_{\a\dot{\a}} \tilde{\l}^{\dot{\a} I} = m \lambda_\a{}^I \,, 
\quad 
p^{\dot{\a}\a} \l_\a{}^I = m \tilde{\l}^{\dot{\a} I}\,.
\label{Dirac-eq}
\el
Comparing this with the textbook convention for Dirac spinors, 
\bl
&
(p_\m \g^\m - m) u(p) = 0 \,,
\quad 
(p_\m \g^\m + m) v(p) = 0 \,, 
\el
leads to the natural identification, 
\bl
u^I (p) = \begin{pmatrix} \l_\a{}^I \\ \tilde{\l}^{\dot{\a} I} \end{pmatrix} \,, 
\quad 
v^I (p) = \begin{pmatrix} \l_\a{}^I \\ - \tilde{\l}^{\dot{\a} I}  \end{pmatrix} \,.
\el
Similarly, for the conjugate Dirac spinors, we have
\bl
&\bar{u}(p) (p_\m \g^\m - m) = 0 \,,
\quad 
\bar{v}(p) (p_\m \g^\m + m) = 0 \,, 
\\
&\bar{u}_I (p) = 
\begin{pmatrix} - \l^{\a}_{~I} & \tilde{\l}_{\dot{\a} I}  \end{pmatrix} ,  
\quad
\bar{v}_I (p) = 
\begin{pmatrix} \l^\a{}_I & \tilde{\l}_{\dot{\a} I}  \end{pmatrix} .
\el

\paragraph{BOLD} 

For a fixed massive particle, the $SU(2)$ Little group is always completely symmetrized. 
Ref.~\cite{Arkani-Hamed:2017jhn} introduced the \bf{BOLD} notation, which suppresses the $SU(2)$ little group indices by means of an auxiliary parameter for each particle. 
For instance, for particle 1, 
\bl
(\l_1)_\a{}^I (t_1)_I  = \ket{1^I} (t_1)_I = \ket{\bf{1}} \,, 
\\
(\tilde{\l}_1)^{\dot{\a}I} (t_1)_I  = \sket{1^I} (t_1)_I = \sket{\bf{1}} \,. 
\el
It is clear how to reinstate the SU(2) index if needed. 

The Dirac equation eq.\eqref{Dirac-eq} can be written in the BOLD bra-ket notation as
\bl
p_k \sket{\bf{k}}= m \ket{\bf{k}} \,, 
\quad 
\bra{\bf{k}} p_k = - m \sbra{\bf{k}} \,.
\label{Dirac-bold}
\el
%
%
In the main text, we define the $x$ factor for a 3pt amplitude:
\bl
x \bra{3} = \frac{\sbra{3} \bar{p}_1 }{m} = - \frac{\sbra{3} \bar{p}_2 }{m} \,, 
\quad 
x^{-1} \sbra{3} 
=  \frac{\bra{3} p_1 }{m} = - \frac{\bra{3} p_2 }{m}
\,.
\el
Decomposing the massive momenta into the spinor helicity variables, we can derive
\bl
x \langle 3 \bf{1} \rangle = + [ 3 \bf{1} ] \,, 
\quad 
x \langle 3 \bf{2} \rangle = - [ 3 \bf{2} ] \,, 
\el
Combining these with the Dirac equation eq.\eqref{Dirac-bold}, we obtain 
a useful identity, 
\bl
\lan \bf{21} \ra = [\bf{21} ] + \frac{[\bf{2}3][3\bf{1}]}{mx} 
= [\bf{2}| \left( \iden + \frac{|3][3|}{m}\right) |\bf{1}] \,.
\el

\subsection{High-Energy limit} \label{sec:HE}

\paragraph{Definition} 
Consider a system of massive particles whose masses are equal or similar to some fixed $m$. 
As in the scattering problem of the main text, 
we assume that the particle number is conserved 
and the mass of each particle is also conserved. 
Let $p_i$ be the {\em incoming} momenta, and $\gamma_{ij} = p_i \cdot p_j / m_i m_j$ $(i \neq j$ be 
the Lorentz invariant measure of the pairwise relative velocity. 
The High Energy (HE) limit is defined such that all $\gamma_{ij}$'s grow arbitrarily large 
while the ratios $\g_{ij}/\g_{kl}$ remain fixed. 

\paragraph{Frame dependence}
In the center of momentum (COM) frame among all incoming momenta, 
it can be shown that $p^0  = E \gg m$ holds for each  particle in the HE limit. 
Suppose 
\bl
p^\mu = (E,0,0, p) 
\quad 
\Longrightarrow 
\quad 
p_{\a\dot{\a}} = 
\begin{pmatrix} 
E-p & 0 \\ 
0 & E+p 
\end{pmatrix}\,,
\label{p-COM}
\el
in the COM frame. The two diagonal matrix elements are well-separated in the HE limit, 
\bl
E + p = 2E \left( 1 - \frac{m^2}{4E^2} + \cdots \right)  \,,
\quad 
E - p = \frac{m^2}{E+p} = \frac{m^2}{2E}\left( 1 + \frac{m^2}{4E^2} + \cdots \right) \,,  
\el
where we suppressed corrections of order $\CO(m/E)^4$.

Unlike the definition of the HE limit, the relation $E \gg m$ depends on the Lorentz frame; 
it does not hold in the particle's own rest frame.
We can specify the frame dependence in a Lorentz covariant way. 
Let $u^\mu$ be the time-like unit vector of the COM frame. Introduce
\bl
(p|u)_\a{}^\b = p_{\a\dot{\a}} u^{\dot{\a} \b} \,,
\quad 
(u|p)^{\dot{\a}}{}_{\dot{\b}} =u^{\dot{\a} \a} p_{\a\dot{\b}} \,. 
\el 
In the COM frame, where $u^\mu = (1,0,0,0)$, 
$u^{\dot{\a}\a}$ is the identity matrix. So, both $(p|u)$ and $(u|p)$ have the same matrix elements 
as $p_{\a\dot{\a}}$  in eq.\eqref{p-COM}.  But, now $(p|u)$ and $(u|p)$ are Lorentz covariant operators acting on spinors. 
Their eigenvalues, which coincide when $E \pm p$ in the COM frame with 
$p^\mu = (E, 0,0,p)$,   
can now be regarded as Lorentz invariant quantities. 

Bearing in mind the frame dependence, we decompose each massive momentum as 
\bl
p = \ket{\l} \sbra{\tilde{\l}}  + \ket{\eta'} \sbra{\tilde{\eta}'} \,, 
\quad 
\lan \l \eta' \ra = m = [ \tilde{\eta}' \tilde{\lambda}] \,.
\el
The first piece corresponds to the large eigenvalue ($E+p$) and the second piece to the small one ($E-p$). It is often convenient to rescale the sub-leading piece 
by $(\eta', \tilde{\eta}') = m (\eta, \tilde{\eta})$, 
\bl
p = \ket{\l} \sbra{\tilde{\l}}  + m^2 \ket{\eta} \sbra{\tilde{\eta}} \,, 
\quad 
\lan \l \eta \ra = 1 = [ \tilde{\eta} \tilde{\lambda}] \,,
\label{p-rescaled-1}
\el
or, to discuss many particles at once, 
\bl
p_i = \ket{i} \sbra{i}  + m^2 \ket{\underline{i}} \sbra{\underline{i}} \,, 
\quad 
\lan i  \underline{i} \ra = 1 = [ \underline{i} i] \,.
\label{p-rescaled-2}
\el
In this notation, the definition of the HE limit can be rewritten as \cite{Arkani-Hamed:2017jhn} 
\bl
\frac{\lan i j \ra}{\sqrt{m_i m_j}} \gg 1 \,,
\quad 
\sqrt{m_i m_j} [ \underline{i} \underline{j}] \ll 1 \,.
\el

\paragraph{Explicit form of helicity spinor variables}


The spinor helicity variable $\l_\a{}^I$ is defined up to actions of the $SL(2,\IC)$ Lorentz group and the $SU(2)$ Little group. For numerical computations, it might be useful to have a prescription to fix both group actions. 

To fix the Lorentz group action, we choose a Lorentz frame (the COM or some other) with a time-like unit vector $u^\mu$. In the $u^\mu = (1,0,0,0)$ frame, we write $p^\mu = (E,\vec{p})$ with $E>0$
and introduce the notations
\bl
\vec{p} = p  \hat{n}  \,, 
\quad 
p = |\vec{p}| \,,
\quad 
\hat{n} \cdot \hat{n} = 1 \,. 
\el 
Choosing a Lorentz frame breaks $SL(2,\IC)$ to $SU(2)$ acting on the 3d space orthogonal to $u^\mu$. 
We temporarily introduce notations adjusted for this $SU(2)$.  
The round ket $\fket{v}$ denotes an $SU(2)$ spinor and $\fbra{v}$ denotes 
the Hermitian conjugagte of $\fket{v}$.  

We start by the familiar eigenvalue problem in SU(2):
\bl
(\hat{n} \cdot \vec{\s}) \, \fket{n^\pm} = \pm \fket{n^\pm} \,, 
\quad 
\hat{n} = (\sin\theta \cos\phi, \sin\theta \sin\phi, \cos\theta)\,. 
\el
We fix the phase ambiguity for the normalized eigenvectors $\fket{n^\pm}$ by setting 
\bl
\fket{{n}^+} := 
\begin{pmatrix} 
\cos \frac{\th}{2} \\ e^{i \phi} \sin \frac{\th}{2} 
\end{pmatrix}
\,, 
\quad 
\fket{{n}^-} := \begin{pmatrix} 
 - e^{-i \phi} \sin \frac{\th}{2} \\ \cos \frac{\th}{2} 
 \end{pmatrix}
\,.
\el
In terms of these SU(2) spinors, we may write
\bl
p_\mu \s^\mu &= E - \vec{p} \cdot \vec{\s} = (E - p) |n^+) (n^+| + (E + p) |n^-)(n^-| \,. 
\el
Comparing it with the Lorentz covariant expression (with $I \in \{ + , -\}$), 
\bl
p_\mu \s^\mu = \ket{\l^+} \sbra{\tilde{\l}_+} + \ket{\l^-} \sbra{\tilde{\l}_-} \,,
\el
leads to the identification
\bl
\ket{\l^\pm} = \sqrt{E\mp p} |n^\pm) \,,
\quad 
\sbra{\tilde{\l}_\pm} = \sqrt{E\mp p} (n^\pm| \,.
\el
To make contact with the HE limit, 
we make simple replacements to recover eq.\eqref{p-rescaled-1}:
\bl
\ket{\l^+} \rightarrow m \ket{\eta} \,,
\;\;
 \sbra{\tilde{\l}_+} \rightarrow m \sbra{\tilde{\eta}} \,,
\qquad
\ket{\l^-} \rightarrow  \ket{\l} \,,
\;\;
 \sbra{\tilde{\l}_-} \rightarrow \sbra{\tilde{\l}} \,. \label{eq:s-hHEreduction}
\el

\paragraph{HE limit of 3pt amplitudes} 
We use the decomposition eq.\eqref{p-rescaled-2} to examine the HE limit of the 3pt amplitudes with two massive particles of the same mass and spin coupled to a massless particle. Without loss of generality, we assume that the massless particle has positive helicity.

It is well-known that the 3pt amplitude for three massless particle can be non-vanishing 
only if the momenta are complex valued and either $\ket{1} \propto \ket{2} \propto \ket{3}$ or $\sket{1} \propto \sket{2} \propto \sket{3}$ holds. We cover the two cases separately. 

\paragraph{Case I : $\ket{1} \propto \ket{2} \propto \ket{3}$.}

Momentum conservation requires that
\bl
0 = \ket{3} \sbra{3} + \ket{1} \sbra{1} +\ket{2} \sbra{2} 
+  m^2 \left(\ket{\underline{1}} \sbra{\underline{1}} 
+ \ket{\underline{2}} \sbra{\underline{2}} \right) \,.
\label{3pt-p-cons}
\el
When $[13]$, $[23]$ and $[12]$ are all comparable and much bigger than $m$, 
up to $\CO(m)$ corrections, 
\bl
\ket{1} \approx - \frac{[32]}{[12]} \ket{3} \,,
\quad 
 \ket{2} \approx - \frac{[31]}{[21]} \ket{3} \,.
\el
To the leading order in $m$, the $x$ factor becomes
\bl
x = \frac{\sbra{3} \bar{p}_1 \ket{\z}}{m \lan 3 \z \ra} 
\approx \frac{[31]\lan 1 \zeta \ra}{m \lan 3 \z \ra}  
\approx \frac{[23][31]}{m [12]} \,.
\el

Recall that the 3pt minimal coupling is
\bl
A_3^\text{(min)} = \frac{m^{h-2s}}{M^{h-1}} x^h \lan \bf{21} \ra^{2s} \,, 
\label{3pt-elem}
\el
where $M$ is a fixed dimensionful coupling such as the Planck mass. In the HE limit, the spin of the massive particles effectively split into helicities of massless particles. In one extreme case with $h_1 = + s$ and $h_2= -s$, we recover the massless 3pt amplitude $A_3(k_1^+,k_2^-,k_3^+)$ as follows
\bl
A_3^\text{(min)} \to 
\frac{m^{h-2s}}{M^{h-1}} x^h \lan 2^- 1^+ \ra^{2s} 
= \frac{1}{M^{h-1}} (mx)^h \lan 2 \underline{1} \ra^{2s} 
\approx \frac{1}{M^{h-1}} \left(\frac{[23][31]}{[12]}\right)^h \left( \frac{[31]}{[23]} \right)^{2s} \,.
\el
Exchanging particle 1 and 2 gives $A_3(k_1^-,k_2^+,k_3^+)$.

\paragraph{Case II : $\sket{1} \propto \sket{2} \propto \sket{3}$.} 
We begin again with the momentum conservation eq.\eqref{3pt-p-cons}. 
When $\lan 13 \ra$, $\lan 23\ra$ and $\lan 12 \ra$ are all comparable 
and much bigger than $m$, up to $\CO(m)$ corrections,
\bl
\sbra{1} \approx - \frac{\lan 23 \ra}{\lan 21 \ra} \sbra{3} \,,
\quad 
\sbra{2} \approx - \frac{\lan 13 \ra}{\lan 12 \ra} \sbra{3} \,.
\el
The $x$ factor is approximately, 
\bl
x = \frac{m [3 \z]}{ \bra{3} p_1 \sket{\z}} 
\approx \frac{m [3\z]}{\lan 31 \ra [1 \z]} 
\approx \frac{m \lan 12 \ra}{\lan 23 \ra \lan 31 \ra} \,.
\el
Starting from the minimal coupling eq.\eqref{3pt-elem} and take the case 
with $h_1 = -s$ and $h_2 = -s$, we recover the massless amplitude  $A_3(k_1^-,k_2^-,k_3^+)$, 
\bl
A_3^\text{(min)} \to \frac{m^{h-2s}}{M^{h-1}} x^h \lan 2^- 1^- \ra^{2s} 
\approx 
\frac{m^{2h-2s}}{M^{h-1}} \left( \frac{\lan 12 \ra}{\lan 23 \ra \lan 31 \ra} \right)^h \lan 21 \ra^{2s} \,.
\el

\subsection{Spin operator} \label{sec:spin-op}
Pauli-Lubanski pseudovector can be considered as an operator acting on the space of spinors. 
The general definition of the operator ($\e^{0123} = +1 = - \e_{0123}$) 
\bl
W_\m &:= m S_\m = - \frac{1}{2} \e_{\m\n\l\s} P^\n J^{\l\s} 
\label{eq:PLvecdef}
\el
and the definition of $J^{\m\n}$ for spinors in eq.\eqc{eq:SL2Celemdef} give 
\bl
m \left( S_\m \right)_{\a}^{~\b} = 
\frac{1}{4} \left[ \s_\m (p \cdot \bar{\s}) - (p \cdot \s) \bar{\s}_\m \right]_{\a}^{~\b} \,.\label{eq:spinopang}
\el
Its action on the spinor-helicity variable $\l_\a{}^I$ for the momentum 
$p_{\a\dot{\a}} = \l_\a{}^I \tilde{\l}_{I\dot{\a}}$ is
\bl
m (S_\m \l^I)_\a &= \frac{1}{4} \left[ m \s_\m \tilde{\l}^I - (p \cdot \s) \bar{\s}_\m \l^I \right]_\a \,.
\el
An analogous statement for the dotted spinors is
\bl
m \left( S_\m \right)^{\dot{\a}}_{~\dot{\b}} 
&= - \frac{1}{2} \e_{\m\n\l\s} P^\n \left( J^{\l\s} \right)^{\dot{\a}}_{~\dot{\b}} 
= - \frac{1}{4} \left[ \bar{\s}_\m (p \cdot \s) - (p \cdot \bar{\s}) \s_\m \right]^{\dot{\a}}_{~\dot{\b}} \,.
\label{eq:spinopsq}
\el
In the helicity basis defined in section \ref{sec:HE}, 
\bl
n^\m S_\m \l_\a^{\pm} &= \mp \frac{E}{2m} \l_\a^{\pm}
\\ n^\m S_\m \bar{\l}^{\dot\a \pm} &= \mp \frac{E}{2m} \bar{\l}^{\dot\a \pm}
\el
where $n^\m = (0,\vec{n})$ is the unit spatial vector pointed towards the direction of particle's momentum. Note that $n^\m S_\m = - (\vec{n} \cdot \vec{S})$; it is natural to associate $\tilde{\l}^{\dot{\a}}_{-}$ to positive helicity states and $\l_\a^{-}$ to negative helicity states. The signs are flipped when ``bra'' vectors are used.
\bl
\l^{\a \pm} n^\m S_\m &= \pm \frac{E}{2m} \l^{\a \pm}
\\ \bar{\l}_{\dot\a}^{\pm} n^\m S_\m &= \pm \frac{E}{2m} \bar{\l}_{\dot\a}^{\pm}
\el

The spin operator for multiple spinor indices follows from the Lie algebra, 
\bl
\left( J^{\m\n} \right)_{\a_1 \a_2 \cdots \a_{2s}}^{~~~~~~\b_1 \b_2 \cdots \b_{2s}} &= \sum_i \left( J^{\m\n} \right)_{\a_i}^{~\b_i} \bar{\iden}_i  \,,
\label{eq:largespinop}
\el
where $\bar{\iden}_i$ is defined as $\bar{\iden}_i = \delta_{\a_1}^{\b_1} \cdots \delta_{\a_{i-1}}^{\b_{i-1}} \delta_{\a_{i+1}}^{\b_{i+1}} \cdots \delta_{\a_{2s}}^{\b_{2s}}$. 
When acting exclusively on the totally symmetric representation, the spinor operator is effectively proportional to the spin, 
\bl
\left( S_\m \right)_{\a_1 \a_2 \cdots \a_{2s}}^{~~~~~~\b_1 \b_2 \cdots \b_{2s}} &= \sum_i \left( S_\m \right)_{\a_i}^{~\b_i} \bar{\iden}_i 
\;\; \sim \;\; 
2s \left( S_\m \right)_{\a_1}^{~\b_1} \bar{\iden}_1 \label{eq:spinophs} \,.
\el
A similar equivalence works for the dotted spinors as well.

\subsection{Polarisation}

\paragraph{Massless case} 
We take the following definitions for the polarisation vectors of photons, 
\bl
\varepsilon^{+}_{\m}(k) := \frac{\sbra{k} \bar{\s}_\m \ket{\z} }{\sqrt{2} \lan k \z \ra}
\,,
\quad
\varepsilon^{-}_{\m}(k) := \frac{\bra{k} {\s}_\m \sket{\z} }{\sqrt{2} [ k \z ]}
\,, \label{eq:spin1polartenmassless}
\el
where $\z$ parametrises the gauge redundancy. The polarisation vectors satisfy  
\bl 
\varepsilon^{\pm} \cdot (\varepsilon^{\pm})^\ast = -1 
\quad 
\mbox{and}
\quad  
\varepsilon^{\pm} \cdot (\varepsilon^{\mp})^\ast = 0 \,.
\el
Alternatively, in the bi-spinor notation, 
\bl
\varepsilon^{+}(k) = \sqrt{2} \frac{\sket{k} \bra{\z} }{\lan k \z \ra}
\,,
\quad
\varepsilon^{-}(k) = \sqrt{2} \frac{\ket{k} \sbra{\z} }{[ k \z ]} \,.
\el
The polarisation tensors for higher-spin particles are constructed as symmetric products of eq.\eqc{eq:spin1polartenmassless}.

\paragraph{Massive case}

For a massive spin 1 particle, we adopt the following definition for the polarisation vector:
\bl 
\varepsilon^{IJ}_\m(p) &:= \frac{1}{\sqrt{2} m} \bra{p^{\{ I}} \s_\m \sket{p^{J \} }} 
= \frac{1}{2 \sqrt{2} m} \left( \bra{p^{I}} \s_\m \sket{p^{J}} + \bra{p^{J}} \s_\m \sket{p^{I}} \right) 
\,. \label{eq:spin1polarten}
\el
They are orthonormal in the sense that  
\bl
\varepsilon^{IJ} \cdot (\varepsilon^{KL})^\ast = - \half(\delta^{I}_{K} \delta^{J}_{L} + \delta^{I}_{L} \delta^{J}_{K}) \,,
\quad
\sum_{I,J} \varepsilon^{IJ}_\m (\varepsilon^{IJ}_\n)^* = - \left( \eta_{\m\n} - \frac{p_\m p_\n}{m^2} \right) \,.
\el
The reduction of massive polarisation vectors to the massless case in the HE limit can be seen by adopting the helicity basis introduced in section \ref{sec:HE}. Inserting the HE spinor-helicity variables eq.\eqc{eq:s-hHEreduction} into eq.\eqc{eq:spin1polarten} and using the defining relations for $\eta$ spinors $\la \l \eta \ra = [ \tilde{\eta} \tilde{\l} ] = 1$ to eliminate the Little group dependence on $\eta$ spinors gives the massless polarisation vectors eq.\eqc{eq:spin1polartenmassless} for the transverse polarisations $I=J=+$ or $I=J=-$.

The polarisation tensors for higher-spin particles are constructed in an analogous way to the massless case.

\section{The normalization of Gravitomagnetic Zeeman coupling}\label{FermiWalkerTransport} 
It is expected that the full gravitational potential $V$ will have ``scalar potential'' coupling $m\Phi$ and Zeeman-like coupling $\a \vec{S} \cdot \vec{B}$ with gravitomagnetic field $\vec{B} := \nabla \times \vec{A}$.
\begin{align}
V &:= m \Phi + \a \vec{S} \cdot \vec{B} \label{eq:gravHamilansatz}
\end{align}
The coefficient $\a$ will be fixed by requiring that the correct time evolution of the spin-operator $\vec{S}$ will be reproduced by the corresponding Hamiltonian. The natural evolution of spin vectors in general relativity required by the equivalence principle is described by what is known as the Fermi-Walker transport:\footnote{{This equation assumes that finite-size effects or tidal effects are negligible. When such effects cannot be neglected spin evolves according to a different set of equations known as Mathisson-Papapetrou-Dixon equations.}}
\begin{align}
\frac{D_F S^\m}{ds} = u^\n \nabla_\n S^\m + \e(u^\m a_\n - a^\m u_\n) S^\n = 0\,.
\end{align}
The vector $u^\m$ is the tangent vector of the curve $\g(s)$ along which $S^\m$ is transported, and is normalised by $u^\m u_\m = \e = \pm 1$. The acceleration vector $a^\m$ is defined as $a^\m := u^\n \nabla_\n u^\m$. Setting $u = \p_0$, Fermi-Walker transport for spin vector gives the following equation.
\begin{align}\label{FermiWalker}
\frac{D_F S^i}{ds} = \p_0 S^i + \G^{i}_{0j} S^j = 0
\end{align}
The Christoffel symbols up to $\CO(h)$ are given by, assuming stationary solutions, i.e. $\p_0 = 0$,
\bl
\bld
\G^{0}_{i0} &= \G^{0}_{0i} = \nabla \Phi
\\ \G^{i}_{00} &= \nabla \Phi
\\ \G^{i}_{0j} &= \G^{i}_{j0} = \frac{1}{2} (\p_j \CA^i - \p_i \CA^j) = - \frac{1}{2} \e^{ijk} ( \nabla \times \vec{\CA} )^k
\\ \G^{0}_{ij} &= - \frac{1}{2} (\p_i \CA^j + \p_j \CA^i)
\\ \G^{i}_{jk} &= - ( \delta^{ij} \Phi_{,k} + \delta^{ik} \Phi_{,j} - \delta^{jk} \Phi_{,i} )\,.
\eld
\el
Substituting the Christoffel symbols, eq.\eqc{FermiWalker} gives an analogue of Larmor precession in electrodynamics.
\begin{align}
\frac{\p}{\p t} \vec{S} = \frac{1}{2} \vec{S} \times \vec{\CB}
\end{align}
Since eq.\eqc{FermiWalker} must be reproduced from eq.\eqc{eq:gravHamilansatz} in the same way as Larmor precession is reproduced from Zeeman coupling, from the relations
\eq
 [S^i , S^j] = i \hbar \epsilon^{ijk} S^k,\quad \frac{\partial}{\partial t} \mathcal{O} = \frac{1}{i \hbar} [\mathcal{O} , H ]
\eqe
one can deduce that $\a = - \half$. 
\section{Some Details of the $t$-channel Matching of the Higher Spin Graviton Compton Amplitude}
Let's go back to eq.\eqref{eq:M_trial}, and take the $t$-channel residue.\footnote{We define $\sm \equiv s_m$, and $\um \equiv u_m$, also factors of $M_{pl}$ will be temporarily suppressed here for simplicity.} Exapanding $\mathcal{F} = \mathcal{F}_1 + \mathcal{F}_2$ yields:
\begin{equation}\label{eq:I_plus_II}
\begin{split}
Res[Ansatz]\Big|_{t=0}
&= - \frac{\MixLeft{3}{p_1}{2}^4}{\sm \um}\mathcal{F}_1^{2s} - \MixLeft{3}{p_1}{2}^2 K^2  \sum_{r=1}^{2s-1} r \binom{2s}{r+1} \mathcal{F}_1^{2s-r-1} \mathcal{F}_2^{r-1} \\
& \qquad - \frac{2s \MixLeft{3}{p_1}{2}^3}{\sm \um}\mathcal{F}_1^{2s-1}\Big( \frac{ C_{\AB{23}} + C_{\SB{23}}}{2} \Big)  \\
& \qquad - \frac{\MixLeft{3}{p_1}{2}^2}{\sm \um}\Big( \frac{ C_{\AB{23}} + C_{\SB{23}}}{2} \Big)^2 \sum_{r=2}^{2s} \binom{2s}{r} \mathcal{F}_1^{2s-r} \mathcal{F}_2^{r-2} \\
& \qquad +  \frac{\MixLeft{3}{p_1}{2}^2}{\sm} K  \Big( \frac{ C_{\AB{23}} + C_{\SB{23}} }{2} \Big) \sum_{r=1}^{2s-1} (r-1)\binom{2s}{r+1} \mathcal{F}_1^{2s-1-r} \mathcal{F}_2^{r-1} \\
\end{split}
\end{equation}
\noindent
where we have used
\begin{equation}\label{eq:Id_3P12_script_A2}
\begin{split}
\MixLeft{3}{p_1}{2}\mathcal{F}_2 = \frac{\AB{3 \boldsymbol{4}} \SB{2 \boldsymbol{1}}}{2m^2}(u-m^2) + \frac{\AB{3 \boldsymbol{1}} \SB{2 \boldsymbol{4}}}{2m^2}(s-m^2) + \frac{1}{2} ( C_{\AB{23}} + C_{\SB{23}} )
\end{split}
\end{equation}
\noindent
to cancel as much $\sm$ and $\um$ as possible until there is no more $\mathcal{F}_2$ in the leading term of each of the summation. Then we identify the last term in the first line of Eq\eqref{eq:I_plus_II} as $Poly$  and the piece that only carries the $\sm$ pole as $Pole_s$, so that:
\begin{equation}\label{eq:I_plus_II_2}
\begin{split}
Res[Ansatz]\Big|_{t=0}
&= - \frac{\MixLeft{3}{p_1}{2}^4}{\sm \um}\mathcal{F}_1^{2s} \\
& \qquad - \frac{2s \MixLeft{3}{p_1}{2}^3}{\sm \um}\mathcal{F}_1^{2s-1}\Big( \frac{ C_{\AB{23}} + C_{\SB{23}}}{2} \Big)  \\
& \qquad - \frac{\MixLeft{3}{p_1}{2}^2}{\sm \um}\Big( \frac{ C_{\AB{23}} + C_{\SB{23}}}{2} \Big)^2 \sum_{r=2}^{2s} \binom{2s}{r} \mathcal{F}_1^{2s-r} \mathcal{F}_2^{r-2} \\
& \qquad + Poly + Pole_s
\end{split}
\end{equation}
Now, we are free to write $\mathcal{F}_1$ in the terms other than $Poly$ and $Poly_s$ as:
\begin{equation}\label{eq:Curly_A1_anti-MHV}
\mathcal{F}_1 = F_1 +\frac{1}{2}(\tilde{F}_1 - F_1)
\end{equation}
\noindent or
\begin{equation}\label{eq:Curly_A1_MHV}
\mathcal{F}_1 = \tilde{F}_1 +\frac{1}{2}(F_1 - \tilde{F}_1)
\end{equation}
\noindent
such that the first term in the expansion of $- \frac{\MixLeft{3}{p_1}{2}^4}{\sm \um}\mathcal{F}_1^{2s}$ matches eq.\eqref{eq:anti_MHV_t_res} if we choose eq.\eqref{eq:Curly_A1_anti-MHV} and it matches eq.\eqref{eq:MHV_t_res} if we choose eq.\eqref{eq:Curly_A1_MHV}. First we consider the $\AB{23} = 0$ case, where $C_{\AB{23}} = 0$. Expanding $\mathcal{F}_1$ with eq.\eqref{eq:Curly_A1_anti-MHV} and apply 
\begin{equation}\label{eq:Id_k3P1k2_AT1_minus_A1}
\MixLeft{3}{p_1}{2} (\tilde{F}_1 - F_1) = \MixLeft{3}{p_1}{2}(F_2 - \tilde{F}_2) = -C_{\SB{23}} + C_{\AB{23}} 
\end{equation}
repeatedly until there is no more $\MixLeft{3}{p_1}{2}$ to be absorbed. On the other hand, since $r=1$ in the summation of $Pole_s$ is zero, we can further apply eq.\eqref{eq:Id_3P12_script_A2} to $Pole_s$ once more. We end up with:
\begin{equation}\label{eq:Smoke_Clear_expression_1}
\begin{split}
Res[Ansatz]\Big|_{\AB{23}=0} 
& \equiv - \frac{\MixLeft{3}{p_1}{2}^4}{\sm \um}F_1^{2s} \\
& \qquad + \Big\{ f_S(4) + g_S(2) - \frac{ \MixLeft{3}{p_1}{2} K^2 C_{\SB{23}} }{2} \binom{2s}{3} \mathcal{F}_1^{2s-3} \Big\} + Poly
\end{split}
\end{equation}
\noindent
where $f_S(n)$ is defined by:
\begin{equation}
\begin{split}
f_S(n) \equiv & - \frac{C_{\SB{23}}^4 }{2^n s_m u_m} \Big\{ \sum_{r=1}^{2s-n} \binom{2s}{r+n} \binom{r+n-1}{n}F_1^{2s-n-r}  \Big( \frac{\tilde{F}_1 - F_1}{2} \Big)^r + \sum_{r=1}^{2s-n}\binom{2s}{r+n} \mathcal{F}_1^{2s-r-n} \mathcal{F}_2^r \Big\} \\
& \qquad + \frac{ K C_{\SB{23}}^3}{2^{n-1} s_m} \sum_{r=1}^{2s-n} r \binom{2s}{r+n} \mathcal{F}_1^{2s-r-n} \mathcal{F}_2^r
\end{split}
\end{equation}
\noindent
and satisfies $f_S(n\geq 2s) = 0$. As long as $f_S(n)$ are present, there are still $\sm$ and $\um$ poles that should be further taken care of. This can be dealt with by applying the recursion relation
\begin{equation}\label{eq:monster_identity}
f_S(n) = f_S(n+2)(\tilde{F}_1- F_1)^2 + h_S(n) (\tilde{F}_1 - F_1) + g_S(n)(\tilde{F}_1- F_1)^2
\end{equation}
\noindent
until $f(n \geq 2s)$ such that the residue is completely local.\footnote{The recursion relation eq.\eqref{eq:monster_identity} is obtained by
$$C_{\SB{23}}\mathcal{F}_2\big|_{\AB{23} = 0} = -\MixLeft{3}{p_1}{2} (\tilde{F_1} - F_1)\mathcal{F}_2 = -(\tilde{F}_1 - F_1) \Big(u_mK + \frac{C_{\SB{23}}}{2} \Big).$$ And there are $\um$ present to cancel with the ones in the denominators, leaving only local terms.} Now we are only left with 
\begin{equation}\label{eq:Smoke_Clear_expression_2}
\begin{split}
Res[Ansatz]\Big|_{\AB{23}=0} 
&= -\frac{\MixLeft{3}{p_1}{2}^4}{\sm \um}F_1^{2s} + Poly \\
& \qquad + \Bigg\{ \sum_{r=0}^{ \lceil s \rceil - 3} h_{S}(4 + 2r) (\tilde{F}_1 - F_1)^{2r+1} +  \sum_{r=0}^{ \lceil s \rceil - 2} g_{S}(2 + 2r) (\tilde{F}_1 - F_1)^{2r} \\
& \qquad \quad - \frac{\MixLeft{3}{p_1}{2} K^2 C_{\SB{23}}}{2} \binom{2s}{3} \mathcal{F}_1^{2s-3} \Bigg\} \\
& \equiv - \frac{\MixLeft{3}{p_1}{2}^4}{\sm \um}F_1^{2s} + Poly + Poly_{\SB{23}}
\end{split}
\end{equation}
\noindent 
which is just the correct residue eq.\eqref{eq:anti_MHV_t_res} plus  pure polynomial terms.

For $\SB{23} = 0$,  we do not need to do the calculations again. Since we are using the variables $\mathcal{F}$, $\mathcal{F}_1$ and $\mathcal{F}_2$ that are symmetric under $F_1 \leftrightarrow \tilde{F}_1$ and $F_2 \leftrightarrow \tilde{F}_2$, we can just simply use the substitutions $F_1 \leftrightarrow \tilde{F}_1$ and $C_{\SB{23}} \leftrightarrow C_{\AB{23}}$. So the $\SB{23} = 0$ residue is:
\begin{equation}\label{eq:Smoke_Clear_SB23_is_0}
\begin{split}
Res[Ansatz]\Big|_{\SB{23}=0} 
&= -\frac{\MixLeft{3}{p_1}{2}^4}{\sm \um}\tilde{F}_1^{2s} + Poly \\
& \qquad + \Bigg\{ \sum_{r=0}^{ \lceil s \rceil - 3} h_{A}(4 + 2r) (F_1 - \tilde{F}_1)^{2r+1} +  \sum_{r=0}^{ \lceil s \rceil - 2} g_{A}(2 + 2r) (F_1 - \tilde{F}_1)^{2r} \\
& \qquad \quad - \frac{ \MixLeft{3}{p_1}{2} K^2 C_{\AB{23}} }{2} \binom{2s}{3} \mathcal{F}_1^{2s-3} \Bigg\} \\
& \equiv - \frac{\MixLeft{3}{p_1}{2}^4}{\sm \um} \tilde{F}_1^{2s} + Poly + Poly_{\AB{23}}
\end{split}
\end{equation}
We again end up with an expression whose leading term already matches the desired residue eq.\eqref{eq:MHV_t_res}. So, all we need to do to match all three channels is subtracting off
\begin{equation}\label{eq:Excess_t_channel}
\frac{ Poly + Poly_{\SB{23}} + Poly_{\AB{23}} }{t}
\end{equation}
\noindent
from $Ansatz$.

\section{Wilson coefficients for black holes} \label{sec:BHWilson}
The Wilson coefficients $C_\#$ for coupling of spin degrees of freedom to spacetime curvature have an interpretation as gravitational multipole moments generated by spin effects. For this purpose it is convenient to introduce the vector $a^\m := \frac{1}{m} S^\m$. The terms linear in $h_{\m\n}$ in the one-body effective action can be recast as follows.
\bl
\bld
L &= - \frac{\k m}{2} \sum_{n=0}^{\infty} \frac{C_{\text{ES}^{2n}}}{(2n)!} \left( - \left( - a \cdot \p \right)^{2} \right)^{n} u_\m u_\n h^{\m\n}
\\ & \phantom{==} + \frac{\k m}{2} \sum_{n=0}^{\infty} \frac{C_{\text{BS}^{2n+1}}}{(2n+1)!} \left( - \left( - a \cdot \p \right)^{2} \right)^{n} u_{(\m} \e_{\n)\a\b\g} u^{\a} a^{\b} \p^{\g} h^{\m\n} + \left( u \cdot \p \right) \left[ \cdots \right] + \CO(h^2)
\eld
\el
The notation $C_{\text{ES}^0} = C_{\text{BS}^1} = 1$ has been adopted to simplify the equations, and covariant SSC was used to express $S_{\m\n}$ as $S_{\m\n} = m \e_{\m\n\l\s} u^\l a^\s$. The round brackets on $\m$ and $\n$ indices on the second line indicates symmetrisation, i.e. $A_{(\m\n)} := \half (A_{\m\n} + A_{\n\m})$. When integrated on the worldline, the terms with $(u \cdot \p)$ can be converted to boundary terms which becomes irrelevant when trying to interpret this Lagrangian as the source term for $h_{\m\n}$. Upon integration by parts, this Lagrangian reduces to the following source term expression\footnote{The coupling constant $\k$ has been absorbed into definition of $h_{\m\n}$.}.
\bg
S_{int} = - \int d^4 x ~ \half h_{\m\n}(x) T^{\m\n}(x)
\\ \bld
T_{\m\n}(x) &= m \int ds \left[ u_\m u_\n \sum_{n=0}^{\infty} \frac{C_{\text{ES}^{2n}}}{(2n)!} \left( - \left( a \cdot \p \right)^{2} \right)^{n} \delta^4[x - x_{\text{wl}}(s) ]
\right.
\\ & \phantom{=} \left.\phantom{=} + u_{(\m} \e_{\n)\a\b\g} u^{\a} a^{\b} \p^{\g} \sum_{n=0}^{\infty} \frac{C_{\text{BS}^{2n+1}}}{(2n+1)!} \left( - \left( a \cdot \p \right)^{2} \right)^{n} \delta^4[x - x_{\text{wl}}(s) ] \right]
\eld
\eg
$x_{\text{wl}}(s)$ is the worldline of the particle, parametrised by $s$.

The solution for $h_{\m\n}$ can be constructed from Green's function method\cite{Vines:2017hyw}.
\bl
h_{\m\n}(x) &= \int d^4 y ~\CP_{\m\n}^{\l\s}(x-y) T_{\l\s}(y)
\\ \CP_{\m\n}^{\l\s}(x) &= 4 G \CP_{\m\n}^{\l\s} ~ \CG_{\text{ret}}(x)
\\ \CG_{\text{ret}}(x) &= \th(x^0) ~ \delta \left( \frac{x^2}{2} \right)
\el
The retarded scalar Green's function $\CG_{\text{ret}}(x)$ is given as the solution to the sourced wave equation $\square ~\CG_{\text{ret}}(x) = - 4 \pi \delta^4(x)$, and has Dirac delta values over the future-directed light cone. The tensor $\CP_{\m\n}^{\l\s} = \delta^{(\l}_{(\m}\delta^{\s)}_{\n)} - \half \eta_{\m\n} \eta^{\l\s}$ is the trace-reverser, which can be factored out to yield a simpler equation for trace-reversed graviton field $\bar{h}_{\m\n} = \CP_{\m\n}^{\a\b} h_{\a\b}$.
\bl
\bar{h}_{\m\n}(x) &= 4G \int d^4 y ~ \CG_{\text{ret}} (x - y) T_{\m\n}(y)
\el
Note that integration by parts identity $\int dy K(x-y) \frac{d}{dy} f(y-z) = \frac{d}{dx} \int dy K(x-y) f(y-z)$ for vanishing boundary contributions can be applied to pull out the derivatives on the source term. Setting the worldline of the particle to lie at the origin, i.e. $x_{\text{wl}}(s) = (s,\vec{0})$, the following expression for the trace-reversed graviton field is obtained.
\bl
\bld
\bar{h}_{\m\n}(x) &= u_\m u_\n \sum_{n=0}^{\infty} \frac{C_{\text{ES}^{2n}}}{(2n)!} \left( - \left( a \cdot \p \right)^{2} \right)^{n} \frac{4Gm}{r}
\\ &\phantom{=} + u_{(\m} \e_{\n)\a\b\g} u^{\a} a^{\b} \p^{\g} \sum_{n=0}^{\infty} \frac{C_{\text{BS}^{2n+1}}}{(2n+1)!} \left( - \left( a \cdot \p \right)^{2} \right)^{n} \frac{4Gm}{r}
\eld \label{eq:WilsoncoeffGeom}
\el
It is known that trace-reversed graviton field for exact Kerr geometry $\bar{h}^{\text{Kerr}}_{\m\n}$ can be put in the following form\cite{Vines:2017hyw}.
\bl
\bld
\bar{h}^{\text{Kerr}}_{\m\n}(x) &= u_\m u_\n \sum_{n=0}^{\infty} \frac{1}{(2n)!} \left( - \left( a \cdot \p \right)^{2} \right)^{n} \frac{4Gm}{r}
\\ &\phantom{=} + u_{(\m} \e_{\n)\a\b\g} u^{\a} a^{\b} \p^{\g} \sum_{n=0}^{\infty} \frac{1}{(2n+1)!} \left( - \left( a \cdot \p \right)^{2} \right)^{n} \frac{4Gm}{r}
\eld \label{eq:KerrGeom}
\el
Comparing eq.\eqc{eq:WilsoncoeffGeom} with eq.\eqc{eq:KerrGeom}, it can be concluded that Wilson coefficients $C_{\#}$ for black holes are unity.

\section{Spin-orbit factor corrections to polarisation tensor contractions} \label{sec:SOcorrections}
Define $p_a^\m = \frac{P_2^\m + P_1^\m}{2}$, $q^\m = P_1^\m - P_2^\m$. In terms of average momentum and momentum transfer, the polarisation tensors can be expressed as follows.
\bl
\varepsilon(P_2) &= \varepsilon(p_a) - \half q^\m \frac{\p}{\p p_a^\m } \varepsilon(p_a) + \cdots
\\ \varepsilon(P_1) &= \varepsilon(p_a) + \half q^\m \frac{\p}{\p p_a^\m } \varepsilon(p_a) + \cdots
\el
Since polarisation tensors are defined in some reference frame and then extended to arbitrary momentum by boosts for massive particles, the polarisation tensor $\varepsilon(p)$ can be schematically be written as follows.
\bl
\varepsilon(p) &= G(p;p_0) \varepsilon(p_0)
\el
Thus, the derivative on polarisation tensor can be represented as
\bl
\frac{\p}{\p p^\m} \varepsilon(p) &= \lim_{\delta p \to 0} \frac{G(p+\delta p;p_0) G^{-1}(p;p_0) - \iden}{\delta p} \varepsilon(p)
\el
In the non-relativistic limit with $p_0 = (m, \vec{0})$, the following relations can be derived which holds at linear order in momentum.
\bg
G(p;p_0) = e^{- i \vec{\l}(\vec{p}) \cdot \vec{K}} \simeq e^{\frac{i}{m} \vec{p} \cdot \vec{K}}
\\ \vec{K} = J^{i0} = S^{i0} 
\\ \frac{\p}{\p p^\m} \varepsilon(p) \simeq \lim_{\delta p \to 0} \frac{e^{\frac{i}{m} (\vec{p} + \delta \vec{p}) \cdot \vec{K}} e^{- \frac{i}{m} \vec{p} \cdot \vec{K}} - \iden}{\delta p} \varepsilon(p) \simeq \frac{i}{m} \vec{K} \varepsilon(p)
\eg
Using NW SSC $S^{\m\n}(p_\n + m \delta^0_{\n})=0$, the following relation can be derived for $S^{i0}$;
\bg
S^{i0} (p_0 + m) = - S^{ij}p_j = \e^{ijk} p^j S^k
\\ \vec{K} = S^{i0} = \frac{1}{p_0 + m} \vec{p} \times \vec{S} \simeq \frac{1}{2m} \vec{p} \times \vec{S}
\eg
Therefore, the derivative can be represented as follows in the non-relativistic limit.
\bl
q^\m \frac{\p}{\p p_a^\m } \varepsilon(p_a) &= \vec{q} \cdot \frac{\p}{\p \vec{p_a} } \varepsilon(p_a) \simeq \vec{q} \cdot \left( \frac{i}{2m^2} \vec{p_a} \times \vec{S_a} \right) \varepsilon(p_a)
\el
Summing up, the polarisation tensors can be represented as
\bl
\varepsilon(P_2) &= \varepsilon(p_a) + \frac{i}{4m_a^2} \vec{S_a} \cdot \left( \vec{p_a} \times \vec{q} \right) \varepsilon(p_a) + \cdots
\\ \varepsilon(P_1) &= \varepsilon(p_a) - \frac{i}{4m_a^2} \vec{S_a} \cdot \left( \vec{p_a} \times \vec{q} \right) \varepsilon(p_a) + \cdots
\\ \varepsilon^\ast (P_2) \varepsilon(P_1) &= \varepsilon^\ast(p_a) \left[ \iden - \frac{i}{2m_a^2} \left( \vec{p_a} \times \vec{q} \right) \cdot \vec{S_a} + \cdots \right] \varepsilon(p_a)
\el
For particle $b$, there is an additional sign factor due to definition of $\vec{q}$, which is consistent with the dictionary provided in \cite{Holstein:2008sx}.
\bl
\varepsilon^\ast (P_4) \varepsilon(P_3) &= \varepsilon^\ast(p_b) \left[ \iden + \frac{i}{2m_b^2} \left( \vec{p_b} \times \vec{q} \right) \cdot \vec{S_b} + \cdots \right] \varepsilon(p_b)
\el

\newpage

%


\bibliography{mybib}{}
\bibliographystyle{JHEP}

\end{document}